\pdfoutput=1
\documentclass[11pt,a4paper]{article}

\def\bea#1\eea{\begin{eqnarray}#1\end{eqnarray}}
\def\be#1\ee{\begin{equation}#1\end{equation}}
\def\ba#1\ea{\begin{align}#1\end{align}}

\usepackage{subcaption}

\usepackage[usenames,dvipsnames]{xcolor}

\usepackage[utf8]{inputenc}
\usepackage{jheppub}
\usepackage{amsmath}
\usepackage{multicol}
\usepackage{bbm}
\usepackage{enumerate}
\usepackage[inline]{enumitem}
\usepackage{bibentry}
\usepackage{comment}
\usepackage{amsthm}
\usepackage{mathrsfs}
\usepackage{upgreek}
\usepackage{amssymb}
\usepackage{bm}
\usepackage{setspace}
\usepackage{array,multirow,arydshln}
\usepackage{bigdelim}
\usepackage{scalerel}
\usepackage{diagbox}
\usepackage{caption}
\usepackage{tabularx}
\usepackage{empheq}

\usepackage{tikz}
\usepackage[compat=1.1.0]{tikz-feynman}

\usetikzlibrary{shapes.geometric,arrows,arrows.meta,decorations.pathmorphing,decorations.markings,patterns}

\newcommand*\widefbox[1]{\fbox{\hspace{0.5em}#1\hspace{0.5em}}}
\newcommand*\widefboxb[1]{{\setlength\fboxsep{8pt}\fbox{\hspace{2em}#1\hspace{2em}}}}
\newcommand*\widefboxc[1]{{\setlength\fboxsep{8pt}\fbox{\hspace{1em}#1\hspace{1em}}}}

\newcommand*{\halfway}{0.5*\pgfdecoratedpathlength+4pt}
\tikzset{vertex/.style={inner sep=0,minimum size=3pt,circle,fill}}

\def\<{\langle}
\def\>{\rangle}

\def\yz#1\yz {{\color{blue} [[YZ: #1]] }}

\usepackage{overpic}
\preprint{UUITP-52/19, LCTP-19-35}

\title{On Positive Geometry and Scattering Forms for Matter Particles}

\author[a]{Aidan Herderschee,}
\emailAdd{aidanh@umich.edu}

\author[b,c]{Song He,}
\emailAdd{songhe@itp.ac.cn}

\author[d]{Fei Teng}
\emailAdd{fei.teng@physics.uu.se}

\author[b,e,c]{and Yong Zhang}
\emailAdd{yongzhang@itp.ac.cn}

\affiliation[a]{Leinweber Center for Theoretical Physics, \\
Randall Laboratory of Physics, Department of Physics, \\
University of Michigan, Ann Arbor, MI 48109, USA}

\affiliation[b]{CAS Key Laboratory of Theoretical Physics, Institute of Theoretical Physics, Chinese Academy of Sciences, Beijing 100190, China}

\affiliation[c]{School of Physical Sciences, University of Chinese Academy of Sciences, No.19A Yuquan Road, Beijing 100049, China}

\affiliation[d]{Department of Physics and Astronomy, Uppsala University, 75108 Uppsala, Sweden}

\affiliation[e]{Perimeter Institute for Theoretical Physics, Waterloo, ON N2L 2Y5, Canada}

\date{\today}

\abstract{We initiate the study of positive geometry and scattering forms for tree-level amplitudes with matter particles in the  (anti-)fundamental representation of the color/flavor group. As a toy example, we study the bi-color scalar theory, which supplements the bi-adjoint theory with scalars in the (anti-)fundamental representations of both groups. Using a recursive construction we obtain a class of unbounded polytopes called open associahedra (or associahedra with certain facets at infinity) whose canonical form computes amplitudes in bi-color theory, for arbitrary number of legs and flavor assignments. 
In addition, we discuss the duality between color factors and wedge products, or ``color is kinematics”, for amplitudes with matter particles as well.
}

\begin{document}

\maketitle
\flushbottom

%\section{Introduction}

\section{Invitation: bi-color scalars and open associahedra}

In~\cite{Arkani-Hamed:2017mur}, a novel geometric understanding, which resembles the amplituhedron for the planar ${\cal N}=4$ SYM \cite{Arkani_Hamed_2014,Arkani_Hamed_2018}, has been proposed for scattering amplitudes in various massless theories in general spacetime dimension. The key idea is, instead of considering amplitudes as functions, they are most naturally thought of as differential forms on the kinematic space spanned by Mandelstam variables. In particular, tree-level amplitudes of bi-adjoint $\phi^3$ theory are given by the \emph{canonical form}~\cite{Arkani-Hamed:2017tmz} of a classic polytope -- the associahedron, living naturally in a subspace of the kinematic space. In this so-called \emph{ABHY (Arkani-Hamed, Bai, He, Yan) realization} of the \emph{kinematic associahedron}, the usual Feynman-diagram expansion in terms of planar cubic tree graphs corresponds to a particular triangulation, and the geometric picture allows us to see hidden symmetry and obtain new formula, even for planar $\phi^3$ amplitudes~\cite{He:2018svj}.

More generally, for tree amplitudes in any massless theories with adjoint particles, such as Yang-Mills theory (YM) or ${\rm U} (N)$ non-linear sigma models (NLSM), the scattering forms on kinematic space are dual to the fully color-dressed amplitudes despite having no explicit color factors. This is due to the fact that wedge products of Mandelstam variables in the scattering forms satisfy the same Jacobi relations as color factors, a statement dubbed ``Color is Kinematics"; relatedly, the usual trace decomposition was shown to be equivalent to pullback of scattering forms to subspace for the associahedron of a planar ordering. All these scattering forms are $d\log$ forms for $\phi^3$ diagrams dressed with kinematic numerators, and requiring them to be well-defined on the projectivized kinematic space leads to a  geometric origin for color-kinematics duality.

A natural question is how general is this geometric picture for scattering amplitudes. For example, can it be extended to more general theories, such as QCD, which have particles in the (anti-)fundamental representations? Since we work in general dimensions where mass can be obtained via dimensional reduction, it is convenient to restrict to the massless case.\footnote{More precisely, the results in this paper will not change if fundamental representation particles acquire mass from a dimensional reduction.} Recently there has been progress in the study of QCD amplitudes in the context of color-kinematics duality and double copy of Bern-Carrasco-Johansson (BCJ)~\cite{Bern:2008qj}. In particular, a generalization of bi-adjoint $\phi^3$ theory has been proposed, which includes scalars in the (anti-)fundamental representations of both groups, \emph{e.g.,} ${\rm U} (N)\times {\rm U}(N')$ \cite{Naculich_2014}. We use $[T^a,T^b]=\tilde{f}^{abc}T^c$ and $\text{tr}(T^aT^b)=\delta^{ab}$ as the normalization of Lie algebra generators and structure constants. The Lagrangian reads:
\begin{align}\label{eq:bicolor}
\mathcal{L}_{\phi^3}=&\;\frac{1}{2}\partial_\mu\phi^{aA}\partial^{\mu}\phi^{aA}+\frac{\lambda}{6}\tilde{f}^{abc}\tilde{f}^{ABC}\phi^{aA}\phi^{bB}\phi^{cC}\nonumber\\
&+\sum_{r=1}^{n_f}\left[\partial_\mu(\varphi_r)_{\underline{iI}}\partial^\mu(\varphi^*_r)^{\overline{iI}}+\lambda\phi^{aA}(\varphi^*_r)^{\overline{iI}}(T^a)_{\underline{i}}^{\overline{j}}(T^A)_{\underline{I}}^{\overline{J}}(\varphi_r)_{\underline{jJ}}\right],
\end{align}
where $\lambda$ is an arbitrary coupling constant, $r$ is the flavor index of the complex scalar $\varphi$, and $n_f$ is the total number of flavors. The real scalar $\phi$ carries adjoint representation indices $(a,A)$ of the symmetry group $U(N)\times U(N')$, while $\varphi$ and $\varphi^*$ carry fundamental representation indices $(\underline{i},\underline{I})$ and anti-fundamental ones $(\overline{i},\overline{I})$ respectively. More details will be given in section~\ref{sec24}. %and $\lambda$ is an arbitrary coupling constant.  
Among other things, amplitudes of such \emph{bi-color scalars} play an important role for KLT-type double copy relations for QCD amplitudes~\cite{Brown:2018wss,Johansson:2019dnu}, just as the original KLT relations have the bi-adjoint $\phi^3$ amplitudes as the central object~\cite{Kawai:1985xq,Cachazo:2013iea}. In this paper,  we initiate systematic studies of a geometric picture for amplitudes with \mbox{(anti-)}fundamental particles by considering tree amplitudes of the bi-color scalar theory. There is a direct generalization of the kinematic associahedron, which computes bi-color scalar amplitudes, and the statement ``color is kinematics'' also extends to these cases. 

Recall that for any assignment of adjoint and (anti-)fundamental particles, flavor decomposition of the amplitude for both $U(N)$ and $U(N')$ lead to a matrix of bi-color scalar (double-partial) amplitudes $m[\alpha|\beta]$, where $\alpha$ and $\beta$ denote the orderings; $m[\alpha|\beta]$ is given by a sum of trivalent tree graphs, which are determined by the collection of all poles compatible with both orderings and the flavor structure for particles in the (anti-)fundamental. For simplicity, we restrict ourselves to the diagonal case with $\alpha=\beta=(12\cdots n)$, $A_n:= m[12 \cdots n\,|\,12 \cdots n]$, since more general cases can be viewed as the intersection of diagonal cases~\cite{Cachazo:2013iea, Brown:2018wss, Mizera_2017, Frost_2018}. In the special case with only bi-adjoint particles, the amplitude is given by the sum of all (Catalan number of) planar cubic tree graphs, which contain all $n(n{-}3)/2$ planar variables since there is no restriction from flavor structures. As we will review shortly, the bi-adjoint $\phi^3$ amplitude is given by the canonical form of a $(n{-}3)$-dim associahedron in kinematic space; the latter is the intersection of (1) the top-dimensional cone with all planar variables being positive, and (2) an $(n{-}3)$-dim subspace given by all non-adjacent $s_{i,j}$ with $1\leqslant i<j<n$ being negative constants ($n(n{-}3)/2-(n{-}3)=(n{-}2)(n{-}3)/2$ constraints in total)~\cite{Arkani-Hamed:2017mur}. 

As a main result of this paper, which will be presented in section \ref{posgeoinkinspace}, we obtain any general bi-color amplitude $A_n$ as the canonical form of an \emph{open associahedron}, \emph{i.e.} an associahedron with faces sent to infinity. In fact, such open associahedra already appear for off-diagonal $m[\alpha|\beta]$ for bi-adjoint cases~\cite{Arkani-Hamed:2017mur}, and our results are a generalization in the presence of $k$ pairs of (anti-)fundamental particles with distinct flavors. Out of all the $n(n{-}3)/2$ planar poles, only $\mathfrak{N}$ of them are allowed by the flavor structure and color ordering while the remaining ones are set to positive constants. Almost identical to the bi-adjoint case, we can construct any open associahedron as the intersection of an $\mathfrak{N}$-dim positive cone with $\mathfrak{N}\leqslant n(n{-}3)/2$ and a $(n{-}3)$-dim subspace given by $\mathfrak{N}-(n{-}3)$ constraints that are still in the form of setting certain Mandelstam variables to negative constants. Very interestingly, these constraints can be given by a recursive procedure that generalizes the ``inverse soft construction'' of the bi-adjoint associahedra. We can fix the entire open associahedra through the analysis of two factorization channels (facets) at each step. The basic observation we have is that the pullback of \emph{planar scattering form} to the subspace gives the canonical form of open associahedron, which consists of only those Feynman diagrams allowed by the flavor structure.

Let's present a few explicit examples to illustrate the simplicity of our construction, and the detailed discussion will be given in section~\ref{sec:reccontr}.  We put labels $1,2,\ldots, n$ on the boundary of a disk, and represent the flavor structure by connecting each pair of particles in the (anti-)fundamental by a directed line. For example, for $n=4$ we have 
%the following
three possibilities with $k=0,1,2$.
Note that $k=1$ case is identical to the bi-adjoint ($k=0$) case, and both consist of two Feynman diagrams with $s=s_{1,2}$, $t=s_{2,3}$ channels,
\begin{align} \label{4ptttt}
&A_4\left[\begin{tikzpicture}[baseline={([yshift=-1ex]current bounding box.center)},every node/.style={font=\footnotesize,},dir/.style={decoration={markings, mark=at position \halfway with {\arrow{Latex}}},postaction={decorate}},scale=0.7]
\draw [thick] (0,0) circle (1cm);
\node (p1) at (-120:1) [vertex,label={[label distance=-2pt]240:{$1$}}] {};
\node (p3) at (120:1) [vertex,label={[label distance=-2pt]120:{$2$}}] {};
\node (p4) at (60:1) [vertex,label={[label distance=-2pt]60:{$3$}}] {};
\node (p6) at (-60:1) [vertex,label={[label distance=-2pt]-60:{$4$}}] {};
\end{tikzpicture}\right]
=\frac{1}{s}+\frac{1}{t}\,,\qquad 
A_4\left[\begin{tikzpicture}[baseline={([yshift=-1ex]current bounding box.center)},every node/.style={font=\footnotesize,},dir/.style={decoration={markings, mark=at position \halfway with {\arrow{Latex}}},postaction={decorate}},scale=0.7]
\draw [thick] (0,0) circle (1cm);
\node (p1) at (-120:1) [vertex,label={[label distance=-2pt]240:{$1$}}] {};
\node (p3) at (120:1) [vertex,label={[label distance=-2pt]120:{$2$}}] {};
\node (p4) at (60:1) [vertex,label={[label distance=-2pt]60:{$3$}}] {};
\node (p6) at (-60:1) [vertex,label={[label distance=-2pt]-60:{$4$}}] {};
\draw [thick,dir] (p1.center) to [bend right=20] (p3.center);
\end{tikzpicture}\right]
=
\frac{1}{s}+\frac{1}{t}\,. 
\end{align}
The disk graphs stand for the sum of all Feynman diagrams with cyclically ordered fundamental, anti-fundamental, and adjoint states. The corresponding associahedron is an interval given by:
%Here we always keep the orderings in both two color groups as the canonical ones $1,2,\cdots,n$. Besides, we assume different pairs of bi-fundamental scalars are of different flavors.   For 4 points, as we all know, if all of them are bi-adjoint scalars, there are 2 Feynman diagrams and the corresponding associahedron is an interval,
\begin{align}
&\pmb{\mathcal{A}}_4\!\left[\,
\begin{tikzpicture}[baseline={([yshift=-1ex]current bounding box.center)},every node/.style={font=\footnotesize,},dir/.style={decoration={markings, mark=at position \halfway with {\arrow{Latex}}},postaction={decorate}},scale=0.7]
\draw [thick] (0,0) circle (1cm);
\node (p1) at (-120:1) [vertex,label={[label distance=-3pt]240:{$1$}}] {};
\node (p3) at (120:1) [vertex,label={[label distance=-3pt]120:{$2$}}] {};
\node (p4) at (60:1) [vertex,label={[label distance=-3pt]60:{$3$}}] {};
\node (p6) at (-60:1) [vertex,label={[label distance=-3pt]-60:{$4$}}] {};
\end{tikzpicture}
\,\right]=
\pmb{\mathcal{A}}_4\!\left[\,\begin{tikzpicture}[baseline={([yshift=-1ex]current bounding box.center)},every node/.style={font=\footnotesize,},dir/.style={decoration={markings, mark=at position \halfway with {\arrow{Latex}}},postaction={decorate}},scale=0.7]
\draw [thick] (0,0) circle (1cm);
\node (p1) at (-120:1) [vertex,label={[label distance=-3pt]240:{$1$}}] {};
\node (p3) at (120:1) [vertex,label={[label distance=-3pt]120:{$2$}}] {};
\node (p4) at (60:1) [vertex,label={[label distance=-3pt]60:{$3$}}] {};
\node (p6) at (-60:1) [vertex,label={[label distance=-3pt]-60:{$4$}}] {};
\draw [thick,dir] (p1.center) to [bend right=20] (p3.center);
\end{tikzpicture}
\,\right]=\{s>0, t>0, u=-s-t = - c\}\,,
\end{align}
where $c$ is a positive constant. The pullback of the planar form $d\log \frac{s}{t}|_{s+t=c}=d s (\frac 1 s+\frac 1 t)$ is the canonical form of the interval, which gives the amplitude. However, for the $k=2$ case with {\it e.g.} \{1,2\} and \{3,4\} as two pairs of bi-fundamental scalars, the $t$-pole is forbidden by conservation of flavors and there is only one Feynman diagram left,
\begin{align} 
A_4\!\left[\begin{tikzpicture}[baseline={([yshift=-1ex]current bounding box.center)},every node/.style={font=\footnotesize,},dir/.style={decoration={markings, mark=at position \halfway with {\arrow{Latex}}},postaction={decorate}},scale=0.7]
\draw [thick] (0,0) circle (1cm);
\node (p1) at (-120:1) [vertex,label={[label distance=-3pt]240:{$1$}}] {};
\node (p3) at (120:1) [vertex,label={[label distance=-3pt]120:{$2$}}] {};
\node (p4) at (60:1) [vertex,label={[label distance=-3pt]60:{$3$}}] {};
\node (p6) at (-60:1) [vertex,label={[label distance=-3pt]-60:{$4$}}] {};
\draw [thick,dir] (p1.center) to [bend right=20] (p3.center);
\draw [thick,dir] (p4.center) to [bend right=20] (p6.center);
\end{tikzpicture}\right]
=\frac{1}{s}\,,
\qquad
\pmb{\mathcal{A}}_4\!\left[\,
\begin{tikzpicture}[baseline={([yshift=-1ex]current bounding box.center)},every node/.style={font=\footnotesize,},dir/.style={decoration={markings, mark=at position \halfway with {\arrow{Latex}}},postaction={decorate}},scale=0.7]
\draw [thick] (0,0) circle (1cm);
\node (p1) at (-120:1) [vertex,label={[label distance=-3pt]240:{$1$}}] {};
\node (p3) at (120:1) [vertex,label={[label distance=-3pt]120:{$2$}}] {};
\node (p4) at (60:1) [vertex,label={[label distance=-3pt]60:{$3$}}] {};
\node (p6) at (-60:1) [vertex,label={[label distance=-3pt]-60:{$4$}}] {};
\draw [thick,dir] (p1.center) to [bend right=20] (p3.center);
\draw [thick,dir] (p4.center) to [bend right=20] (p6.center);
\end{tikzpicture}
\,\right]= \{s>0,t=b>0\}\,.
\end{align} 
Here we see the simplest example of open associahedron, a half line from $s=0$ to infinity, which corresponds to setting the other boundary $t=0$ to infinity. The pullback of $d\log s/t$ to the space gives the canonical form $d\log s=d s/s$ (and the correct amplitude). In this case, the forbidden pole $t$ is set to a positive constant $b$, and we have only $\mathfrak{N}=1$ pole $s$ so there is no need to further specify a subspace. This amplitude and open associahedron are identical to those for the bi-adjoint off-diagonal case $m[1234|1243]$~\cite{Arkani-Hamed:2017mur}. 

To give further examples of our results, we need to introduce some notations. Throughout the paper, we use the following definition of Mandelstam variables:
\begin{align}
s_{i,j,k,l,\ldots}:=(p_i+p_j+p_k+p_l+\ldots)^2\,.
\end{align}
The subscript can also be sets:
\begin{align}
s_{\mathsf{A},\mathsf{B},\mathsf{C},\mathsf{D},\ldots}:=\Big(\sum_{i\in\mathsf{A}\cup\mathsf{B}\cup\mathsf{C}\cup\mathsf{D}\cup\ldots}p_i\Big)^2\,.
\end{align}
The planar Mandelstam variable $X_{i,j}$ is defined as
\begin{align}
& X_{i,j}:=s_{i,i+1,\ldots,j-1}=(p_i+p_{i+1}+\ldots+p_{j-1})^2\,,
\end{align}
such that on the support of momentum conservation $X_{i,j}=X_{j,i}$.

As we will see shortly, it is convenient to first consider amplitudes with only bi-fundamental scalars, \emph{i.e.}, the case with $n=2 k$. These represent the most non-trivial part of our construction, with adding bi-adjoint scalars relatively easier. Among such pure bi-fundamental scalar amplitudes, the simplest one correspond to the case where the $k$ pairs of particles form parallel lines. As shown in eq.~\eqref{para11}, they correspond to two pairs at two ends, $(1, 2)$, $(k{+}1, k{+}2)$, and $k{-}2$ pairs in between, $(i, n{+}3{-}i)$ for $i=3,\cdots, k$. It is straightforward to see that the flavor structure allows $k{-}1$ propagators of the bi-adjoint type and $2(k-2)$ ones of the bi-fundamental type ($\mathfrak{N}=3k{-}5$ poles in total); there are $2^{k{-}2}$ Feynman diagrams as follows:
\begin{align} \label{para11}
A_n\!\left[\begin{tikzpicture}[baseline={([yshift=-1.ex]current bounding box.center)},every node/.style={font=\footnotesize,},dir/.style={decoration={markings, mark=at position \halfway with {\arrow{Latex}}},postaction={decorate}}]
\draw [thick] (0,0) circle (1cm);
\node (pnm1) at (-90:1) [vertex,label={[label distance=-2pt]below:{$n{-}1$}}] {};
\node (pn) at (-180+3*180/8:1) [vertex,label={[label distance=-3pt]-180+3*180/8:{$n$}}] {};
\node (p1) at (-180+180/8:1) [vertex,label={[label distance=-3pt]-180+180/8:{$1$}}] {};
\node (p2) at (-180-1*180/8:1) [vertex,label={[label distance=-3pt]-180-1*180/8:{$2$}}] {};
\node (p3) at (-180-3*180/8:1) [vertex,label={[label distance=-3pt]-180-3*180/8:{$3$}}] {};
\node (p4) at (90:1) [vertex,label={[label distance=-3pt]above:{$4$}}] {};
\node (pk) at (-180-6*180/8:1) [vertex,label={[label distance=-3pt]-180-6*180/8:{$k$}}] {};
\node (pkp1) at (-180-7*180/8:1) [vertex,label={[label distance=-3pt]-180-7*180/8:{$k{+}1$}}] {};
\node (pkp2) at (-180-9*180/8:1) [vertex,label={[label distance=-3pt]-180-9*180/8:{$k{+}2$}}] {};
\node (pkp3) at (-180-10*180/8:1) [vertex,label={[label distance=-2pt]-180-10*180/8:{$k{+}3$}}] {};
\draw [thick,dir] (p1.center) to [bend right=20] (p2.center);
\draw [thick,dir] (p3.center) to [bend right=20] (pn.center);
\draw [thick,dir] (p4.center) to [bend right=20] (pnm1.center);
\draw [thick,dir] (pk.center) to [bend right=20] (pkp3.center);
\draw [thick,dir] (pkp1.center) to [bend right=20] (pkp2.center);
\node at (.2,0) {$\cdots$};
%\node at (-180-4.5*180/8:1.2)  {$\cdot$};
%\node at (-180-4.2*180/8:1.2)  {$\cdot$};
%\node at (-180-4.8*180/8:1.2)  {$\cdot$};
%\node at (180+4.5*180/8:1.2)  {$\cdot$};
%\node at (180+4.2*180/8:1.2)  {$\cdot$};
%\node at (180+4.8*180/8:1.2)  {$\cdot$};
\end{tikzpicture}\right]=\frac{1}{X_{1,3}}\prod_{i=3}^{k}\left(\frac{1}{X_{i,n+3-i}}+\frac{1}{X_{i+1,n+4-i}}\right)\frac{1}{X_{i+1,n+3-i}}
%&\frac{1}{s_{1,2}} \left(\frac{1}{s_{1,2,3}}+\frac{1}{s_{1,2,n}} \right) \frac{1}{s_{1,2,3,n}}
%\left(\frac{1}{s_{1,2,3,n,4}}+\frac{1}{s_{1,2,3,n,n-1}} \right) %\frac{1}{s_{1,2,3,4,n-1,n}}
%\nonumber \\ &\times \cdots \times  \frac{1}{s_{k,k+1,k+2,k+3}} \left(\frac{1}{s_{k+1,k+2,k}}+\frac{1}{s_{k+1,k+2,k+3}} \right) 
%\frac{1}{s_{k+1,k+2}}\,.
\,.
\end{align}  
In other words, the amplitude factorizes as the product of $k{-}1$ factors of $\frac 1 s$ and $k{-}2$ factors of $\frac 1 s+\frac 1 s$. The corresponding open associahedron must be the direct product of $k{-}1$ half-lines and $k{-}2$ intervals. It is nice that the subspace is given by setting the $k{-}2=\mathfrak{N}-(n{-}3)$ Mandelstam variables for the non-adjacent pairs,  {\it i.e.} $s_{i, n{+}3{-}i}$, to negative constants: 
% The corresponding associahedron is given by setting all  Mandelstam variables corresponding to the non-adjacent bi-fundamental pairs as negative constants 
\begin{align}
\pmb{H}_n\!\left[\begin{tikzpicture}[baseline={([yshift=-1ex]current bounding box.center)},every node/.style={font=\footnotesize,},dir/.style={decoration={markings, mark=at position \halfway with {\arrow{Latex}}},postaction={decorate}}]
\draw [thick] (0,0) circle (1cm);
\node (pnm1) at (-90:1) [vertex,label={[label distance=-2pt]below:{$n{-}1$}}] {};
\node (pn) at (-180+3*180/8:1) [vertex,label={[label distance=-3pt]-180+3*180/8:{$n$}}] {};
\node (p1) at (-180+180/8:1) [vertex,label={[label distance=-3pt]-180+180/8:{$1$}}] {};
\node (p2) at (-180-1*180/8:1) [vertex,label={[label distance=-3pt]-180-1*180/8:{$2$}}] {};
\node (p3) at (-180-3*180/8:1) [vertex,label={[label distance=-3pt]-180-3*180/8:{$3$}}] {};
\node (p4) at (90:1) [vertex,label={[label distance=-3pt]above:{$4$}}] {};
\node (pk) at (-180-6*180/8:1) [vertex,label={[label distance=-3pt]-180-6*180/8:{$k$}}] {};
\node (pkp1) at (-180-7*180/8:1) [vertex,label={[label distance=-3pt]-180-7*180/8:{$k{+}1$}}] {};
\node (pkp2) at (-180-9*180/8:1) [vertex,label={[label distance=-3pt]-180-9*180/8:{$k{+}2$}}] {};
\node (pkp3) at (-180-10*180/8:1) [vertex,label={[label distance=-2pt]-180-10*180/8:{$k{+}3$}}] {};
\draw [thick,dir] (p1.center) to [bend right=20] (p2.center);
\draw [thick,dir] (p3.center) to [bend right=20] (pn.center);
\draw [thick,dir] (p4.center) to [bend right=20] (pnm1.center);
\draw [thick,dir] (pk.center) to [bend right=20] (pkp3.center);
\draw [thick,dir] (pkp1.center) to [bend right=20] (pkp2.center);
\node at (.2,0) {$\cdots$};
%\node at (-180-4.5*180/8:1.2)  {$\cdot$};
%\node at (-180-4.2*180/8:1.2)  {$\cdot$};
%\node at (-180-4.8*180/8:1.2)  {$\cdot$};
%\node at (180+4.5*180/8:1.2)  {$\cdot$};
%\node at (180+4.2*180/8:1.2)  {$\cdot$};
%\node at (180+4.8*180/8:1.2)  {$\cdot$};
\end{tikzpicture}\right]
=\bigcup_{i=3}^{k}\left\{-s_{i,n+3-i}=c_{i,n+3-i}>0\right\}.
%\{s_{3,n},s_{4,n-1}, \cdots,s_{k,k+3} \text{~are~set~as~negative~constants} \} \,.
\end{align}
Requiring the $\mathfrak{N}$ allowed poles to be positive, together with these subspace constraints, gives the open associahedron whose canonical form is the correct amplitude. 

For $n=2k$, the situation gets more interesting when we go beyond the parallel case, and the first such example is with three adjacent bi-fundamental pairs for $n=6$, where we have $\mathfrak{N}=6$ allowed poles and four Feynman diagrams
\be 
A_6\!\left[\! \begin{tikzpicture}[baseline={([yshift=-1.ex]current bounding box.center)},every node/.style={font=\footnotesize,},dir/.style={decoration={markings, mark=at position \halfway with {\arrow{Latex}}},postaction={decorate}},scale=0.75]
\draw [thick] (0,0) circle (1cm);
\node (p1) at (-120:1) [vertex,label={[label distance=-3pt]240:{$1$}}] {};
\node (p2) at (-1,0) [vertex,label={[label distance=-1pt]180:{$2$}}] {};
\node (p3) at (120:1) [vertex,label={[label distance=-3pt]120:{$3$}}] {};
\node (p4) at (60:1) [vertex,label={[label distance=-3pt]60:{$4$}}] {};
\node (p5) at (1,0) [vertex,label={[label distance=-1pt]0:{$5$}}] {};
\node (p6) at (-60:1) [vertex,label={[label distance=-3pt]-60:{$6$}}] {};
\draw [thick,dir] (p1.center) to [bend right=20] (p2.center);
\draw [thick,dir] (p3.center) to [bend right=20] (p4.center);
\draw [thick,dir] (p5.center) to [bend right=20] (p6.center);
\end{tikzpicture}\! \right ] =\frac{1}{X_{1,3} X_{3,5} X_{1,5}}+\frac{1}{X_{1,3} X_{3,5} X_{3,6}}+\frac{1}{X_{1,3} X_{1,4}
   X_{1,5}}+\frac{1}{X_{3,5} X_{2,5} X_{1,5}}\,. 
\ee 
Remarkably, the open associahedron is given by the subspace with $N-3=3$ constraints, which set three-particle Mandelstam variables to negative constants:
\begin{align}\label{eq:H6a}
\pmb{H}_6\!\left[
\begin{tikzpicture}[baseline={([yshift=-1ex]current bounding box.center)},every node/.style={font=\footnotesize,},dir/.style={decoration={markings, mark=at position \halfway with {\arrow{Latex}}},postaction={decorate}},scale=0.75]
\draw [thick] (0,0) circle (1cm);
\node (p1) at (-120:1) [vertex,label={[label distance=-3pt]240:{$1$}}] {};
\node (p2) at (-1,0) [vertex,label={[label distance=-1pt]180:{$2$}}] {};
\node (p3) at (120:1) [vertex,label={[label distance=-3pt]120:{$3$}}] {};
\node (p4) at (60:1) [vertex,label={[label distance=-3pt]60:{$4$}}] {};
\node (p5) at (1,0) [vertex,label={[label distance=-1pt]0:{$5$}}] {};
\node (p6) at (-60:1) [vertex,label={[label distance=-3pt]-60:{$6$}}] {};
\draw [thick,dir] (p1.center) to [bend right=20] (p2.center);
\draw [thick,dir] (p3.center) to [bend right=20] (p4.center);
\draw [thick,dir] (p5.center) to [bend right=20] (p6.center);
\end{tikzpicture}\right]= \{ s_{2,5,6},s_{3,5,6},s_{3,4,6} \text{~set~to~negative~constants}\}\,.
\end{align}
The construction for open associahedron becomes more complicated as $k$ and $n$ increase. However, there exists a nice recursive procedure that allows us to insert a bi-adjoint scalar or an adjacent pair of bi-fundamental scalars. For example, for the following configurations with $n=7$ and $n=8$, we have $\mathfrak{N}=11$ and $\mathfrak{N}=14$ allowed poles respectively, $\{X_{1,3},X_{1,4},X_{1,5},X_{1,6},X_{2,5},X_{2,7},X_{3,5},X_{3,6},X_{3,7},X_{4,7},X_{5,7}\}$,  $\{X_{1,3},X_{1,4},X_{1,5},X_{1,6},X_{1,7},$\\ 
$X_{2,5}, X_{2,7},X_{3,5},X_{3,6},X_{3,7},X_{3,8},X_{4,7},X_{5,7},X_{5,8}\}$; here we present their subspaces with $7$ and $9$ constraints respectively: 
\begin{align} \label{eq:H7a}
&
\pmb{H}_7\!\left[
\begin{tikzpicture}[baseline={([yshift=-1ex]current bounding box.center)},every node/.style={font=\footnotesize,},dir/.style={decoration={markings, mark=at position \halfway with {\arrow{Latex}}},postaction={decorate}},scale=0.8]
\draw [thick] (0,0) circle (1cm);
\node (p1) at (-112.5:1) [vertex,label={[label distance=-3pt]240:{$1$}}] {};
\node (p2) at (-157.5:1) [vertex,label={[label distance=-3pt]-157.5:{$2$}}] {};
\node (p3) at (157.5:1) [vertex,label={[label distance=-3pt]157.5:{$3$}}] {};
\node (p4) at (112.5:1) [vertex,label={[label distance=-3pt]112.5:{$4$}}] {};
\node (p5) at (67.5:1) [vertex,label={[label distance=-3pt]67.5:{$5$}}] {};
\node (p6) at (22.5:1) [vertex,label={[label distance=-3pt]22.5:{$6$}}] {};
\node (p7) at (-45:1) [vertex,label={[label distance=-3pt]-45:{$7$}}] {};
\draw [thick,dir] (p1.center) to [bend right=20] (p2.center);
\draw [thick,dir] (p3.center) to [bend right=20] (p4.center);
\draw [thick,dir] (p5.center) to [bend right=20] (p6.center);
\end{tikzpicture}
\right]=\left\{
\begin{array}{c}
s_{2,5,6},s_{3,5,6},s_{3,4,6},s_{2,7},s_{3,7},s_{4,7},s_{5,7} \\ \text{~set~to~negative~constants}
\end{array}
\right\}
\,,\\
\label{eq:H8a}
&\pmb{H}_8\!\left[
\begin{tikzpicture}[baseline={([yshift=-1.ex]current bounding box.center)},every node/.style={font=\footnotesize,},dir/.style={decoration={markings, mark=at position \halfway with {\arrow{Latex}}},postaction={decorate}},scale=0.8]
\draw [thick] (0,0) circle (1cm);
\node (p1) at (-112.5:1) [vertex,label={[label distance=-3pt]-112.5:{$1$}}] {};
\node (p2) at (-157.5:1) [vertex,label={[label distance=-3pt]-157.5:{$2$}}] {};
\node (p3) at (157.5:1) [vertex,label={[label distance=-3pt]157.5:{$3$}}] {};
\node (p4) at (112.5:1) [vertex,label={[label distance=-3pt]112.5:{$4$}}] {};
\node (p5) at (67.5:1) [vertex,label={[label distance=-3pt]67.5:{$5$}}] {};
\node (p6) at (22.5:1) [vertex,label={[label distance=-3pt]22.5:{$6$}}] {};
\node (p7) at (-22.5:1) [vertex,label={[label distance=-3pt]-22.5:{$7$}}] {};
\node (p8) at (-67.5:1) [vertex,label={[label distance=-3pt]-67.5:{$8$}}] {};
\draw [thick,dir] (p1.center) to [bend right=20] (p2.center);
\draw [thick,dir] (p3.center) to [bend right=20] (p4.center);
\draw [thick,dir] (p5.center) to [bend right=20] (p6.center);
\draw [thick,dir] (p7.center) to [bend right=20] (p8.center);
\end{tikzpicture}
\right]=\left\{
\begin{array}{c} 
   s_{2,5,6},s_{3,5,6},s_{3,4,6},s_{2,7,8},s_{3,7,8},s_{4,7,8},s_{5,7,8},s_{3,4,8},s_{5,6,8} \\ \text{~set~to~negative~constants}
\end{array}
\right\} \,.
\end{align} 
We will show in section~\ref{posgeoinkinspace} how to derive these constraints from the recursive procedure.

As our last example in the introduction, here we present a more involved case for $n=11$, $k=4$, which has $\mathfrak{N}=22$ allowed poles, $\{ X_{1,3},X_{1,4},X_{1,6},X_{1,7},X_{1,8},X_{1,9},X_{1,10},X_{2,8},X_{3,6},$ $X_{3,7},X_{3,8},X_{3,9},X_{
   3,10},X_{3,11},X_{4,6},X_{4,7},X_{4,8},X_{5,7},X_{6,8},X_{8,10},X_{8,11},X_{9,11} \}$, and the subspace is given by $14$ constraints 
\begin{align}\label{eq:H_11}
\pmb{H}_{11}\!\left[
\begin{tikzpicture}[baseline={([yshift=-1ex]current bounding box.center)},every node/.style={font=\footnotesize,},dir/.style={decoration={markings, mark=at position \halfway with {\arrow{Latex}}},postaction={decorate}},scale=0.8]
\draw [thick] (0,0) circle (1cm);
\node (p1) at (-90-180/11:1) [vertex,label={[label distance=-2pt]below:{$1$}}] {};
\node (p2) at (-90-3* 180/11:1) [vertex,label={[label distance=-3pt]-90-3*180/11:{$2$}}] {};
\node (p3) at (-90-5 *180/11:1) [vertex,label={[label distance=-3pt]left:{$3$}}] {};
\node (p4) at (-90-7* 180/11:1) [vertex,label={[label distance=-3pt]-90-7*180/11:{$4$}}] {};
\node (p5) at (-90-9 *180/11:1) [vertex,label={[label distance=-3pt]-90-9*180/11:{$5$}}] {};
\node (p6) at (-90-11*180/11:1) [vertex,label={[label distance=-2pt]-90-11*180/11:{$6$}}] {};
\node (p7) at (-90-13*180/11:1) [vertex,label={[label distance=-3pt]-90-13*180/11:{$7$}}] {};
\node (p8) at (-90-15*180/11:1) [vertex,label={[label distance=-3pt]-90-15*180/11:{$8$}}] {};
\node (p9) at (-90-17*180/11:1) [vertex,label={[label distance=-3pt]right:{$9$}}] {};
\node (p10) at (-90-19*180/11:1) [vertex,label={[label distance=-3pt]-90-19*180/11:{$10$}}] {};
\node (p11) at (-90-21*180/11:1) [vertex,label={[label distance=-2pt]below:{$11$}}] {};
\draw [thick,dir] (p1.center) to [bend right=20] (p2.center);
\draw [thick,dir] (p3.center) to [bend right=20] (p7.center);
\draw [thick,dir] (p4.center) to [bend right=20] (p5.center);
\draw [thick,dir] (p8.center) to [bend right=20] (p11.center);
\end{tikzpicture}
\right]= 
\left\{
\begin{array}{c}
 s_{3,6},s_{4,6},s_{3,7},s_{4,5,7},s_{8,10},s_{2,8,9,10,11},s_{3,8,9,10,11} , \\
 s_{4,5,8,9,10,11} ,
  s_{6,8,9,10,11},s_{3,4,5,6,7,11},s_{8,11},s_{9,11} , \\
  s_{3,4,5,6,7,9},s_{3,4,5,6,7,10}
  \\
 \text{~set~to~negative~constants}
\end{array}
\right\}\,. 
\end{align}

The outline of this paper is as follows. We first review kinematic associahedron and bi-color theory in section \ref{review}.  Then we derive the subspace for above examples, ending up with the constructions of subspace for most general cases of bi-color theory in section \ref{posgeoinkinspace}.  Some deformed versions of subspace and explicit factorization examples are put in appendix \ref{sec:defsoftlimit} and \ref{sec:factorizationExamples}.  In section \ref{sec4}, we show how ``color is kinematic" for the color-dressed amplitudes of bi-color theory,  with some details put in appendix \ref{sec:derivatofdim} and \ref{explciitformofHa}.

\section{Review}\label{review}

\subsection{Kinematic Associahedron}
A prime example of an amplitude that is the canonical form of a polytope is the case of the associahedron for bi-adjoint scalar theory, as discovered in \cite{Arkani-Hamed:2017mur}. Here we give a brief review of it.  In large enough spacetime dimensions, the kinematic space of $n$ massless particles, ${\cal K}_n$, can be spanned by all independent $s_{i,j}$'s, thus it has dimension $d=n(n{-}3)/2$. 
Since there are  $n(n-3)/2$ planar poles $X_{i,j}$ with $i+1<j$ in a cyclic ordering in the bi-adjoint scalar amplitude $m[12 \cdots n\,|\,12 \cdots n]$,  we can start with the top cone where  all planar poles are positive. The remaining objective is to find a $(n-3)$-dimension hyperplane whose intersection with the cone gives the associahedron.

 The hyperplane can be expressed by $d-(n-3)=(n-2)(n-3)/2$ constraints.  
As described in \cite{Arkani-Hamed:2017mur}, one way to 
construct the hyperplane is to set  all  $-s_{i,j}=X_{i,j}+X_{i+1,j+1}-X_{i,j+1}-X_{i+1,j}$ with $1\leqslant i<  i+1<j\leqslant n-1$ to positive constants.
For example,  for the four-point case~\eqref{4ptttt},  the 2-dim cone is  $\{s>0,t>0\}$ and the 1-dim subspace is $\{u=-s-t=-c\}$. Their intersection is an interval,  which is a 1-dim associahedron. 

In the following paper, we will often see the intersection of a subspace and a cone or another subspace, 
  \begin{align}
P=Q\cap R\,,
\end{align}
where $P,Q,R$ can be described by  sets of constraints.
 In geometry, polytope $P$ is indeed an intersection of two others.  However, algebraically,  we can say the set of constraints $\pmb{P}$ for the polytope $P$  is the union  of those of  $Q$ and $R$,
\begin{align}
\pmb{P}=\pmb{Q}\cup\pmb{R} \,.
\end{align}
For example, for the 4-point case \eqref{4ptttt}, we have 
\begin{align}
&\pmb{\mathcal{A}}_{4}\left[\,
\begin{tikzpicture}[baseline={([yshift=-1ex]current bounding box.center)},every node/.style={font=\footnotesize,},dir/.style={decoration={markings, mark=at position \halfway with {\arrow{Latex}}},postaction={decorate}},scale=0.7]
\draw [thick] (0,0) circle (1cm);
\node (p1) at (-120:1) [vertex,label={[label distance=-3pt]240:{$1$}}] {};
\node (p3) at (120:1) [vertex,label={[label distance=-3pt]120:{$2$}}] {};
\node (p4) at (60:1) [vertex,label={[label distance=-3pt]60:{$3$}}] {};
\node (p6) at (-60:1) [vertex,label={[label distance=-3pt]-60:{$4$}}] {};
\end{tikzpicture}
\,\right]=\{s>0, t>0\}\cup\{u=-s-t = - c\}\,.
\end{align}
The cone in the construction ensures the boundaries of the associahedron  correspond to some vanishing planar poles. Furthermore, the choice of the constraints for the subspace makes sure that 
the associahedron  factorizes correctly, which means each codim-1 bounary of the associahedron is a direct product of two lower dimension associahedra, as proved in the same paper \cite{Arkani-Hamed:2017mur}. \\
\indent The canonical form of the associahedron, which gives the bi-adjoint scalar amplitude $m[12\ldots n|12\ldots n]$, can be obtained by the pullback of the \emph{planar scattering form}
\begin{align}\label{eq:planarSF}
\Omega_n[1,2,\cdots,n]:=\sum_{\text{planar }g}\text{sign}(g)\bigwedge_{a=1}^{n-3}d\log X_{i_a,j_a}\,,
\end{align}
where the summation is over all the planar cubic graphs following the ordering $[1,2,\ldots n]$, and $X_{i_a,j_a}$'s are the propagators of the graph $g$ which become facets of the associahedron. The sign function $\text{sign}(g)$ can be uniquely fixed (up to an overall $\pm$) by requiring the planar scattering form being locally projective. We refer the readers to~\cite{Arkani-Hamed:2017mur} for a more detailed definition. In addition to the four-point form $d\log\frac{s}{t}$ discussed in the introduction, at five points we have
\begin{align}
\label{eq:Omega5}
\Omega_5[1,2,3,4,5]=&
~d\log X_{1,4} \wedge d\log X_{1,3}+d\log X_{1,3} \wedge d\log X_{3,5}+d\log X_{3,5} \wedge d\log X_{2,5}\nonumber\\
&+d\log X_{2,5} \wedge d\log X_{2,4}+d\log X_{2,4} \wedge d\log X_{1,4}\,.
\end{align}
Similarly, we can define the \emph{$\alpha$-planar scattering form} as
\begin{align}\label{eq:alphaSF}
\Omega_n[\alpha]:=\sum_{\text{$\alpha$-planar }g}\text{sign}(g|\alpha)\bigwedge_{a=1}^{n-3}d\log X_{\alpha(i_a),\alpha(j_a)}
\end{align}
which can be obtained from eq.~\eqref{eq:planarSF} by a permutation $\alpha$. Locality and unitarity is manifest in the associahedron. In addition, ``color is kinematics'' in this representation, as we will review next.

%As will be reviewed in the next subsection, 
%   the scattering form of the bi-adjoint scalar theory in the canonical ordering can be written as 
%\begin{align}
%\Omega_n[1,2,\cdots,n]=\sum_{\text{ }g}\text{sign}(g)\bigwedge_{a=1}^{n-3}d\log X_{i_a,j_a}\,,
%\end{align}
%where the sum is over all planar cubic graphs.
%A simple way to determine the relative sign 
%$\text{sign}(g)$ up to an overall sign is to require that there is no pole at infinity.  See more in the next subsection. 
%In addition to the four-point example, for five points, we have 
%\begin{align}
%\Omega_5[1,2,3,4,5]=&
%~d\log X_{1,4} \wedge d\log X_{1,3}+d\log X_{1,3} \wedge d\log X_{3,5}+d\log X_{3,5} \wedge d\log X_{2,5}\nonumber\\
%&+d\log X_{2,5} \wedge d\log X_{2,4}+d\log X_{2,4} \wedge d\log X_{1,4}\,.
%\end{align}

%Interestingly, the pullback of the scattering form to the associahedron, which gives the amplitude, is the canonical form of the associahedron.  Locality and unitarity is manifest in the associahedron. In addition, ``color is kinematics'' in this representation.

\subsection{Scattering Forms and ``Color is Kinematics''}\label{scatteringformsrevieew}
Now we review the definition of the \emph{full scattering form}, which is the natural generalization of eq.~(\ref{eq:alphaSF}). The full scattering form encodes nontrivial kinematic numerators and all the color orderings. For convenience, we denote  the collection of all cubic tree Feynman diagrams with $n$ external legs as $\Gamma_n$.  Each $g\in \Gamma_n$ is specified by $n{-}3$ mutually compatible propagators. We denote them as $s_{I}$, where $I\in g$ are associated with the internal propagators. We define their wedge product as:
%\be
%W(g|\alpha_g):=\text{sign}(g|\alpha_g)\,d s^{(g)}_1 \wedge d s^{(g)}_2 \wedge \cdots \wedge d s^{(g)}_{n{-}3} \,,
%\ee
\be
W(g|\alpha_g):=\text{sign}(g|\alpha_g)\bigwedge_{I\in g} ds_{I}
\ee
where $\alpha_g$ is a color ordering compatible with $g$. The overall sign depends on ordering of the $d s $'s. Both $W(g|\alpha_g)$ and $\text{sign}(g|\alpha_g)$ satisfy the \emph{mutation and vertex flip rule}~\cite{Arkani-Hamed:2017mur}.
 
The full scattering form is defined as an $(n{-}3)$-form in ${\cal K}_n$: a linear combination of $d\log$'s of propagators for each diagram,
\be\label{form}
\Omega_{n}[N]:=\sum_{g\in \Gamma_n}~N(g|\alpha_g)W(g|\alpha_g)\prod_{I\in g}\frac{1}{s_{I}}\,,
\ee
where for any three graphs as in figure~\ref{figjacobi}, we require their numerators satisfy 
\be\label{Jacobi2}
%N_{g_S}+ N_{g_T}+ N_{g_U}=0\,.
N(g_S|I_1I_2I_3I_4)+N(g_T|I_1I_4I_2I_3)+N(g_U|I_1I_3I_4I_2)=0\,.
\ee
This requirement guarantees that 
the scattering form is \emph{projective}~\cite{Arkani-Hamed:2017mur}, {\it i.e.} it is invariant under a GL$(1)$ transformation $s_I \to \Lambda (s) s_I$ for all subsets $I$ (with $\Lambda(s)$ depending on $s$).{\footnote {Here the GL$(1)$ transformation will not be applied to the numerators $N(g|\alpha_g)$.  A restrict description is to use another kind of variables  in the so-called  big  kinematic space \cite{Arkani-Hamed:2017mur}. We postpone it to section~\ref{sec:gensmallkinespace}}

An explicit example of the scattering form~\eqref{form} is given by eq.~\eqref{eq:alphaSF} where the numerators are simply $\textrm{sign}(g|\alpha)$ if the diagram is compatible with $\alpha$ and $0$ otherwise. Its pullback to a subspace is the canonical form of an associahedron. More examples of differential forms whose pullback are the canonical form of polytopes are given in \cite{Arkani-Hamed:2017mur,He:2018pue,Gao:2017dek}.  Note that a linear combination of scattering forms is still a scattering form. For example, one can construct the scattering forms for YM and NLSM this way~\cite{Arkani-Hamed:2017mur}. 
 
For any triplet $g_{S}, g_{T}, g_{U}$ of graphs that differ only by one propagator, as shown in figure~\ref{figjacobi},
 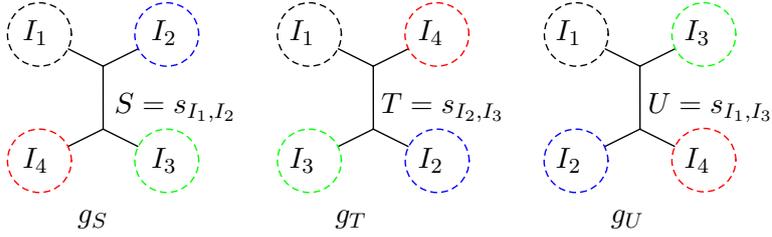
\begin{figure}[t]
\begin{center}
\begin{overpic}[width=.8\linewidth]{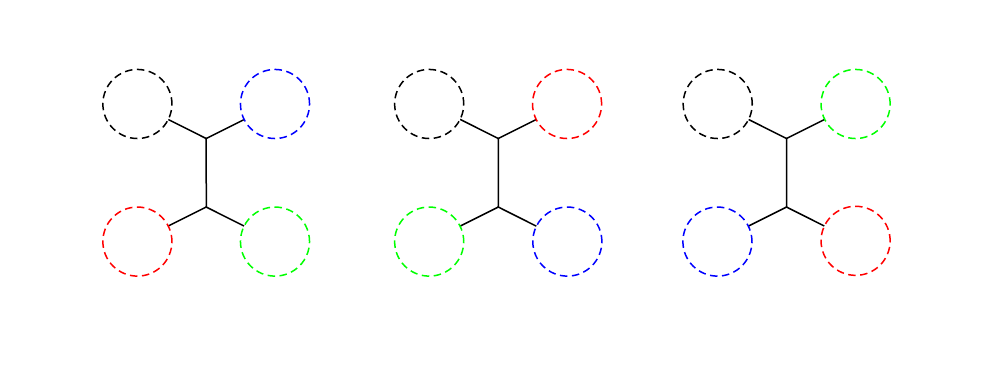}
%\label{ggst7}
\put(12,26){$I_1$}\put(26,26){$I_2$}\put(26,12){$I_3$}\put(12,12){$I_4$}
\put(41,26){$I_1$}\put(55,26){$I_4$}\put(55,12){$I_2$}\put(41,12){$I_3$}
\put(70,26){$I_1$}\put(84,26){$I_3$}\put(84,12){$I_4$}\put(70,12){$I_2$}
%\put(43,26){$I_4$}\put(56,26){$I_1$}\put(56,12){$I_2$}
\put(22,18){$S=s_{I_1,I_2}$}\put(51,18){$T=s_{I_2,I_3}$}\put(80,18){$U=s_{I_1,I_3}$}
\put(18,6){$g_S$}\put(46,6){$g_T$}\put(76,6){$g_U$}
\end{overpic}
\end{center}
\caption{A triplet of three cubic tree graphs that differ by one propagator.}
%\caption{A four set partition $I_1\sqcup I_2\sqcup I_3\sqcup I_4$ of the external labels and the three corresponding channels. The three graphs $g_s,g_t,g_u$ are identical except for a 4-point subgraph.}
\label{figjacobi}
\end{figure}
 there is a so-called seven-term identity implied by momentum conservation, 
\begin{align}\label{eq:7term}
s_{I_1,I_2}+s_{I_2,I_3}+s_{I_1,I_3}=s_{I_1}+s_{I_2}+s_{I_3}+s_{I_4}\,,
\end{align}
where the four propagators connecting to the four subgraphs are denoted as $s_{I_1}, \ldots, s_{I_4}$. This leads to an identity that their wedge products satisfy the Jacobi identity that is equivalent to those of color factors,
\be\label{Jacobi}
%W_{g_{S}} + W_{g_{T}}+ W_{g_{U}}=\cdots \wedge (d S+ d T + d U)\wedge \cdots=0\,,
W(g_S|I_1I_2I_3I_4)+W(g_T|I_1I_4I_2I_3)+W(g_U|I_1I_3I_4I_2)=\cdots \wedge (d S+ d T + d U)\wedge \cdots=0\,,
\ee
where the distinct Mandelstam variables are $S, T, U$, respectively, and ``$\cdots$'' denotes the wedge products of the remaining $n{-}4$ propagators shared by the three graphs. Eq.~\eqref{Jacobi} implies a  duality between color factors and differential forms on kinematic space ${\cal K}_n$:
\be\label{eq:color_dual}
C(g|\alpha_g)\qquad\leftrightarrow \qquad W(g|\alpha_g)\,.
\ee
Hence ``color is kinematics''. 
\iffalse
For example, for $n{=}4$, there are three color factors dual to 1-forms:
\begin{align}
C_s=f^{a_1a_2b}f^{ba_3a_4}\qquad\leftrightarrow&\qquad ds\nonumber\\
C_t=f^{a_1a_4b}f^{ba_2a_3}\qquad\leftrightarrow&\qquad dt\nonumber\\
C_u=f^{a_1a_3b}f^{ba_4a_2}\qquad\leftrightarrow&\qquad du\nonumber
\end{align}
\fi 
Considering a color-dressed amplitude ${\bf M}_n$,
\be\label{eq:amp_M}
{\bf M}_n[N]=\sum_{g\in \Gamma_n}N(g|\alpha_g)C(g|\alpha_g)\prod_{I\in g }\frac{1}{s_{I}}\,,
\ee
the duality \eqref{eq:color_dual} leads naturally to a  duality between color-dressed amplitudes and scattering forms,
\be\label{eq:color_dual_amp}
{\bf M}_n[N]\quad\leftrightarrow\quad \Omega_{n}[N]\,.
\ee
In addition, a color-ordered amplitude can be obtained by pulling back the scattering form $\Omega_{n}$ onto an appropriate subspace~\cite{Arkani-Hamed:2017mur}.

\subsection{Color Ordered Amplitudes with both Adjoint and Fundamental Particles}
A color ordered amplitude $A_n[\alpha]$ only receives contributions from the Feynman diagrams  that contain no crossing edges when the external particles are put on a circle according to the ordering $\alpha$. In particular, each particle in fundamental ($\mathfrak{f}$) representation is connected to its anti-fundamental ($\mathfrak{af}$) partner by a directed line representing the color flow. For example, we have
\begin{align}\label{eq:colorOrderExamples}
A_6[\underline{1},2,\overline{3},\underline{4},5,\overline{6}]=
A_6\!\left[\!\begin{tikzpicture}[baseline={([yshift=-1.ex]current bounding box.center)},every node/.style={font=\footnotesize,},dir/.style={decoration={markings, mark=at position \halfway with {\arrow{Latex}}},postaction={decorate}},scale=0.95]
\draw [thick] (0,0) circle (1cm);
\node (p1) at (-120:1) [vertex,label={[label distance=-2pt]240:{$1$}}] {};
\node (p2) at (-1,0) [vertex,label={[label distance=-1pt]180:{$2$}}] {};
\node (p3) at (120:1) [vertex,label={[label distance=-2pt]120:{$3$}}] {};
\node (p4) at (60:1) [vertex,label={[label distance=-2pt]60:{$4$}}] {};
\node (p5) at (1,0) [vertex,label={[label distance=-1pt]0:{$5$}}] {};
\node (p6) at (-60:1) [vertex,label={[label distance=-2pt]-60:{$6$}}] {};
\draw [thick,dir] (p1.center) to [bend right=20] (p3.center);
\draw [thick,dir] (p4.center) to [bend right=20] (p6.center);
\end{tikzpicture}\!\right], & &
A_8[\underline{1},\underline{2},\overline{3},4,\overline{5},\underline{6},7,\overline{8}]=
A_8\!\left[\!\begin{tikzpicture}[baseline={([yshift=-1.ex]current bounding box.center)},every node/.style={font=\footnotesize,},dir/.style={decoration={markings, mark=at position \halfway with {\arrow{Latex}}},postaction={decorate}},scale=0.95]
\draw [thick] (0,0) circle (1cm);
\node (p1) at (-135:1) [vertex,label={[label distance=-2pt]-135:{$1$}}] {};
\node (p2) at (-1,0) [vertex,label={[label distance=-1pt]180:{$2$}}] {};
\node (p3) at (135:1) [vertex,label={[label distance=-2pt]135:{$3$}}] {};
\node (p4) at (0,1) [vertex,label={[label distance=-1pt]90:{$4$}}] {};
\node (p5) at (45:1) [vertex,label={[label distance=-2pt]45:{$5$}}] {};
\node (p6) at (1,0) [vertex,label={[label distance=-1pt]0:{$6$}}] {};
\node (p7) at (-45:1) [vertex,label={[label distance=-2pt]-45:{$7$}}] {};
\node (p8) at (0,-1) [vertex,label={[label distance=-1pt]-90:{$8$}}] {};
\draw [thick,dir] (p1.center) -- (p5.center);
\draw [thick,dir] (p6.center) to [bend right=20] (p8.center);
\draw [thick,dir] (p2.center) to [bend right=80] (p3.center);
\end{tikzpicture}\!\right],
\end{align}
where we use underscores (bars) to denote $\mathfrak{f}$ ($\mathfrak{af}$) particles.\footnote{When the ordering of states is arbitrary, we will denote flavor pairs using capital letters, like $\bar{A}$ and $\underline{A}$.} We further require that the $\mathfrak{f}$~particles always come before their $\mathfrak{af}$ partners in $\alpha$. This is because we can always use these color ordered amplitudes to linearly expand those with flipped pairs~\cite{Melia:2013bta}. \emph{For most parts of this paper, we assume each $\mathfrak{f}$-$\mathfrak{af}$ pair carries a distinct flavor.} The single flavor case can be recovered by averaging over different flavor assignments.

By associating each $\mathfrak{f}$ particle with an open parenthesis and each $\mathfrak{af}$ one with a closed parenthesis, we can convert an ordering $\alpha$ into a Dyck word with adjoint-representation ($\mathfrak{adj}$) particle insertions, the whole set of which is denoted as $\text{Dyck}_{n,k}$ for $n$ external particles and $k$ $\mathfrak{f}$-$\mathfrak{af}$ pairs. We call an $\mathfrak{f}$-$\mathfrak{af}$ pair $(l,r)$ \emph{adjacent} in $\alpha$ if there exists an $i$ that $\alpha(i)=l$ and $\alpha(i+1)=r$. The orderings given in eq.~\eqref{eq:colorOrderExamples} can thus be written as $[(1,2,3),(4,5,6)]$ and $[(1,(2,3),4,5),(6,7,8)]$. In the second ordering, the $\mathfrak{f}$-$\mathfrak{af}$ pair $(2,3)$ is adjacent. In this work, we study the color orderings that are given by $\text{Dyck}_{n,k}$ modulo cyclicity:
\begin{align}
\pmb{M}_{n,k}=\text{Dyck}_{n,k}/\{\text{cyclic permutations}\},
\end{align}
namely, two color orderings are considered the same if they differ by a cyclic permutation that respects the parenthesis structure. For example, both orderings in eq.~\eqref{eq:colorOrderExamples} have two equivalent representations:
\begin{align}
[(1,2,3),(4,5,6)]&\sim[(4,5,6),(1,2,3)]\,,\nonumber\\
[(1,(2,3),4,5),(6,7,8)]&\sim[(6,7,8),(1,(2,3),4,5)]\,.
\end{align}
At $k=0$, $\pmb{M}_{n,k}$ returns to the usual trace basis for $\mathfrak{adj}$ particles. The size of $\pmb{M}_{n,k}$ is $(n-1)!/k!$~\cite{Kalin:2017oqr}.

As a basis, $\pmb{M}_{n,k}$ is redundant since there exist linear relations between color ordered amplitudes. If there are no $\mathfrak{f}$-$\mathfrak{af}$ pairs, the minimal basis under the color Lie algebra is the Del Duca-Dixon-Maltoni (DDM) basis~\cite{DelDuca:1999rs}, given by $A_n[1,2,\sigma]$ with $\sigma$ a permutation of the rest $n{-}2$ particles. For $k\geqslant 1$, we can by convention label one $\mathfrak{f}$-$\mathfrak{af}$ pair as $(1,2)$. The minimal basis under the color Lie algebra is the Melia basis, in which the pair $(1,2)$ always comes first~\cite{Melia:2013bta,Melia:2013epa,Johansson:2015oia}: 
\begin{align}
%\left\{\alpha=[(1,2),\sigma]\,\middle |\,\sigma\in\text{Dyck}_{k-1}\times\{\mathfrak{adj}\text{ particle insertions}\}_{n-2k}\right\}.
\left\{\alpha=[(1,2),\sigma]\,\middle |\,\sigma\in\text{Dyck}_{n-2,k-1}\right\}.
\end{align}
The size of Melia basis is $(n-2)!/k!$.
We can write an ordering $\alpha$ in the Melia basis as
\begin{align}
\alpha=[(1,2),\mathsf{B}_2,\mathsf{B}_3,\ldots,\mathsf{B}_m]=[\mathsf{B}_1,\mathsf{B}_2,\mathsf{B}_3,\ldots,\mathsf{B}_m]\,,
\end{align}
where each $\mathsf{B}_i$ is a \emph{block}, and we always fix $\mathsf{B}_1=(1,2)$ in the Melia basis. The block $\mathsf{B}_i$ can either be an $\mathfrak{adj}$ block, which contains a single $\mathfrak{adj}$ particle $g_i$: $\mathsf{B}_i=g_i$, or an $\mathfrak{f}$-$\mathfrak{af}$ block, which is defined as a Dyck word (with $\mathfrak{adj}$ particle insertions) that is enclosed by an overall parenthesis. The simplest $\mathfrak{f}$-$\mathfrak{af}$ block contains just an adjacent $\mathfrak{f}$-$\mathfrak{af}$ pair $\mathsf{B}_i=(l_i,r_i)$.  In general, it has substructures:
\begin{align}\label{eq:block}
\mathsf{B}_i=(l_i,\mathsf{B}_{i_1},\mathsf{B}_{i_2},\ldots,\mathsf{B}_{i_s},r_i)\,,\qquad l_i\in\{\mathfrak{f}\}\text{ and }r_i\in\{\mathfrak{af}\}\,,
\end{align}
where each $\mathsf{B}_{i_\ell}$ is again a block, but for future convenience we call it a \emph{sub-block} of $\mathsf{B}_i$. The definition of a block is thus recursive and it terminates when we reach an $\mathfrak{adj}$ block or an adjacent $\mathfrak{f}$-$\mathfrak{af}$ pair. We define $\text{sub}[\mathsf{B}_i]$ as the collection of all the sub-blocks of $\mathsf{B}_i$. For example, if $\mathsf{B}_i$ is given by eq.~\eqref{eq:block}, we have
\begin{align}\label{eq:sub-block}
\text{sub}[\mathsf{B}_i]=\{\mathsf{B}_{i_1},\mathsf{B}_{i_2},\ldots,\mathsf{B}_{i_s}\}\,.
\end{align}
It is also convenient to view an $\mathfrak{adj}$ block $\mathsf{B}_i=g_i$ as a degenerate limit of an $\mathfrak{f}$-$\mathfrak{af}$ block, in which $l_i=r_i=g_i$ and $\text{sub}[\mathsf{B}_i]=\emptyset$. Pictorially, a block $\mathsf{B}_i$ is represented by all the structures bounded by the line $(l_i,r_i)$.

%We can arrange any $\alpha\in\pmb{M}_{n,k}$ on the boundary of a disk following the planar ordering $\mathcal{O}(\alpha)$, and connect each $\mathfrak{f}$-$\mathfrak{af}$ pair by a solid line. A block $\mathsf{B}_i$ is represented by all the structures bounded by the line $(l_i,r_i)$.

\subsection{Bi-color \texorpdfstring{$\phi^3$}{phi3} Amplitudes}\label{sec24}

The above discussion applies to generic color ordered amplitudes, for example, QCD. Now we move on to some features peculiar to the amplitudes of the bi-color scalar theory  whose Lagrangian are given by eq.~\eqref{eq:bicolor}. 
%In that formula, $\lambda$ is an arbitrary coupling constant. The normalization of Lie algebra generators and structure constants is $[T^a,T^b]=\tilde{f}^{abc}T^c$ and $\text{tr}(T^aT^b)=\delta^{ab}$. The real scalar $\phi$ carries adjoint representation indices $(a,A)$ of the symmetry group $U(N)\times U(N')$, while $\varphi$ and $\varphi^*$ carry fundamental representation indices $(\underline{i},\underline{I})$ and anti-fundamental ones $(\overline{i},\overline{I})$ respectively. 
We introduce a flavor function that returns the abstract flavor symbol of the fields:
\begin{align}
f(\varphi_r)=f_r\,, & & f(\varphi^*_r)=-f_r\,,& & f(\phi)=0\,,
\end{align}
where each $f_r$ is non-numeric and distinct in the sense that $f_a\pm f_b$ with $a\neq b$ is kept unevaluated. With the help of this flavor function, we can define 
\begin{align}\label{eq:theta}
\vartheta_{I}=\left\{\begin{array}{lll}
1 & \hphantom{aa} & \sum_{s\in I} f(s)=0\text{ or a single term }\pm f_r \\ 0 & & \text{otherwise}
\end{array}\right.\,,
\end{align}
such that a propagator $1/s_I$ is allowed by flavor conservation if and only if $\vartheta_I=1$.

We can expand the full color-dressed amplitude of the theory~\eqref{eq:bicolor} by doubly color-ordered amplitudes $m[\alpha|\beta]$. In this paper, we will mainly study the diagonal component $A_n[\alpha]:= m[\alpha|\alpha]$. The major simplification in this scalar theory is that the kinematic numerator is trivial. As a result, color ordered amplitudes do not distinguish particles and anti-particles, for example,
$A_6[\underline{1},2,\overline{3},\underline{4},5,\overline{6}]=A_6[\overline{1},2,\underline{3},\underline{4},5,\overline{6}]$,
%In other words, we can ignore the arrow on the line connecting an $\mathfrak{f}$-$\mathfrak{af}$ pair.
although the color factors of these two cases are different. While this feature does not change the size of $\pmb{M}_{n,k}$, and the minimal Melia basis is still the same, each ordering in $\pmb{M}_{n,k}$ potentially gets more equivalent representations. This is because we can flip the parentheses if necessary. For example, we now have
\begin{align}
[(1,2,3),(4,5,6)]&\sim [2,(3,(4,5,6),1)]\,,\text{ etc,}\nonumber\\
[(1,(2,3),4,5),(6,7,8)]&\sim[(2,3),4,(5,(6,7,8),1)]\,,\text{ etc,}
\end{align}
both of which are not valid for generic color ordered amplitudes like QCD. Nevertheless, for the bi-color scalar theory~\eqref{eq:bicolor}, We can use this enlarged cyclic freedom to represent any $\alpha\in\pmb{M}_{n,k}$ as
\begin{align}\label{eq:genericOrder}
\alpha=[\mathsf{B}_1,\mathsf{B}_2,\mathsf{B}_3,\ldots,\mathsf{B}_m]\,,\text{ where either }\mathsf{B}_1=(l_1,r_1)\text{ or }\mathsf{B}_1=g_1\,.
\end{align}
The $\mathsf{B}_1=g_1$ case is always possible if $\mathfrak{adj}$ particles are present. Otherwise, there must exist at least one adjacent $\mathfrak{f}$-$\mathfrak{af}$ pair, which we can identify as $\mathsf{B}_{1}=(l_1,r_1)$. If all $\mathfrak{f}$-$\mathfrak{af}$ pairs are adjacent in $\alpha$, then for each block $\mathsf{B}_i$ in $\alpha$ we either have $\mathsf{B}_i=(l_i,r_i)$ or $\mathsf{B}_i=g_i$. For generic configurations, the representation is usually not unique, but any one of the equivalence class will work for our purpose.
%When It also allows us to flip the parenthesis if necessary.

%For future convenience, we define $\pmb{M}_{n,k}^{\text{planar}}$ as a subset of $\pmb{M}_{n,k}$ in which the external particles follow the canonical planar ordering and $\mathsf{B}_1=(1,2)$: 
%\begin{align}\label{eq:planarMelia}
%\pmb{M}^{\text{planar}}_{n,k}=\left\{\alpha=[(1,2),\sigma]\in\pmb{M}_{n,k}\text{ and }\mathcal{O}(\alpha)=(1,2,\ldots,n)\right\}
%\end{align}
%where the function $\mathcal{O}(\alpha)$ gives the planar ordering of $\alpha\in\pmb{M}_{n,k}$. Essentially, it removes all the parentheses in $\alpha$. When constructing the positive geometry, we will give explicit examples for the orderings in this subset.

\section{Open Associahedra in Kinematic Space}
\label{posgeoinkinspace}
We now construct the kinematic polytope associated to the tree-level $\phi^3$ amplitudes of the bi-color theory~\eqref{eq:bicolor}. The discussion here applies to amplitudes with a pair of the same ordering $\alpha\in\pmb{M}_{n,k}$.
%, denoted as $A_n[\alpha]:= A_n[\alpha|\alpha]$. This section is a generalization to the pure bi-adjoint associahedron first given in~\cite{Arkani-Hamed:2017mur}.

To start with, we define $\Delta_n[\alpha]$ as a positive region in kinematic space where all the $\alpha$-planar variables forbidden by the flavor assignment are set to positive constants. It can be written as ${\Delta}_n[\alpha]={P}_n[\alpha]\cap {F}_n[\alpha]$, where the subspace ${P}_n$ and ${F}_n$ are given by the constraints
\begin{align}\label{eq:PnFn}
&(\text{positivity constraints}) & &\pmb{P}_n[\alpha]=\left\{X_{\alpha(i),\alpha(j)}\geqslant 0\text{ for all }1\leqslant i<j\leqslant n\right\}, \\
&(\text{flavor constraints}) & &\pmb{F}_n[\alpha]=\Big\{X_{\alpha(i),\alpha(j)}=b_{\alpha(i),\alpha(j)}>0\text{ if }\vartheta_{\alpha(i),\ldots,\alpha(j-1)}=0 \Big\},\nonumber
\end{align}
and $\vartheta$ is given in eq.~\eqref{eq:theta}. The constraints that carve out the region $\Delta_n[\alpha]$ are given by $\pmb{\Delta}_n[\alpha]=\pmb{P}_n[\alpha]\cup\pmb{F}_n[\alpha]$.
%\begin{align}
%%& X_{\alpha(i),\alpha(j)}\geqslant 0\quad\text{if}\quad\sum_{s=i}^{j}q_{\alpha(s)}=0\text{ or a single term,} & &(\text{positivity constraints})\nonumber\\
%%& X_{\alpha(i),\alpha(j)}=b_{\alpha(i),\alpha(j)}>0\quad\text{otherwise,} & &(\text{flavor constraints})
%\end{align}
%for all $1\leqslant i<j\leqslant n$. 
The flavor constraints, present when there are more than two $\mathfrak{f}$-$\mathfrak{af}$ pairs, remove all the Feynman diagrams that violate the flavor conservation from the $\alpha$-planar scattering form~\eqref{eq:alphaSF}. These forbidden poles can easily be visualized in the polygon dual to Feynman diagrams. For each $\mathfrak{f}$-$\mathfrak{af}$ pair $\big(\alpha(i),\alpha(j)\big)$, we draw a line connecting the edge $E_{\alpha(i),\alpha(i+1)}$ and $E_{\alpha(j),\alpha(j+1)}$, called a \emph{flavor line}. 
%we draw a diagonal connecting vertex $\alpha(i)$ and $\alpha(j{+}1)$, called a \emph{flavor line}. 
%A given vertex $\alpha(i)$ can be shared by at most two flavor lines: $\big(\alpha(i),\alpha(j)\big)$ and $\big(\alpha(k),\alpha(i-1)\big)$ with $k<i<j$ in the cyclic sense. When we draw a diagonal $X_{\alpha(i),\alpha(\ell)}$, it is now important to make sure that \emph{in the vicinity of the vertex $\alpha(i)$}, the flavor line $\big(\alpha(i),\alpha(j)\big)$, if exists, is on the $\alpha$-ascending side of $X_{\alpha(i),\alpha(\ell)}$, while the flavor line $\big(\alpha(k),\alpha(i-1)\big)$, if exists, is on the $\alpha$-descending side of $X_{\alpha(i),\alpha(\ell)}$. The same rule also applies to the other end point $\alpha(\ell)$, if it is connected to flavor lines.
%\footnote{Two different $\mathfrak{f}$-$\mathfrak{af}$ pairs might correspond to the same diagonal, but they are treated as two different flavor lines. As an example, see the left panel of figure~\ref{fig:forbiddenPoles}.} 
Then a pole $X_{a,b}$ is forbidden by flavor conservation if the diagonal $(a,b)$ crosses two or more flavor lines. Two examples are given in figure~\ref{fig:forbiddenPoles}.
%In particular, we have
%\begin{align}
%\pmb{F}_{n}[\alpha,n]=\pmb{F}_{n-1}[\alpha]\;\text{ for }\;\alpha\in\pmb{M}_{n-1,k}\,,
%\end{align}
%since adding a flavor neutral $\mathfrak{adj}$ particle $n$ does not forbids additional planar poles. 
The dimension of ${\Delta}_n[\alpha]$ is
\begin{align}
n-3\leqslant\text{dim}{\Delta}_{n}[\alpha]\leqslant\frac{n(n-3)}{2}-\frac{k(k-1)}{2}\,,
\end{align}
where $k$ is the number of $\mathfrak{f}$-$\mathfrak{af}$ pairs. The first equality only holds when $n{=}4$ and $k{=}2$. The maximal dimension of $\Delta_n[\alpha]$ is reached 
%while the second equality is reached 
%In particular, we have
\begin{align}\label{eq:F_nIS}
\text{if all the $\mathfrak{f}$-$\mathfrak{af}$ pairs are adjacent in $\alpha$:}\quad\text{dim}{\Delta}_n[\alpha]=\frac{n(n-3)}{2}-\frac{k(k-1)}{2}\,.
\end{align}
%which is the maximal dimension of $\Delta_n[\alpha]$. 
To obtain the $(n{-}3)$-dimensional kinematic polytope ${\mathcal{A}}_n[\alpha]$, we need to further restrict ${\Delta}_n[\alpha]$ to a subspace ${H}_n[\alpha]$:
\begin{align}
(\text{subspace})& & {\mathcal{A}}_n[\alpha]={H}_n[\alpha]\cap{\Delta}_n[\alpha]\,,\nonumber\\
(\text{set of constraints})& & \pmb{\mathcal{A}}_n[\alpha]=\pmb{H}_n[\alpha]\cup\pmb{\Delta}_n[\alpha]\,.
\end{align}
%Without losing generality, we only study the $\alpha$ in the planar Melia basis~\eqref{eq:planarMelia}. 
For $k=1$, since the amplitude is the same as the pure bi-adjoint one, the subspace $H_n[\alpha]$ is also the same, given by the constraints: 
\begin{align}
\pmb{H}_n[(1,2),3,\ldots,n]=\left\{-s_{i,j}=c_{i,j}>0\text{ for non-adjacent pairs in }2\leqslant i<j\leqslant n\right\}.
\end{align} 
The resultant kinematic polytope ${\mathcal{A}}_n[\alpha]$ is just the associahedron~\cite{Arkani-Hamed:2017mur}. The restriction equations for the original ABHY associahedra can be derived in a number of ways, each with their own generalizations \cite{Arkani-Hamed:2017mur, baziermatte2018abhy,Salvatori:2018fjp}. We will focus on the inverse soft construction of associahedra~\cite{He:2018pue}, as this construction has a natural generalization for open associahedra. Although it was originally derived by considering how CHY integrands behave under soft limits, the inverse-soft construction is still applicable to bi-color amplitudes with both adjoint and (anti-)fundamental states, which currently lack a clear CHY picture, due to its geometric interpretation. Starting from $\pmb{H}_{n-1}$, the original inverse-soft construction for closed associahedra fixes $\pmb{H}_{n}$ by imposing 
\begin{itemize}
\item The facet of the $X_{1,n-1}=s_{n-1,n}\rightarrow 0$ factorization channel of $\mathcal{A}_{n}$ must be $\mathcal{A}_{n-1}$. This can be done by inheriting all the constraints in $\pmb{H}_{n-1}$.
\item Additional constraints in $\pmb{H}_n$ must ensure that the facet $X_{1,n-1}$ does not intersect the facets of the channels incompatible to it. This can be done by setting the non-planar Mandelstam variables in certain seven-term identities to a negative constant.
%Constraints of $\mathcal{A}_{n}$ must be given by setting non-planar Mandelstam variables in certain 7-term identities to a negative constants.
\end{itemize} 
In the case of closed associahedra, the inverse soft construction yields~\cite{He:2018pue}
\begin{equation}\label{eq:IS}
\pmb{H}_{n}:=\pmb{H}_{n-1}\cup \{ -s_{i,n}=c_{i,n}>0\,|\,2\leqslant i\leqslant n-2\}\,.
%\label{factorizationss}
\end{equation}
Together with $\pmb{H}_{3}=\emptyset$ it is enough to fix the restriction equations of $\mathcal{A}_{n}$. Using the explicit forms of the restriction equations, one can directly prove that $\mathcal{A}_{n}$ has the correct structure.

Since we are now working with (anti-)fundamental states and open associahedra, the inverse soft construction of subspaces requires a generalization. If state $n$ is an $\mathfrak{adj}$ state, the inverse soft factorization condition is the same. If we instead want to add an $\mathfrak{f}$-$\mathfrak{af}$ block $\mathsf{B}_m$, the new inverse soft factorization conditions are:
\begin{itemize}
\item The facet of the $X_{l_1,l_m}=P_m^2\rightarrow 0$ and $X_{l_m,r_m}=L_m^2\rightarrow 0$ factorization channels of $\mathcal{A}_{n}$ must be a direct product of $$\mathcal{A}[\mathsf{B}_1,\ldots P_m]\times\mathcal{A}[-P_m,\mathsf{B}_m] \,\text{ and }\, \mathcal{A}[\mathsf{B}_1,\ldots,(L_m,r_m)]\times\mathcal{A}[(-L_m,l_m),\cdots]$$ 
respectively, where in $\mathcal{A}[(-L_m,r_m),\ldots]$ the ``$\ldots$'' are the sub-blocks of $\mathsf{B}_m$.
\item Additional constraints in $\pmb{H}_n$ must ensure that the facets of the channels incompatible to $X_{l_1,l_m}$ and $X_{l_m,r_m}$ do not intersect them. \emph{We assume that these constraints are still given by setting the non-planar Mandelstam variable in certain seven-term identities to a negative constant.}\footnote{Since incompatible factorization channels must overlap, we can always write them as $s_{I_1,I_2}$ and $s_{I_2,I_3}$, where $I_2$ is the overlap. Then setting the non-planar Mandelstam variable $s_{I_1,I_3}=-c_{I_1,I_3}<0$  can forbid $s_{I_1,I_2}$ and $s_{I_2,I_3}$ being zero at the same time, since the right hand side of the seven-term identity~\eqref{eq:7term} is manifestly positive in the subspace.}
%$\mathcal{A}_{n-1}$ \textit{and} the facet of the $s_{\mathsf{B}_{m}}\rightarrow 0$ factorization channel of $\mathcal{A}[\mathsf{B}_{1},\ldots,\mathsf{B}_{m-1},\mathsf{B}_{m}]$ must be $\mathcal{A}[\mathsf{B}_{1},\ldots,\mathsf{B}_{m-1}]\times \mathcal{A}[\mathsf{B}_{m}]$ 
%\item Constraints of $\mathcal{A}_{n}$ must be given by setting non-planar Mandelstam variables in certain 7-term identities to a negative constants.
\end{itemize} 
Remarkably, we again find that $\mathcal{A}_{n}$ has the correct form when these conditions are imposed! It is very interesting that analysis of only two factorization channels seems to be enough to recursively enforce good behavior of the entire open associahedra.

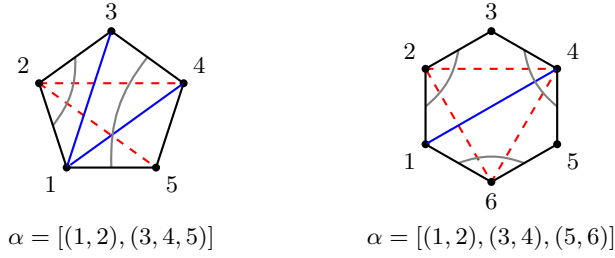
\begin{figure}[t]
\centering
\begin{tikzpicture}[every node/.style={font=\footnotesize,},dir/.style={decoration={markings, mark=at position \halfway with {\arrow{Latex}}},postaction={decorate}}]
\coordinate (p2) at (162:1);
\coordinate (p4) at (18:1);
\coordinate (p5) at (-54:1);
\draw [thick,red,dashed] (p2) -- (p5) (p2) -- (p4);
\node (p1) at (-126:1) [vertex,label={[label distance=-2pt]-126:{$1$}}] {};
\node at (p2) [vertex,label={[label distance=-2pt]162:{$2$}}] {};
\node (p3) at (0,1)  [vertex,label={[label distance=-1pt]90:{$3$}}] {};
\node at (p4) [vertex,label={[label distance=-2pt]18:{$4$}}] {};
\node at (p5) [vertex,label={[label distance=-2pt]-54:{$5$}}] {};
\draw [thick,blue] (p1.center) -- (p3.center) (p1) -- (p4.center);
\path (p1.center) -- (p2.center) node (f1) [pos=0.5] {};
\path (p2.center) -- (p3.center) node (f2) [pos=0.5] {};
\path (p3.center) -- (p4.center) node (f3) [pos=0.5] {};
\path (p5.center) -- (p1.center) node (f5) [pos=0.5] {};
\draw [thick,gray] (f1.center) to [bend right=20] (f2.center) (f3.center) to [bend right=20] (f5.center);
\draw [thick] (p1.center) -- (p2.center) -- (p3.center) -- (p4.center) -- (p5.center) -- cycle;
%\draw [thick] (p1.center) to [bend left=30] (p3.center) (p1.center) to [bend right=30] (p3.center);
\node at (0,-1.75) [] {$\alpha=[(1,2),(3,4,5)]$};
\begin{scope}[xshift=5cm]
\coordinate (p1) at (-150:1);
\coordinate (p2) at (150:1);
\coordinate (p3) at (0,1);
\coordinate (p4) at (30:1);
\coordinate (p5) at (-30:1);
\coordinate (p6) at (0,-1);
\draw [red,thick,dashed] (p2) -- (p4) -- (p6) -- cycle;
\node at (p1)  [vertex,label={[label distance=-2pt]-150:{$1$}}] {};
\node at (p2)  [vertex,label={[label distance=-2pt]150:{$2$}}] {};
\node at (p3) [vertex,label={[label distance=-1pt]90:{$3$}}] {};
\node at (p4) [vertex,label={[label distance=-2pt]30:{$4$}}] {};
\node at (p5) [vertex,label={[label distance=-2pt]-30:{$5$}}] {};
\node at (p6) [vertex,label={[label distance=-1pt]-90:{$6$}}] {};
\draw [blue,thick] (p1.center) -- (p4.center);
\path (p1.center) -- (p2.center) node (f1) [pos=0.5] {};
\path (p2.center) -- (p3.center) node (f2) [pos=0.5] {};
\path (p3.center) -- (p4.center) node (f3) [pos=0.5] {};
\path (p4.center) -- (p5.center) node (f4) [pos=0.5] {};
\path (p5.center) -- (p6.center) node (f5) [pos=0.5] {};
\path (p6.center) -- (p1.center) node (f6) [pos=0.5] {};
\draw [thick,gray] (f1.center) to [bend right=20] (f2.center) (f3.center) to [bend right=20] (f4.center) (f5.center) to [bend right=20] (f6.center);
\draw [thick] (p1) -- (p2) -- (p3) -- (p4) -- (p5) -- (p6) -- cycle;
%\draw [thick] (p1) -- (p3) -- (p5) -- cycle;
\node at (0,-1.75) [] {$\alpha=[(1,2),(3,4),(5,6)]$};
\end{scope}
\end{tikzpicture}
\caption{The flavor lines given by the $\mathfrak{f}$-$\mathfrak{af}$ pairs are shown in gray, and the red dashed diagonals denote planar poles forbidden by flavor conservation. The blue solid lines are a few examples of the planar poles allowed by flavor conservation.}
%Note that in the left panel, the $\mathfrak{f}$-$\mathfrak{af}$ pair $(1,2)$ and $(3,5)$ correspond to the same diagonal but counted as two different flavor lines.
\label{fig:forbiddenPoles}
\end{figure}

\subsection{Special Color Orderings}
At $n=4$, there are two planar amplitudes $A_4[(1,2),3,4]$ and $A_4[(1,2),(3,4)]$. They can be obtained by pulling the planar scattering form $\Omega_4[1,2,3,4]=d\log s-d\log t$ back to
the following positive geometries respectively:
\begin{align}\label{eq:4pspace}
\pmb{\mathcal{A}}_4[(1,2),3,4]&=\pmb{H}_4[(1,2),3,4]\cup\pmb{\Delta}_4[(1,2),3,4]=\{-u=c_{2,4}>0\}\cup\{s>0\,,t>0\}\,, \nonumber\\
\pmb{\mathcal{A}}_4[(1,2),(3,4)]&=\pmb{H}_4[(1,2),(3,4)]\cup\pmb{\Delta}_4[(1,2),(3,4)]=\emptyset\cup\{s>0\,,t=b_{2,3}>0\}\,.
\end{align}
The former is the same as the bi-adjoint case, while for the latter, we have $t=\text{const}$ in $\pmb{\Delta}_4[(1,2),(3,4)]$ as the flavor conservation forbids this channel. Since now the subspace $\Delta_4$ is already one dimensional, no more constraints are needed so $\pmb{H}_4=\emptyset$ and $H_4$ is simply the full two-dimensional plane $\mathbb{R}^2$ spanned by $s$ and $t$. One can easily show that indeed, 
\begin{align}\label{eq:4pamp}
& \Omega_4[1,2,3,4]\Big|_{{\mathcal{A}}_4[(1,2),3,4]}=\Big(\frac{1}{s}+\frac{1}{t}\Big)ds=A_4[(1,2),3,4]ds\,,\nonumber\\
& \Omega_4[1,2,3,4]\Big|_{{\mathcal{A}}_4[(1,2),(3,4)]}=\frac{ds}{s}=A_4[(1,2),(3,4)]ds\,.
\end{align}

Starting from eq.~\eqref{eq:4pspace}, we show that certain five-point subspaces can be obtained by an inverse soft construction. While the subspace for the ordering $[(1,2),3,4,5]$ is given in eq.~\eqref{eq:IS}, we begin with $[(1,2),(3,4),5]$ as an example. Adding the $\mathfrak{adj}$ particle $5$ to $[(1,2),(3,4)]$ does not lead to any new flavor constraints, so we have
\begin{align}
\pmb{F}_5[(1,2),(3,4),5]=\pmb{F}_4[(1,2),(3,4)]=\left\{s_{2,3}=b_{2,3}>0\right\},
\end{align}
which makes $\Delta_5[(1,2),(3,4),5]$ a four dimensional subspace. We thus need two more constraints in $\pmb{H}_5$ to make $\mathcal{A}_5$ a two dimensional kinematic polytope in which the facet $X_{1,4}=0$ gives the previous four-point sub-geometry $\{X_{1,3}>0\}$. Thus the two additional constraints need to ensure that the facets $X_{2,5}=0$ and $X_{3,5}=0$ do not intersect $X_{1,4}=0$ in the positive region. 
One can check that the following two constraints can do the job:
%Introducing the $\mathfrak{adj}$ particle $5$ gives rise to more factorization channels, and the two constraints in $\pmb{H}_5$ have to forbid incompatible factorization channels being reached simultaneously, namely,\footnote{Working with the $(X_{2,5},X_{1,3})$ and $(X_{3,5},X_{1,4})$ pair is sufficient to guarantees that $X_{2,5}$ and $X_{1,4}$ cannot both equal to zero because of eq.~\eqref{eq:5pfc2}.}
%\begin{align}\label{eq:5pfc1}
%& X_{2,5}\text{ and }X_{1,3}\text{ cannot both equal to zero,}\nonumber\\
%& X_{3,5}\text{ and }X_{1,4}\text{ cannot both equal to zero.}
%\end{align}
%Since incompatible factorization channels must overlap, we can always write them as $S_{I_1,I_2}$ and $S_{I_2,I_3}$, where $I_2$ is the overlap. Then the seven-term identity~\eqref{eq:7term} implies that $-S_{I_1,I_3}=c_{I_1,I_3}>0$ can do the job: since the right hand side of eq.~\eqref{eq:7term} is manifestly positive in the subspace, we cannot set $S_{I_1,I_2}$ and $S_{I_2,I_3}$ to zero at the same time. Applying this general consideration to the requirement~\eqref{eq:5pfc1} leads to $-s_{2,5}=c_{2,5}>0$ and $-s_{3,5}=c_{3,5}>0$. Indeed, these two constraints can be rewritten as %$-c_{2,5}+X_{2,5}+X_{1,3}=X_{3,5}$ and $-c_{3,5}+X_{3,5}+X_{1,5}=X_{1,3}$.
\begin{align}\label{eq:5pfc2}
-c_{2,5}+X_{2,5}+X_{1,3}=X_{3,5}\,,& & -c_{3,5}+X_{3,5}+X_{1,4}=X_{1,3}\,.
\end{align}
%Since all the $X_{i,j}$ are non-negative as imposed by $\pmb{\Delta}_5[(1,2),(3,4),5]$, the requirement~\eqref{eq:5pfc1} is satisfied. 
Now we have already reduced the dimension of the subspace to two, no more constraints can be added. We thus have 
\begin{align}\label{eq:H5_a}
\pmb{H}_5[(1,2),(3,4),5]&=\left\{-s_{2,5}=c_{2,5}>0,-s_{3,5}=c_{3,5}>0\right\}\nonumber\\
&=\pmb{H}_4[(1,2),(3,4)]\cup\pmb{C}_1[(1,2),(3,4),5]\,.
\end{align}
Since $\pmb{H}_4[(1,2),(3,4)]=\emptyset$, the set $\pmb{C}_1[(1,2),(3,4),5]$ is just the right hand side of the first line. If we choose $X_{1,3}$ and $X_{1,4}$ as the basis, the kinematic polytope $\mathcal{A}_5[(1,2),(3,4),5]$ is bounded by the following inequalities:
\begin{align}
& X_{1,3}>0\,,\qquad X_{2,5}=-X_{1,4}+c_{2,5}+c_{3,5}>0\,,\nonumber\\
& X_{1,4}>0\,,\qquad X_{3,5}=X_{1,3}-X_{1,4}+c_{3,5}>0\,.
\end{align}
One can verify that by pulling the five-point planar scattering form~\eqref{eq:Omega5} back to the subspace ${\mathcal{A}}_5[(1,2),(3,4),5]$, we get the correct amplitude:
\begin{align}
\Omega_5[1,2,3,4,5]\Big|_{{\mathcal{A}}_5[(1,2),(3,4),5]}&=\Big(\frac{1}{X_{1,3}X_{1,4}}+\frac{1}{X_{1,3}X_{3,5}}+\frac{1}{X_{2,5}X_{3,5}}\Big)d^2X\nonumber\\
&=A_5[(1,2),(3,4),5]\,d^2X\,.
\end{align}
The construction of~\eqref{eq:H5_a} follows the recursive inverse soft pattern~\eqref{eq:IS}. The shape of $\mathcal{A}_5[(1,2),(3,4),5]$ is shown in the left panel of figure~\ref{fig:A5}.

In fact, the above calculation is a special case of a more general result: %\emph{if all the $\mathfrak{f}$-$\mathfrak{af}$ pairs are adjacent in} $\alpha\in\pmb{M}_{n-1,k}^{\text{planar}}$, \emph{then $\pmb{H}_n[\alpha,n]$ is given by:}
\emph{if} $\alpha\in\pmb{M}_{n-1,k}$ \emph{
does not contain nested $\mathfrak{f}$-$\mathfrak{af}$ pairs, then $\pmb{H}_{n}[\alpha,n]$ is given by:}
\begin{align}\label{eq:H_nIS}
\pmb{H}_{n}[\alpha,n]&=\pmb{H}_{n-1}[\alpha]\cup\pmb{C}_1[\alpha,n]\,,\nonumber\\
\pmb{C}_1[\alpha,n]&=\left\{-s_{\alpha(i),n}=c_{\alpha(i),n}>0\text{ for }2\leqslant i\leqslant n-2\right\}.
\end{align}
In other words, we require that the $\mathfrak{f}$-$\mathfrak{af}$ blocks in $\alpha$ have only $\mathfrak{adj}$ sub-blocks. We can understand this result by a simple factorization analysis. 
%eq.~\eqref{eq:F_nIS} indicates that if all the $\mathfrak{f}$-$\mathfrak{af}$ pairs in $\alpha$ are adjacent, there are $n{-}3$ more constraints in $\pmb{H}_n[\alpha,n]$ than $\pmb{H}_{n-1}[\alpha]$ in order for $\mathcal{A}_n[\alpha,n]$ to have the right dimension.\footnote{This statement holds for the more general case that there are no nested $\mathfrak{f}$-$\mathfrak{af}$ pairs.} These additional constraints 
In the kinematic polytope $\mathcal{A}_n[\alpha,n]$ carved out by the constraints $\pmb{H}_n[\alpha,n]\cup\pmb{F}_n[\alpha,n]$, the facet $X_{\alpha(1),\alpha(n-1)}=0$ should reduce to the $(n{-}1)$-point kinematic polytope $\mathcal{A}_{n-1}[\alpha]$. This can be achieved if $\pmb{H}_{n-1}[\alpha]$ is included in $\pmb{H}_n[\alpha,n]$ while the additional constraints $\pmb{C}_1[\alpha,n]$ are automatically satisfied on the facet $X_{1,n-1}=0$. 
%On the other hand, eq.~\eqref{eq:F_nIS} indicates that if all the $\mathfrak{f}$-$\mathfrak{af}$ pairs in $\alpha$ are adjacent, there are $n{-}3$ more constraints in $\pmb{H}_n[\alpha,n]$ than $\pmb{H}_{n-1}[\alpha]$, namely, $|\pmb{C}_1[\alpha,n]|=n{-}3$. 
On the other hand, if $\alpha$ contains no nested $\mathfrak{f}$-$\mathfrak{af}$ pairs, there are $n{-}3$ more constraints in $\pmb{H}_n[\alpha,n]$ than $\pmb{H}_{n-1}[\alpha]$, namely, $|\pmb{C}_1[\alpha,n]|=n{-}3$. 
It is thus natural to devote them to ensure that the $n{-}3$ facets $X_{\alpha(i),n}=0$ with $2\leqslant i\leqslant n{-}2$, which correspond to the factorization channels incompatible with $X_{\alpha(1),\alpha(n-1)}=0$, do not intersect the facet $X_{\alpha(1),\alpha(n-1)}=0$ in the positive region. Indeed, the $\pmb{C}_1[\alpha,n]$ in eq.~\eqref{eq:H_nIS} can do the job. We can rewrite them as
\begin{align}\label{eq:IS2}
X_{\alpha(i),n}+X_{\alpha(1),\alpha(n-1)}=X_{\alpha(1),\alpha(i)}+\sum_{k=\alpha(i)}^{\alpha(n-2)}c_{k,n}\,,\qquad 2\leqslant i\leqslant n-2\,,
\end{align}
which guarantee that $X_{\alpha(i),n}$ is strictly positive if $X_{\alpha(1),\alpha(n-1)}=0$.
%\footnote{We note that the statement~\eqref{eq:H_nIS} actually holds for the more general case that there are no nested $\mathfrak{f}$-$\mathfrak{af}$ pairs in $\alpha$. One can easily check that $|\pmb{C}_1[\alpha,n]|=n{-}3$ is still true, and then the same reasoning follows. } 
%Since now the dimension of ${\mathcal{A}}_n[\alpha,n]={H}_n[\alpha,n]\cap{\Delta}_n[\alpha,n]$ is $n{-}3$, no more constraints can be added. 
%One can then recursively go to lower points until the last block turns out to be an $\mathfrak{f}$-$\mathfrak{af}$ pair, which we will resolve next.
One can then recursively go to lower points until the last block turns out to be an $\mathfrak{f}$-$\mathfrak{af}$ block.

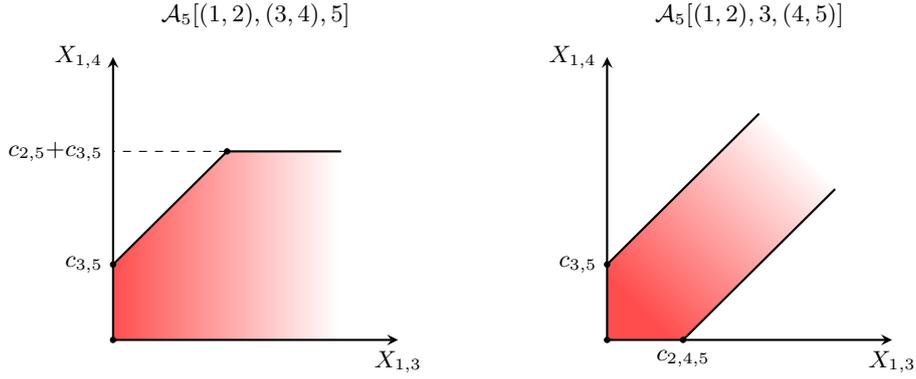
\begin{figure}[t]
	\centering
	\begin{tikzpicture}[every node/.style={font=\footnotesize}]
	\node at (1.875,4.3) {$\mathcal{A}_5[(1,2),(3,4),5]$};
	%\fill [fill=red!50!white] (3,0) -- (0,0) -- (0,1) -- (1.5,2.5) -- (3,2.5);
	\shade [left color=red!70!white,right color=white] (3,0) -- (0,0) -- (0,1) -- (1.5,2.5) -- (3,2.5);
	\draw [thick,-stealth] (0,0) -- (3.75,0) node[below=0pt]{$X_{1,3}$};
	\draw [thick,-stealth] (0,0) -- (0,3.75) node[left=0pt]{$X_{1,4}$};
	\draw [thick] (0,1) -- (1.5,2.5) -- (3,2.5);
	\draw [dashed] (1.5,2.5) -- (0,2.5) node[left=0pt]{$c_{2,5}{+}c_{3,5}$};
	\filldraw (0,0) circle (1pt) (1.5,2.5) circle (1pt)  (0,1) circle (1pt) node[left=0pt]{$c_{3,5}$};
	%\node at (0.5,0) [below=0pt]{$\underbrace{\hphantom{aaaa}}_{c_{3,5}}$};
	\begin{scope}[xshift=6.5cm]
	\node at (1.875,4.3) {$\mathcal{A}_5[(1,2),3,(4,5)]$};
	%\fill [fill=red!50!white] (1,0) ++(2,2) -- (1,0) -- (0,0) -- (0,1) -- ++(2,2);
	\shade [left color=red!70!white,right color=white,shading angle=135] (1,0) ++(2,2) -- (1,0) -- (0,0) -- (0,1) -- ++(2,2);
	\draw [thick,-stealth] (0,0) -- (3.75,0) node[below=0pt]{$X_{1,3}$};
	\draw [thick,-stealth] (0,0) -- (0,3.75) node[left=0pt]{$X_{1,4}$};
	\draw [thick] (1,0) -- ++(2,2) (0,1) -- ++(2,2);
	\filldraw (0,0) circle (1pt) (1,0) circle (1pt) node[below=0pt]{$c_{2,4,5}$} (0,1) circle (1pt);
	\node at (0,1) [left=0pt]{$c_{3,5}$};
	\end{scope}
	\end{tikzpicture}
	\caption{Two examples of the kinematic polytope at $n=5$.}
	\label{fig:A5}
\end{figure}

We now further restrict ourselves to the case that all the $\mathfrak{f}$-$\mathfrak{af}$ pairs in $\alpha$ are \emph{adjacent}, namely, no substructures are allowed in an $\mathfrak{f}$-$\mathfrak{af}$ block.
We start with some special cases of the form $[\alpha,(n{-}1,n)]$. The most trivial one is of course $[(1,2),(3,4)]$ as we have studied in eq.~\eqref{eq:4pspace} and~\eqref{eq:4pamp}. We now move on to $n=5$, considering $[(1,2),3,(4,5)]$. Flavor conservation sets $s_{1,5}=b_{1,5}>0$, and we need two additional constraints in $\pmb{H}_5$. We can obtain this ordering by gluing a three-point vertex $[-q,(4,5)]$ to $[(1,2),3,q]$. On the facet $X_{1,4}=0$ the subspace should return to that of $[(1,2),3,q]$, which is given by $-s_{2,q}=\text{const}$. We thus get the first constraint $-s_{2,q}\rightarrow-s_{2,4,5}=\text{const}$ for $\pmb{H}_5$ from this ``inverse factorization'' consideration. The second constraint needs to guarantee that the facet $X_{3,5}=0$ does not intersect $X_{1,4}=0$, which can be achieved by $-s_{3,5}=c_{3,5}>0$.
%\begin{align}
%X_{1,4}\text{ and }X_{3,5}\text{ cannot both equal to zero,}
%\end{align}
%which leads to the second constraint $-s_{3,5}=c_{3,5}>0$. 
Therefore, we have
\begin{align}
\pmb{H}_{5}[(1,2),3,(4,5)]&=\left\{-s_{2,4,5}=c_{2,4,5}>0,-s_{3,5}=c_{3,5}>0\right\}\nonumber\\
%%&=H_3[(1,2),3]\cup C_1[(1,2),3,(4,5)]\cup C_2[(1,2),3,(4,5)]\,,\nonumber\\
\pmb{F}_{5}[(1,2),3,(4,5)]&=\left\{s_{1,5}=b_{1,5}>0\right\}\,.
\end{align}
%where $C_1[(1,2),3,(4,5)]=\{-s_{2,4,5}=c_{2,4,5}>0\}$ and $C_2[(1,2),3,(4,5)]=\{-s_{3,5}=c_{3,5}>0\}$. 
If we still choose $\{X_{1,3},X_{3,5}\}$ as the basis, the kinematic polytope $\mathcal{A}_5[(1,2),3,(4,5)]$ is bounded by the following inequalities:
\begin{align}
& X_{1,3}>0\,,\qquad X_{3,5}=X_{1,3}-X_{1,4}+c_{3,5}>0\,,\nonumber\\
& X_{1,4}>0\,,\qquad X_{2,4}=X_{1,4}-X_{1,3}+c_{2,4,5}>0\,.
\end{align}
It is then straightforward to check that the pullback of the form~\eqref{eq:Omega5} indeed gives the correct amplitude:
\begin{align}
\Omega_5[1,2,3,4,5]\Big|_{{\mathcal{A}}_5[(1,2),3,(4,5)]}&=\Big(\frac{1}{X_{1,3}X_{1,4}}+\frac{1}{X_{1,4}X_{2,4}}+\frac{1}{X_{1,3}X_{3,5}}\Big)d^2X\nonumber\\
&=A_5[(1,2),3,(4,5)]\,d^2X\,.
\end{align}
The shape of $\mathcal{A}_5[(1,2),3,(4,5)]$ is shown in the right panel of figure~\ref{fig:A5}.

We can generalize the above calculation to the following statement: \emph{if all the $\mathfrak{f}$-$\mathfrak{af}$ pairs are adjacent in} $\alpha\in\pmb{M}_{n-2,k}$, \emph{then $\pmb{H}_n[\alpha,(n-1,n)]$ is given by:} %\emph{if} $\alpha\in\pmb{M}_{n-2,k}^{\text{planar}}$ \emph{contains only adjacent $\mathfrak{f}$-$\mathfrak{af}$ pairs, we have the following recursive construction:}
\begin{align}\label{eq:H_nIF}
\pmb{H}_n[\alpha,(n{-}1,n)]&=\pmb{H}_{n-1}[\alpha,q]\Big|_{s_{q,\ldots}\rightarrow s_{n-1,n,\ldots}}\cup\pmb{C}_2[\alpha,(n{-}1,n)]\nonumber\\
&=\pmb{H}_{n-2}[\alpha]\cup\pmb{C}_1[\alpha,(n{-}1,n)]\cup\pmb{C}_2[\alpha,(n{-}1,n)]\,.
\end{align}
The second equality holds since we can write $\pmb{H}_{n-1}[\alpha,q]=\pmb{H}_{n-2}[\alpha]\cup \pmb{C}_1[\alpha,q]$, and the $q$ only appears in $\pmb{C}_1$. Thus the replacement $s_{q,\ldots}\rightarrow s_{n-1,n,\ldots}$ only affects $\pmb{C}_1$: 
\begin{align}\label{eq:replace1}
\pmb{C}_1[\alpha,(n{-}1,n)]=\pmb{C}_1[\alpha,q]\Big|_{s_{q,\ldots}\rightarrow s_{n-1,n,\ldots}}\,.
\end{align}
It ensures that on the facet $X_{\alpha(1),n-1}=q^2=0$, the constraints land back on $\pmb{H}_{n-1}[\alpha,q]$. To reach an $(n{-}3)$-dimensional kinematic polytope, a simple counting from eq.~\eqref{eq:F_nIS} shows that we need $|\pmb{C}_2[\alpha,(n{-}1,n)]|=n{-}k{-}2$. The additional constraints should be automatically satisfied on $X_{\alpha(1),n-1}=0$.
%\begin{align}
%H_{n}[\alpha,(n{-}1,n)]=&\;H_{n-1}[\alpha,q]\Big|_{s_{q,\ldots}\rightarrow s_{n-1,n,\ldots}}\nonumber\\
%%&\cup\left\{\begin{array}{c}
%-s_{i,n}=c_{i,n}>0\text{ for each }\mathfrak{adj}\text{ particle }i, \\
%-s_{i,i+1,n}=c_{i,i+1,n}>0\text{ for each }\mathfrak{f}\text{-}\mathfrak{af}\text{ pair }(i,i{+}1)
%\end{array}\right\},
%\end{align}
%for $2\leqslant i\leqslant n-2$, and $q$ is an $\mathfrak{adj}$ particle. The replacement $s_{q,\ldots}\rightarrow s_{n-1,n,\ldots}$ ensures that on the factorization channel $X_{1,n-1}=0$, the subspace lands back on $H_{n-1}[\alpha,q]$. 
Therefore, we use them to ensure that the $n{-}k{-}2$ incompatible factorization channels at $X_{\alpha(1),n-1}=0$,
\begin{align}\label{eq:requirement1}
X_{\alpha(i),n}\text{ for each $\mathfrak{adj}$ particle $i$ or each $\mathfrak{f}$-$\mathfrak{af}$ pair $(i,i+1)$ with $3\leqslant i\leqslant n-2$,}
\end{align}
cannot reach zero.
%As before, we choose them as the independent necessary conditions for the new factorization channels being consistent: the second line of eq.~\eqref{eq:H_nIF} ensures that for each $\mathfrak{adj}$ particle $i$, $X_{i,n}$ and $X_{1,i+1}$ cannot both equal to zero  and for each $\mathfrak{f}$-$\mathfrak{af}$ pair $(i,i{+}1)$, $X_{i,n}$ and $X_{1,i+2}$ cannot both equal to zero.
%\begin{align}
%X_{i,n}\text{ and }\left\{\begin{array}{l}
%X_{1,i+1}\text{ if }i\in\{\mathfrak{adj}\} \\
%X_{1,i+2}\text{ if }i\in\{\mathfrak{f}\}
%\end{array}\right.\text{ cannot both equal to zero.}
%\end{align}
Since all these factorization channels carry the flavor $f_{n-1}$, they must contain other $\mathfrak{f}$-$\mathfrak{af}$ pairs as a whole due to flavor conservation. We can thus effectively treat the pairs in $\alpha$ as a single off-shell $\mathfrak{adj}$ particle and write
%One natural choice that resembles $\pmb{C}_1$ is
\begin{align}\label{eq:C2special}
\pmb{C}_2[\alpha,(n{-}1,n)]=\left\{\begin{array}{ll}
-s_{\alpha(i),n}=c_{\alpha(i),n}>0 & ~\mathfrak{adj}\text{ particle }\alpha(i), \\
-s_{\alpha(i),\alpha(i+1),n}=c_{\alpha(i),\alpha(i+1),n}>0 &~\mathfrak{f}\text{-}\mathfrak{af}\text{ pair }\big(\alpha(i),\alpha(i{+}1)\big)
\end{array}\right\},
\end{align}
for $3\leqslant i\leqslant n-2$. 
%Comparing with the $\pmb{C}_1$ in eq.~\eqref{eq:H_nIS}, one can clearly see the resemblance. 
After rewriting these constraints into the form
\begin{align}
X_{\alpha(i),n}+X_{\alpha(1),n-1}=X_{\alpha(1),\alpha(i)}+\sum_{k=i}^{n-2}\left\{\begin{array}{lll}
c_{\alpha(k),n} &~& \mathfrak{adj}\text{ particle }\alpha(k) \\
c_{\alpha(k),\alpha(k+1),n} &~& \mathfrak{f}\text{-}\mathfrak{af}\text{ pair } \big(\alpha(k),\alpha(k+1)\big)
\end{array}\right\},
\end{align}
one can clearly see that the requirement~\eqref{eq:requirement1} is satisfied.
%It would reduce to the $\pmb{C}_1$ given in eq.~\eqref{eq:H_nIS} if the $\mathfrak{f}$-$\mathfrak{af}$ pairs were $\mathfrak{adj}$ particles.
%The reason for using the constraints derived from these crossing pairs is similar to that given below eq.~\eqref{eq:constraint1}.
%One can then easily check that these constraints are actually sufficient to forbid all the new crossing factorization channels.

Using eq.~\eqref{eq:H_nIS} and~\eqref{eq:H_nIF}, we can recursively generate the subspace for all the orderings in $\pmb{M}_{n,k}$ in which the $\mathfrak{f}$-$\mathfrak{af}$ pairs are adjacent.
%and we have checked up to a sufficiently high multiplicity. The algorithm works as follows: starting from a lower-point ordering, we can add an $\mathfrak{adj}$ particle using eq.~\eqref{eq:H_nIS}. If we want to add an $\mathfrak{f}$-$\mathfrak{af}$ pair, we first use eq.~\eqref{eq:H_nIS} to add an $\mathfrak{adj}$ particle and then use eq.~\eqref{eq:H_nIF} to turn it into a $\mathfrak{f}$-$\mathfrak{af}$ pair. 
\begin{figure}
\centering
  \includegraphics[scale=0.6]{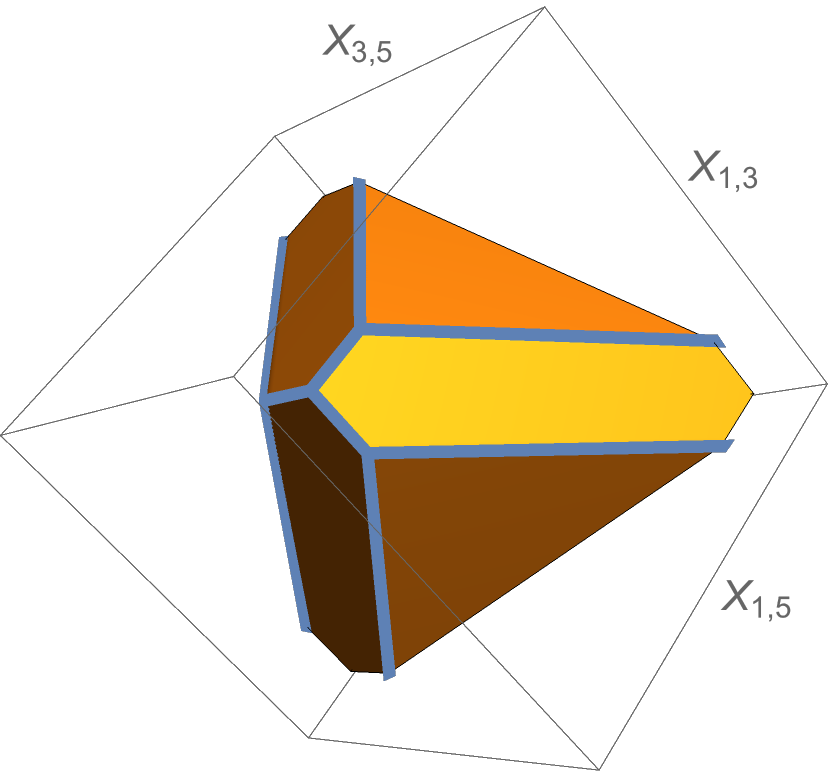}
  \caption{A visualization of the $\mathcal{A}[(1,2),(3,4),(5,6)]$ open associahedra. It can be interpreted as an infinite cone with additional structure. The geometry is unbounded and the thick gray lines correspond to the edges.}
  \label{fig:6pointexample}
\end{figure}
We now give more examples with higher multiplicities. Starting from eq.~\eqref{eq:H5_a} and following eq.~\eqref{eq:H_nIF}, we reproduces eq.~\eqref{eq:H6a},
\begin{align}
\pmb{H}_6[(1,2),(3,4),(5,6)]&=\pmb{H}_5[(1,2),(3,4),q]\Big|_{s_{q,\ldots}\rightarrow s_{5,6,\ldots}}\cup\left\{-s_{3,4,6}=c_{3,4,6}>0\right\}\\
&=\left\{-s_{2,5,6}=c_{2,5,6}>0\,,-s_{3,5,6}=c_{3,5,6}>0\,,-s_{3,4,6}=c_{3,4,6}>0\right\}.\nonumber
\end{align}
%To simplify the notation, in future we write instead
%\begin{align}\label{eq:H6}
%\pmb{H}_{6}[(1,2),(3,4),(5,6)]=\{s_{2,5,6}\,,s_{3,5,6}\,,s_{3,4,6}\}\,.
%\end{align}
%Namely, all the Mandelstam variables are understood to be set to a negative constant.
This polytope is shown in figure~\ref{fig:6pointexample}. If we add another $\mathfrak{adj}$ particle following eq.~\eqref{eq:H_nIS}, we get eq.~\eqref{eq:H7a},
\begin{align}
\pmb{H}_7[(1,2),(3,4),(5,6),7]&=\pmb{H}_6[(1,2),(3,4),(5,6)]\cup\left\{-s_{i,7}=c_{i,7}>0\text{ for }2\leqslant i\leqslant 5\right\}\nonumber\\
%&=\{s_{2,5,6}\,,s_{3,5,6}\,,s_{3,4,6}\,,s_{2,7}\,,s_{3,7}\,,s_{4,7}\,,s_{5,7}\}\,.
&=\left\{\begin{array}{c}
s_{2,5,6}\,,s_{3,5,6}\,,s_{3,4,6}\,,s_{2,7}\,,s_{3,7}\,,s_{4,7}\,,s_{5,7} \\
\text{set to negative constants}\end{array}\right\}\,.
\end{align}
Finally, we can turn the last $\mathfrak{adj}$ particle into a $\mathfrak{f}$-$\mathfrak{af}$ pair $(7,8)$ using eq.~\eqref{eq:H_nIF}, which leads to eq.~\eqref{eq:H8a},
\begin{align}\label{eq:H8b}
&\pmb{H}_8[(1,2),(3,4),(5,6),(7,8)] \nonumber\\
&=\pmb{H}_7[(1,2),(3,4),(5,6),q]\Big|_{s_{q,\ldots}\rightarrow s_{7,8\ldots}}\cup\left\{\begin{array}{c}
s_{3,4,8}\,,s_{5,6,8} \\ 
\text{set to negative constants}
\end{array}\right\}\\
&=\left\{\begin{array}{c}
s_{2,5,6}\,,s_{3,5,6}\,,s_{3,4,6}\,,s_{2,7,8}\,,s_{3,7,8}\,,s_{4,7,8}\,,s_{5,7,8}\,,s_{3,4,8}\,,s_{5,6,8} \\
\text{set to negative constants}
\end{array}\right\}.\nonumber
\end{align}
We have checked up to $n=20$ and ten adjacent $\mathfrak{f}$-$\mathfrak{af}$ pairs that the algorithm indeed generates the correct subspace.
%We note that there is another planar ordering $[(1,2),(3,4,5)]$ at $n=5$. However, the $\mathfrak{f}$-$\mathfrak{af}$ pair $(3,5)$ is not adjacent in this ordering. This happens when a $\mathfrak{f}$-$\mathfrak{af}$ block has substructures. We postpone the discussion of such cases to a later section.

%We now move on to some typical planar orderings at $n=6$ in which all the $\mathfrak{f}$-$\mathfrak{af}$ pairs are adjacent.

%The subspace of this ordering is
%\begin{align}
%%& H_5[(1,2),(3,4,5)]=\left\{-s_{3,5}=c_{3,5}>0\right\},\nonumber\\
%%& F_5[(1,2),(3,4,5)]=\left\{s_{2,3}=b_{2,3}>0,s_{1,5}=b_{1,5}>0\right\}.
%\end{align}
%This subspace is better understood by a more general ``inverse factorization'' prescription in which we glue a four-point amplitude $[-q,(3,4,5)]$ to $[(1,2),q]$. We postpone the discussion to a later subsection.
\subsection{Generic Recursive Construction}
\label{sec:reccontr}
%Although the previous subsection only deals with a special subclass of $\pmb{M}_{n,k}^{\text{planar}}$, it illustrates the idea of our recursive construction. 

We now consider a generic $\alpha\in\pmb{M}_{n,k}$, which can be put in the form of~\eqref{eq:genericOrder} by the enlarged cyclic freedom. We can then read off the last block $\mathsf{B}_m$ and write
\begin{align}
\alpha=[\mathsf{B}_1,\mathsf{B}_2,\ldots,\mathsf{B}_{m-1},\mathsf{B}_m]=[\beta,\mathsf{B}_m]\,,\text{ where }\mathsf{B}_1=(l_1,r_1)\text{ or }g_1\,. 
\end{align}
Suppose we already know the subspace constraints for the lower-point ordering $\beta$, our goal is to obtain the constraints $\pmb{H}_n[\alpha]$ through a recursion. The examples in the previous subsection illustrate the idea for a more general construction.

%While the previous subsection has solved this problem for a special sub-class of $\pmb{M}_{n,k}$, it illustrates the idea for a more general construction.

%We now consider a generic $\alpha\in\pmb{M}_{n,k}$, and find a recursive construction of the constraints $\pmb{H}_n$. We first use the enlarged cyclic freedom to put $\alpha$ into the form of~\eqref{eq:genericOrder}. Then we can read off the last block $\mathsf{B}_m$ and write $\alpha=[\beta,\mathsf{B}_m]$. Suppose we already know the subspace constraints for $\beta$

We first study the case that $\mathsf{B}_m=n$ is a single $\mathfrak{adj}$ particle and try to generalize the inverse soft construction~\eqref{eq:H_nIS}. When adding an $\mathfrak{adj}$ particle $n$ to an $(n{-}1)$-point ordering $\beta$, we expect to write the result as
\begin{align}\label{eq:H_nISgeneric}
\pmb{H}_n[\alpha]=\pmb{H}_n[\beta,n]=\pmb{H}_{n-1}[\beta]\cup\pmb{C}_1[\beta,n]\,,
%\qquad \beta\in\pmb{M}_{n-1,k}\,,
\end{align}
%where all the $n$-dependence only appears in $\pmb{C}_1$. 
where $\pmb{H}_{n-1}[\beta]$ is the set of constraints for the kinematic polytope $\mathcal{A}_{n-1}[\beta]$. Comparing with eq.~\eqref{eq:H_nIS}, the difference is that each block of $\beta=[\mathsf{B}_1,\mathsf{B}_2,\ldots,\mathsf{B}_{m-1}]$, except for $\mathsf{B}_1$, may contain $\mathfrak{f}$-$\mathfrak{af}$ sub-blocks as well as $\mathfrak{adj}$ ones. On the facet $X_{l_1,r_{m-1}}=0$, the constraints should reduce to $\pmb{H}_{n-1}[\beta]$, while those in $\pmb{C}_1[\beta,n]$ are automatically satisfied in the positive region of the $(n{-}1)$-point kinematic subspace $\mathcal{K}_{n-1}$ by the strict positivity of the incompatible channels. Very crucially, these incompatible factorization channels are all of the form $\{X_{l_i,n},X_{r_i,n},X_{l_{\mathsf{I}},n}\}$ for each $\mathsf{B}_i$ and $\mathsf{I}\in\text{sub}[\mathsf{B}_i]$, where $l_{\mathsf{I}}$ is the first particle in the sub-block~$\mathsf{I}$. For a single $\mathsf{B}_i$, they are depicted in figure~\ref{fig:Xjn}. In other words, the particle $n$ does not ``see'' any further substructures in $\mathsf{I}$. The reason is that a planar propagator $X_{j,n}$ with $j\in\mathsf{I}\in\text{sub}[\mathsf{B}_i]$ must cross the flavor line of $(l_i,r_i)$ and thus carry its flavor charge. Consequently, it is not allowed to cross any other flavor lines, which means we have to include the entire set $\mathsf{I}$ into the propagator and thus we must have $j=l_{\mathsf{I}}$. 
%The number of constraints in $\pmb{C}_1$ must equal to the number of incompatible channels, since geometrically these constraints ensure that the facets $\{X_{l_i,n},X_{r_i,n},X_{l_{\mathsf{I}},n}\}$ do not intersect $X_{l_1,r_{m-1}}$.
%However, the particle $n$ only talks to a sub-block $\mathsf{I}\subset\text{sub}[\mathsf{B}_i]$ as a whole, no matter whether $\mathsf{I}$ is an $\mathfrak{adj}$ particle or an $\mathfrak{f}$-$\mathfrak{af}$ block. It does not ``see'' any further substructures in $\mathsf{I}$ in the sense that any $\alpha$-planar pole $s_{\mathsf{A}}$ that contains both $n$ and any number of particles in $\mathsf{I}$ is admissible by flavor conservation only if $\mathsf{I}\subset\mathsf{A}$. 
Therefore, we can treat each sub-block of $\mathsf{B}_i$ as if it were an $\mathfrak{adj}$ particle (albeit off-shell), and write the constraints as the following set of Mandelstam variables being set to negative constants:
%\footnote{Here, we use the notations introduced in eq.~\eqref{eq:block} and~\eqref{eq:sub-block} for blocks. We also use the notation first introduced in eq.~\eqref{eq:H6} for constraints: $s_{i,j,k,\ldots}$ stands for $-s_{i,j,k,\ldots}=c_{i,j,k,\ldots}>0$.}
%\begin{align}\label{eq:C1_n}
%\pmb{C}_1[\beta,n]=\left\{\begin{array}{rl}\displaystyle\{s_{n,r_1}\}\bigcup_{i=2}^{m-1}\left\{s_{n,l_i},s_{n,r_i},s_{n,\mathsf{I}}\,\middle|\,\mathsf{I}\in\text{sub}[\mathsf{B}_i]\right\} &\text{ for }\;\beta=[(l_1,r_1),\mathsf{B}_2,\ldots,\mathsf{B}_{m-1}] \\ 
%\displaystyle\bigcup_{i=2}^{m-1}\left\{s_{n,l_i},s_{n,r_i},s_{n,\mathsf{I}}\,\middle|\,\mathsf{I}\in\text{sub}[\mathsf{B}_i]\right\} & \text{ for }\;\beta=[g_1,\mathsf{B}_2,\ldots,\mathsf{B}_{m-1}]
%\end{array}\right.,
%\end{align}
\begin{empheq}[box=\widefboxb]{align}\label{eq:C1_n}
\pmb{C}_1[\beta,n]=\left.\bigcup_{i=1}^{m-1}\bigcup_{\mathsf{I}\in\text{sub}[\mathsf{B}_i]}
\left\{\begin{array}{c}-s_{n,l_i}=c_{n,l_i}>0 \\
-s_{n,r_i}=c_{n,r_i}>0 \\
-s_{n,\mathsf{I}}=c_{n,\mathsf{I}}>0
\end{array}
\right\}\middle\backslash\{s_{n,l_1},s_{n,r_{m-1}}\}\right.,
%\left\{\begin{array}{c} s_{n,l_i},s_{n,r_i},s_{n,\mathsf{I}} \\
%\text{equal to negative constants}
%\end{array}\right\}\middle\backslash\{s_{n,l_1},s_{n,r_{m-1}}\}\right..
\end{empheq}
where the constraints on the planar Mandelstam variable $s_{n,l_1}$ and $s_{n,r_{m-1}}$ have to be deleted by hand. For $\mathsf{B}_1=(l_1,r_1)$ (or $\mathsf{B}_1=g_1$), this prescription ensures that $l_1$ (or $g_1$) does not appear in the Mandelstam variables set to constants by $\pmb{C}_1$. Had all the sub-blocks $\mathsf{I}$ been $\mathfrak{adj}$ particles, eq.~\eqref{eq:C1_n} would become exactly eq.~\eqref{eq:H_nIS}. 

The constraints in eq.~\eqref{eq:C1_n} must one-to-one correspond to the incompatible channels $\{X_{l_i,n},X_{r_i,n},X_{l_\mathsf{I},n}\}$ to ensure that these facets do not intersect $X_{l_1,r_{m-1}}$ in the positive region. To show this, one can rewrite these constraints into 
$X_{\bullet,n}+X_{l_1,r_{m-1}}=X_{l_1,\bullet}+C$ where $\bullet\in\{l_i,l_\mathsf{I},r_i\}$  and  $C$ is a positive constant.
They do not contribute to any boundaries when $X_{l_1,r_{m-1}}=0$, since the incompatible factorization channels $\{X_{l_i,n},X_{r_i,n},X_{l_\mathsf{I},n}\}$ cannot reach zero. Therefore, we can drop the $\pmb{C}_1$ part and the constraints~\eqref{eq:H_nISgeneric} give correct factorization behavior 
\begin{align}
\pmb{H}_n[\alpha]=\pmb{H}_n[\beta,n]\rightarrow\pmb{H}_{n-1}[\beta]\Big|_{r_{m-1}\rightarrow q}\cup\pmb{H}_3\left[\begin{array}{c} (-q,r_{m-1}),n \\
\text{or} \\
-q,r_{m-1},n
\end{array}\right]=\pmb{H}_{n-1}[\beta]\Big|_{r_{m-1}\rightarrow q}\,,
\end{align}
depending on whether $\mathsf{B}_{m-1}$ is an $\mathfrak{f}$-$\mathfrak{af}$ block or an $\mathfrak{adj}$ particle. Since the three-point kinematic space is zero-dimensional, we have $\pmb{H}_3=\emptyset$ for both cases.

\begin{figure}[t]
\centering
\begin{tikzpicture}[every node/.style={font=\footnotesize},scale=1]
\draw [pattern=north west lines] (0.5,0) ellipse (0.5 and 0.2);
\path (-0.5,0) node (l2) {} -- ++(0.5,0) node (f2l) [midway] {};
\path (1,0) node (r2) {} -- ++(0.5,0) node (f2r) [midway] {};
\filldraw (0,0) circle (1pt);
\draw [thick,gray] (f2l.center) to [bend right=60] (f2r.center);
\draw [thick] (1,0) -- ++(0.5,0) (-0.5,0) -- ++(0.5,0);
\begin{scope}[xshift=1.5cm,rotate=-30]
\draw [pattern=north west lines] (1,0) ellipse (0.5 and 0.2);
\path (0,0) node (l3) {} -- ++(0.5,0) node (f3l)[midway] {};
\filldraw (0.5,0) circle (1pt);
\path (1.5,0) node (r3) {} -- ++(0.5,0) node (f3r) [midway] {};
\draw [thick,gray] (f3l.center) to [bend right=60] (f3r.center);
\node (ri) at (2,0) {};
\draw (0,0) -- ++(0.5,0) (1.5,0) -- ++(0.5,0);
\end{scope}
\begin{scope}[xshift=-0.5cm,rotate=30]
\draw [pattern=north west lines] (-1,0) ellipse (0.5 and 0.2);
\path (-2,0) node (l1) {} -- ++(0.5,0) node (f1l) [midway] {};
\path (-0.5,0) node (r1) {} -- ++(0.5,0) node (f1r) [midway] {};
\draw [thick,gray] (f1l.center) to [bend right=60] (f1r.center);
\draw [thick] (-2,0) -- ++(0.5,0) (-0.5,0) -- ++ (0.5,0);
\end{scope}
\begin{scope}[yshift=-3.5cm,xshift=0.5cm]
\draw [thick,red] (-1,0) node (a1) {} -- (1,0) node (an1) {};
\draw [thick] (-1,0) -- (0,-0.5) node (an) {} -- (1,0);
\end{scope}
\path (l1.center) -- ++(0,-0.5) node (fil) {};
\path (ri.center) -- ++(0,-0.5) node (fir) {};
\draw [thick,gray] (fil.center) -- (fir.center);
\draw [thick] (l1.center) -- ++(0,-1) node (li) {} (ri.center) -- ++(0,-1) node (lj) {};
%\draw [thick] (l1.center) to [bend right=80] (l2.center);
%\draw [thick] (l2.center) to [bend right=80] (l3.center);
%\draw [thick] (l3.center) to [bend right=80] (ri.center);
%\draw [thick] (li.center) -- (lj.center);
\draw [thick] (li.center) -- (a1.center) node [midway,below,sloped]{$\cdots\cdots$};
\draw [thick] (lj.center) -- (an1.center) node [midway,below,sloped]{$\cdots\cdots$};
\draw [thick,red,dashed] (li.center) -- (an.center) -- (l1.center) (l2.center) -- (an.center) -- (l3.center) (ri.center) -- (an.center);
\filldraw (l2.center) circle (1pt) node [above=1.5pt]{$l_{i_2}$} (r2.center) circle (1pt) node [above=1.5pt]{$r_{i_2}$};
\filldraw (l3.center) circle (1pt) node [above=0pt]{$l_{i_3}$} (r3.center) circle (1pt) node [above right=-1pt]{$r_{i_3}$};
\filldraw (l1.center) circle (1pt) node [left=0pt]{$l_{i_1}$} (r1.center) circle (1pt) node [above=0pt]{$r_{i_1}$};
\filldraw (ri.center) circle (1pt) node [right=0pt]{$r_i$};
\filldraw (li.center) circle (1pt) node[left=0pt]{$l_i$} (lj.center) circle (1pt) node[right=0pt]{$l_{i+1}$};
\filldraw (a1.center) circle (1pt) node [left=0pt]{$l_1$} (an1.center) circle (1pt) node [right=1pt]{$r_{m-1}$} (an.center) circle (1pt) node[below=0pt]{$n$};
\end{tikzpicture}
\caption{For a given $\mathsf{B}_i$, the diagonals $\{X_{l_i,n},X_{r_i,n},X_{l_{\mathsf{I}},n}\}$, shown in red dashed lines, are allowed by flavor conservation but incompatible with $X_{l_1,r_{m-1}}$ (red solid line). Flavor lines are shown in gray and each shaded region represents a sub-block $\mathsf{B}_{i_\ell}$ of $\mathsf{B}_i$. Planar propagators are allowed by flavor conservation if they only cross as most one flavor line.}
\label{fig:Xjn}
\end{figure}
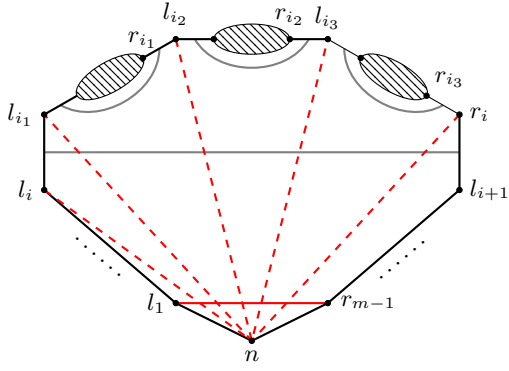
Next, we consider the case that $\mathsf{B}_m=(l_m,r_m)$ is an adjacent $\mathfrak{f}$-$\mathfrak{af}$ pair. Generalizing from eq.~\eqref{eq:H_nIF}, we expect the result to be
\begin{empheq}[box=\widefboxc]{align}\label{eq:H_nIF2}
\pmb{H}_n[\alpha]&=\pmb{H}_n[\beta,(l_m,r_m)]=\pmb{H}_{n-2}[\beta]\cup\pmb{C}_1[\beta,(l_m,r_m)]\cup\pmb{C}_2[\beta,(l_m,r_m)]\,,
%\quad\beta\in\pmb{M}_{n-2,k-1}\,.
\end{empheq}
where $\pmb{H}_{n-2}[\beta]$ is the set of constraints for the kinematic polytope $\mathcal{A}_{n-2}[\beta]$. Following the same reasoning for eq.~\eqref{eq:replace1}, we require that the constraints land back on $\pmb{H}_{n-1}[\beta,q]=\pmb{H}_{n-2}[\beta]\cup\pmb{C}_1[\beta,q]$ on the facet $X_{l_1,l_m}=q^2=0$, which naturally leads to the following replacement on the $\pmb{C}_1$ in eq.~\eqref{eq:C1_n}:
\begin{empheq}[box=\widefboxb]{align}\label{eq:C1lmrm}
\pmb{C}_1[\beta,(l_m,r_m)]=\pmb{C}_1[\beta,q]\Big|_{s_{q,\ldots}\rightarrow s_{l_m,r_m,\ldots}}\,.
\end{empheq}
The constraints $\pmb{C}_2$, on the other hand, have to be automatically satisfied when $X_{l_1,l_m}=0$ due to the positivity of incompatible factorization channels $\{X_{l_i,r_m}\}$ with $2\leqslant i\leqslant m{-}1$. Again, flavor conservation requires that if $X_{j,r_m}$ with $j\in\mathsf{B}_i$ is an admissible planar propagator, we must have $j=l_i$ since $X_{j,r_m}$ already carries the flavor of the pair $(l_m,r_m)$. In other words, the particle $r_m$ only ``sees'' other blocks $\mathsf{B}_i$ as a single off-shell $\mathfrak{adj}$ particle. This leads to the generalization from eq.~\eqref{eq:C2special}:
\begin{empheq}[box=\widefboxb]{align}\label{eq:C2_n}
\pmb{C}_2[\beta,(l_m,r_m)]=\bigcup_{i=2}^{m-1}\{-s_{\mathsf{B}_i,r_m}=c_{\mathsf{B}_i,r_m}>0\}\,.
%\{s_{\mathsf{B}_i,r_m}\text{ equals to negative constant}\}\,.
\end{empheq} 
We note that the leg $l_m$ does not appear in the Mandelstam variables set to constants by $\pmb{C}_2$. There are $m{-}2$ constraints in $\pmb{C}_2$, which equals to the number of incompatible channels $\{X_{l_i,r_m}\}$. If all the $\mathsf{B}_i$'s are either adjacent $\mathfrak{f}$-$\mathfrak{af}$ pairs or $\mathfrak{adj}$ particles, these constraints reduce to eq.~\eqref{eq:C2special}.
By rewriting eq.~\eqref{eq:C2_n} as
\begin{align}
X_{l_i,r_m}+X_{l_1,l_m}=X_{l_1,l_i}+\sum_{k=i}^{m-1}\left(X_{l_k,l_{k+1}}+c_{\mathsf{B}_k,r_m}\right)\,,\qquad 2\leqslant i\leqslant m-1\,,
\end{align}
one can easily see that when $X_{l_1,l_m}=0$, they do not carve out any boundary in the positive region of the kinematic space $\mathcal{K}_{n-2}$ since $X_{l_i,r_m}$ cannot reach zero, such that they can simply be dropped. Therefore, we have proved that the constraints~\eqref{eq:H_nIF2} have the correct factorization behavior at $X_{l_1,l_m}=q^2=0$,
\begin{align}
\pmb{H}_n[\alpha]=\pmb{H}_{n}[\beta,(l_m,r_m)]\rightarrow\pmb{H}_{n-1}[\beta,q]\cup\pmb{H}_3[-q,(l_m,r_m)]=\pmb{H}_{n-1}[\beta,q]\,,
\end{align}
where we have used $\pmb{H}_3=\emptyset$.

We now generalize eq.~\eqref{eq:H_nIF2} and consider $\mathsf{B}_m$ be an $\mathfrak{f}$-$\mathfrak{af}$ block with substructures $\mathsf{B}_m=(l_m,\mathsf{B}_{m_1},\ldots,\mathsf{B}_{m_s},r_m)$. Similar to the previous special cases, we can obtain a recursive construction by studying the factorization involving the last block. We first consider $X_{l_m,r_m}=L_m^2=0$, on which the amplitude should factorize as
\begin{align}\label{eq:ALAR1}
A_n[\alpha]=A_n[\beta,\mathsf{B}_m]\longrightarrow A^L[\beta,(L_m,r_m)]\,\frac{1}{X_{l_m,r_m}}\,A^R[(-{L}_m,l_m),\mathsf{B}_{m_1},\ldots,\mathsf{B}_{m_s}]\,,
\end{align}
where $L_m$ satisfies $(L_m+r_m)^2=s_{\mathsf{B}_m}=X_{l_1,l_m}$. We thus require that the constraints $\pmb{H}_n[\alpha]$ reduce to those for $A^L$ and $A^R$ on the facet $X_{l_m,r_m}=0$. It is natural to include the subspace constraints for the left and right sub-amplitudes into $\pmb{H}_n[\alpha]$ and write
\begin{empheq}[box=\widefboxb]{align}\label{eq:H_ngeneric}
\pmb{H}_n[\alpha]=\pmb{H}_n[\beta,\mathsf{B}_m]=&\;\pmb{H}_{n-|\mathsf{B}_m|+2}[\beta,(L_m,r_m)]\cup\pmb{C}_3[\beta,\mathsf{B}_m]\nonumber\\
&\cup\pmb{H}_{|\mathsf{B}_m|}[(-{L}_m,l_m),\mathsf{B}_{m_1},\ldots,\mathsf{B}_{m_s}]\,.
%\quad\beta\in\pmb{M}_{n-|\mathsf{B}_m|,k-1}\,.
\end{empheq}
While the first and last piece of $\pmb{H}_n[\alpha]$ are defined recursively, the $\pmb{C}_3$ part contains the constraints that automatically drop out when $X_{l_m,r_m}=0$ due to the strict positivity of the incompatible channels $\{X_{l_i,l_\mathsf{I}}\}$ with $1\leqslant i\leqslant m{-}1$ and $\mathsf{I}\in\text{sub}[\mathsf{B}_m]$. These incompatible channels are shown in the left panel of figure~\ref{fig:X1lm}. The number of constraints in $\pmb{C}_3$ should equal to the number of these incompatible channels. 

On the other hand, we can write the recursion~\eqref{eq:H_ngeneric} into a form that better fits to the factorization channel $X_{l_1,l_m}=P_m^2=0$, on which the amplitude behaves as
\begin{align}\label{eq:ALAR2}
A_n[\beta,\mathsf{B}_m]\rightarrow A^L[\beta,P_m]\frac{1}{X_{l_1,l_m}}A^R[-P_m,\mathsf{B}_m]\,.
\end{align}
Starting from eq.~\eqref{eq:H_ngeneric}, we use eq.~\eqref{eq:H_nIF2} to further expand $\pmb{H}_{n-|\mathsf{B}_m|+2}[\beta,(L_m,r_m)]$,
\begin{align}\label{eq:ra1}
\pmb{H}_{n-|\mathsf{B}_m|+2}[\beta,(L_m,r_m)]=\pmb{H}_{n-|\mathsf{B}_m|}[\beta]\cup\pmb{C}_1[\beta,(L_m,r_m)]\cup\pmb{C}_2[\beta,(L_m,r_m)]\,.
%\quad\beta\in\pmb{M}_{n-2,k-1}\,.
\end{align}
Using $(L_m+r_m)^2=P_m^2$, we can rewrite the $\pmb{C}_1$ part as
\begin{align}
\pmb{C}_1[\beta,(L_m,r_m)]=\pmb{C}_1[\beta,(L_m,r_m)]\Big|_{s_{L_m,r_m,\ldots}\rightarrow s_{P_m,\ldots}}=\pmb{C}_1[\beta,P_m]\,,
\end{align}
such that the combination $\pmb{H}_{n-|\mathsf{B}_m|}[\beta]\cup\pmb{C}_1[\beta,(L_m,r_m)]$ becomes
\begin{align}
\pmb{H}_{n-|\mathsf{B}_m|}[\beta]\cup\pmb{C}_1[\beta,(L_m,r_m)]=\pmb{H}_{n-|\mathsf{B}_m|}[\beta]\cup\pmb{C}_1[\beta,P_m]=\pmb{H}_{n-|\mathsf{B}_m|+1}[\beta,P_m]\,.
\end{align}
The amplitude $A^R[-P_m,\mathsf{B}_m]$ is given by $\pmb{H}_{|\mathsf{B}_m|+1}[-P_m,\mathsf{B}_m]$.
%which only depends on the right kinematic subspace $\mathcal{K}_{|\mathsf{B}_m|+1}$. 
By using recursively eq.~\eqref{eq:H_ngeneric}, we can write it as
\begin{align}
\pmb{H}_{|\mathsf{B}_m|+1}[-P_m,\mathsf{B}_m]=\pmb{C}_3[-P_m,\mathsf{B}_m]\cup\pmb{H}_{|\mathsf{B}_m|}[(-L_m,l_m),\mathsf{B}_{m_1},\ldots,\mathsf{B}_{m_s}]\,.
\end{align}
where $\pmb{C}_3[-P_m,\mathsf{B}_m]$ must be a subset of $\pmb{C}_3[\beta,\mathsf{B}_m]$ that have support in $\mathcal{K}_{|\mathsf{B}_{m}|+1}$. Therefore, it consists of the constraints that automatically drop out when $X_{l_m,r_m}=0$ due to the strict positivity of $\{X_{l_1,l_\mathsf{I}}\}$, since these channels are incompatible to $X_{l_m,r_m}$ but live in $\mathcal{K}_{|\mathsf{B}_m|+1}$. We can thus divide $\pmb{C}_3$ into two parts, 
\begin{empheq}[box=\widefboxb]{align}\label{eq:ra2}
\pmb{C}_3[\beta,\mathsf{B}_m]=\pmb{C}_3^a[-P_m,\mathsf{B}_m]\cup\pmb{C}_3^b[\beta,\mathsf{B}_m]\,,
\end{empheq}
where $\pmb{C}_3[-P_m,\mathsf{B}_m]=\pmb{C}_3^a[-P_{m},\mathsf{B}_m]$ and $\pmb{C}_3^b[-P_m,\mathsf{B}_m]=\emptyset$. The constraints in $\pmb{C}_3^b$ automatically drop out at both $X_{l_m,r_m}=0$ and $X_{l_1,l_m}=0$ due to the strict positivity of $\{X_{l_i,l_\mathsf{I}}\}$ with $2\leqslant i\leqslant m{-}1$, while those in $\pmb{C}_3^a$ only drop out at the first factorization channel. The joint effect of eq.~\eqref{eq:ra1} to \eqref{eq:ra2} rearranges the recursion~\eqref{eq:H_ngeneric} into the following equivalent form,
\begin{empheq}[box=\widefboxb]{align}\label{eq:H_ngenericalt}
\pmb{H}_n[\beta,\mathsf{B}_m]=&\;\pmb{H}_{n-|\mathsf{B}_m|+1}[\beta,P_m]\cup\pmb{C}_2[\beta,(L_m,r_m)]\cup\pmb{C}_3^b[\beta,\mathsf{B}_m]\nonumber\\
&\cup\pmb{H}_{|\mathsf{B}_m|+1}[-P_m,\mathsf{B}_m]\,,
\end{empheq}
where both $\pmb{C}_2$ and $\pmb{C}_3^b$ drop out automatically when $X_{l_1,l_m}=0$.

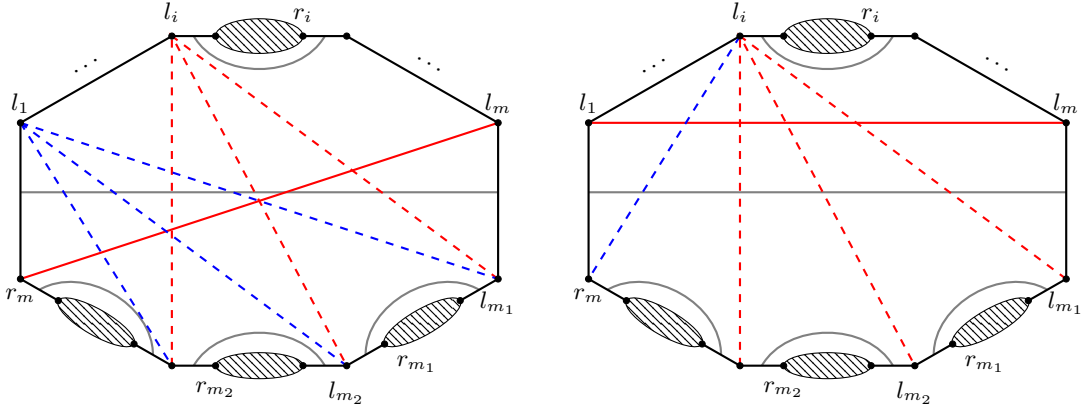
\begin{figure}[t]
\centering
\begin{tikzpicture}[every node/.style={font=\footnotesize},scale=1.15]
	\begin{scope}[xshift=0cm,yshift=0cm]
	\draw [pattern=north west lines] (0.5,0) ellipse (0.5 and 0.2);
	\path (-0.5,0) node (l2) {} -- ++(0.5,0) node (f2l) [midway] {};
	\path (1,0) node (r2) {} -- ++(0.5,0) node (f2r) [midway] {};
	\filldraw (0,0) circle (1pt);
	\draw [thick,gray] (f2l.center) to [bend right=60] (f2r.center);
	\draw [thick] (1,0) -- ++(0.5,0) (-0.5,0) -- ++(0.5,0);
	\end{scope}
	\begin{scope}[xshift=1.5cm]
	\draw [thick] (-30:2) node (ri) {} -- (0,0) node [midway,above,sloped] {$\cdots$};
	\node (l3) at (0,0) {};
	\end{scope}
	\begin{scope}[xshift=-0.5cm]
	\draw [thick] (0,0) -- (-150:2) node [midway,above,sloped] {$\cdots$};
	\node (l1) at (-150:2) {};
	\end{scope}
	\begin{scope}[xshift=0cm,yshift=-3.8cm]
	\draw [pattern=north west lines] (0.5,0) ellipse (0.5 and 0.15);
	\path (0,0) node (rm2) {} -- ++(-0.5,0) node (fm2r) [midway] {};
	\path (1,0) -- ++(0.5,0) node (fm2l) [midway] {};
	\node (lm2) at (1.5,0) {};
	\draw [thick,gray] (fm2l.center) to [bend right=60] (fm2r.center);
	\filldraw (1,0) circle (1pt);
	\draw [thick] (rm2.center) -- ++(-0.5,0) (lm2.center) -- ++(-0.5,0);
	\end{scope}
	\begin{scope}[xshift=1.5cm,yshift=-3.8cm,rotate=30]
	\draw [pattern=north west lines] (1,0) ellipse (0.5 and 0.15);
	\path (0.5,0) node (rm1) {} -- ++(-0.5,0) node (fm1r) [midway] {};
	\path (2,0) node (lm1) {} -- ++ (-0.5,0) node (fm1l) [midway] {};
	\draw [thick,gray] (fm1l.center) to [bend right=60] (fm1r.center);
	\filldraw (1.5,0) circle (1pt);
	\draw [thick] (2,0) -- ++(-0.5,0) (0.5,0) -- ++(-0.5,0);
	\end{scope}
	\begin{scope}[xshift=-0.5cm,yshift=-3.8cm,rotate=-30]
	\draw [pattern=north west lines] (-1,0) ellipse (0.5 and 0.15);
	\path (-1.5,0) node (rm3) {} -- ++(-0.5,0) node (rm) {};
	\path (rm3.center) -- ++(-0.25,0) node (fm3r) {};
	\path (0,0) node (lm3) {} -- ++(-0.5,0) node (fm3l) [midway] {};
	\filldraw (-0.5,0) circle (1pt);
	\draw [thick,gray] (fm3l.center) to [bend right=60] (fm3r.center);
	\draw [thick] (rm3.center) -- (rm.center) (lm3.center) -- ++(-0.5,0);
	\end{scope}
	\path (lm1.center) -- ++(0,1) node (lm) {} (rm.center) -- ++(0,1) node (a1) {};
	\draw [thick,gray] (a1.center) -- (lm.center);
	\draw [thick,red] (rm.center) -- (ri.center); 
	%\draw [thick,violet] (l1.center) -- (ri.center);
	%\draw [thick,violet,dashed] (l2.center) -- (rm.center);
	\draw [thick,red,dashed] (lm1.center) -- (l2.center) -- (lm2.center) (lm3.center) -- (l2.center);
	\draw [thick,blue,dashed] (l1.center) -- (lm1.center) (l1.center) -- (lm2.center) (l1.center) -- (lm3.center);
	\draw [thick] (ri.center) -- (lm1.center);
	\draw [thick] (l1.center) -- (rm.center);
	\filldraw (l2.center) circle (1pt) node [above=1.5pt]{$l_{i}$} (r2.center) circle (1pt) node [above=1.5pt]{$r_{i}$};
	\filldraw (l3.center) circle (1pt) node [above=0pt]{};
	\filldraw (lm2.center) circle (1pt) node[below=2pt]{$l_{m_2}$} (rm2.center) circle (1pt) node[below=2pt]{$r_{m_2}$};
	\filldraw (lm1.center) circle (1pt) node [below=1pt]{$l_{m_1}$} (rm1.center) circle (1pt) node [below right=1pt]{$r_{m_1}$}; 
	\filldraw (ri.center) circle (1pt) node[above=0pt]{$l_m$} (rm.center) circle (1pt) node[below=0pt]{$r_m$} (l1.center) circle (1pt) node[above=0pt]{$l_1$};
	\filldraw (lm3.center) circle (1pt) (rm3.center) circle (1pt);
\begin{scope}[xshift=6.5cm]
	\begin{scope}[xshift=0cm,yshift=0cm]
	\draw [pattern=north west lines] (0.5,0) ellipse (0.5 and 0.2);
	\path (-0.5,0) node (l2) {} -- ++(0.5,0) node (f2l) [midway] {};
	\path (1,0) node (r2) {} -- ++(0.5,0) node (f2r) [midway] {};
	\filldraw (0,0) circle (1pt);
	\draw [thick,gray] (f2l.center) to [bend right=60] (f2r.center);
	\draw [thick] (1,0) -- ++(0.5,0) (-0.5,0) -- ++(0.5,0);
	\end{scope}
	\begin{scope}[xshift=1.5cm]
	\draw [thick] (-30:2) node (ri) {} -- (0,0) node [midway,above,sloped] {$\cdots$};
	\node (l3) at (0,0) {};
	\end{scope}
	\begin{scope}[xshift=-0.5cm]
	\draw [thick] (0,0) -- (-150:2) node [midway,above,sloped] {$\cdots$};
	\node (l1) at (-150:2) {};
	\end{scope}
	\begin{scope}[xshift=0cm,yshift=-3.8cm]
	\draw [pattern=north west lines] (0.5,0) ellipse (0.5 and 0.15);
	\path (0,0) node (rm2) {} -- ++(-0.5,0) node (fm2r) [midway] {};
	\path (1,0) -- ++(0.5,0) node (fm2l) [midway] {};
	\node (lm2) at (1.5,0) {};
	\draw [thick,gray] (fm2l.center) to [bend right=60] (fm2r.center);
	\filldraw (1,0) circle (1pt);
	\draw [thick] (rm2.center) -- ++(-0.5,0) (lm2.center) -- ++(-0.5,0);
	\end{scope}
	\begin{scope}[xshift=1.5cm,yshift=-3.8cm,rotate=30]
	\draw [pattern=north west lines] (1,0) ellipse (0.5 and 0.15);
	\path (0.5,0) node (rm1) {} -- ++(-0.5,0) node (fm1r) [midway] {};
	\path (2,0) node (lm1) {} -- ++ (-0.5,0) node (fm1l) [midway] {};
	\draw [thick,gray] (fm1l.center) to [bend right=60] (fm1r.center);
	\filldraw (1.5,0) circle (1pt);
	\draw [thick] (2,0) -- ++(-0.5,0) (0.5,0) -- ++(-0.5,0);
	\end{scope}
	\begin{scope}[xshift=-0.5cm,yshift=-3.8cm,rotate=-30]
	\draw [pattern=north west lines] (-1,0) ellipse (0.5 and 0.15);
	\path (-1.5,0) node (rm3) {} -- ++(-0.5,0) node (rm) {};
	\path (rm3.center) -- ++(-0.25,0) node (fm3r) {};
	\path (0,0) node (lm3) {} -- ++(-0.5,0) node (fm3l) [midway] {};
	\filldraw (-0.5,0) circle (1pt);
	\draw [thick,gray] (fm3l.center) to [bend right=60] (fm3r.center);
	\draw [thick] (rm3.center) -- (rm.center) (lm3.center) -- ++(-0.5,0);
	\end{scope}
	\path (lm1.center) -- ++(0,1) node (lm) {} (rm.center) -- ++(0,1) node (a1) {};
	\draw [thick,gray] (a1.center) -- (lm.center);
	%\draw [thick,red] (rm.center) -- (ri.center); 
	\draw [thick,red] (l1.center) -- (ri.center);
	\draw [thick,blue,dashed] (l2.center) -- (rm.center);
	\draw [thick,red,dashed] (lm1.center) -- (l2.center) -- (lm2.center) (lm3.center) -- (l2.center);
	%\draw [thick,blue,dashed] (l1.center) -- (lm1.center) (l1.center) -- (lm2.center) (l1.center) -- (lm3.center);
	\draw [thick] (ri.center) -- (lm1.center);
	\draw [thick] (l1.center) -- (rm.center);
	\filldraw (l2.center) circle (1pt) node [above=1.5pt]{$l_{i}$} (r2.center) circle (1pt) node [above=1.5pt]{$r_{i}$};
	\filldraw (l3.center) circle (1pt) node [above=0pt]{};
	\filldraw (lm2.center) circle (1pt) node[below=2pt]{$l_{m_2}$} (rm2.center) circle (1pt) node[below=2pt]{$r_{m_2}$};
	\filldraw (lm1.center) circle (1pt) node [below=1pt]{$l_{m_1}$} (rm1.center) circle (1pt) node [below right=1pt]{$r_{m_1}$}; \filldraw (ri.center) circle (1pt) node[above=0pt]{$l_m$} (rm.center) circle (1pt) node[below=0pt]{$r_m$} (l1.center) circle (1pt) node[above=0pt]{$l_1$};
	\filldraw (lm3.center) circle (1pt) (rm3.center) circle (1pt);
\end{scope}
\end{tikzpicture}
\caption{Left: The diagonals of the form $\{X_{l_i,l_{\mathsf{I}}}\}$ with $\mathsf{I}\in\text{sub}[\mathsf{B}_m]$ and $2\leqslant i\leqslant m{-}1$ are shown in red dashed lines. The diagonals $\{X_{l_1,l_{\mathsf{I}}}\}$ are shown in blue dashed lines. Both of them are allowed by flavor conservation but incompatible to $X_{l_m,r_m}$ (red solid line). Right: $\{X_{l_i,l_\mathsf{I}}\}$ with $2\leqslant i\leqslant m{-}1$ (red dashed lines) are also incompatible to $X_{l_1,l_m}$ (red solid line). The diagonals $\{X_{l_i,r_m}\}$ (blue dashed line) are incompatible to $X_{l_1,l_m}$ but not $X_{l_m,r_m}$,}
\label{fig:X1lm}
\end{figure}

We next show that the following definitions of $\pmb{C}_3^a$ and $\pmb{C}_3^b$ guarantee the correct factorization behavior at both $X_{l_m,r_m}=0$ and $X_{l_1,l_m}=0$,
\begin{empheq}[box=\widefbox]{align}\label{eq:C3_n}
%&\pmb{C}_3[\beta,\mathsf{B}_m]=\pmb{C}_3^a[\mathsf{B}_m]\cup\pmb{C}_3^b[\beta,\mathsf{B}_m]\,,\nonumber\\
\pmb{C}_3^a[-P_m,\mathsf{B}_m]&=\{-s_{l_m,r_m}=c_{l_m,r_m}>0\}\bigcup_{\mathsf{I}\in\text{sub}[\mathsf{B}_m]\backslash\mathsf{B}_{m_s}}\left\{-s_{\mathsf{I},r_m}=c_{\mathsf{I},r_m}>0\right\},\nonumber\\
%\{s_{l_m,r_m}\}\bigcup_{\mathsf{I}\in\text{sub}[\mathsf{B}_m]\backslash\mathsf{B}_{m_s}}\big\{s_{\mathsf{I},r_m}\text{ equals to negative constant}\big\}\,,\nonumber\\
\pmb{C}_3^b[\beta,\mathsf{B}_m]&=\bigcup_{i=2}^{m-1}\bigcup_{\mathsf{I}\in\text{sub}[\mathsf{B}_m]}\left\{
-s_{\mathsf{B}_i,\mathsf{I}}=c_{\mathsf{B}_i,\mathsf{I}}>0\right\}\,.
%\bigcup_{i=2}^{m-1}\bigcup_{\mathsf{I}\in\text{sub}[\mathsf{B}_m]}\left\{
%s_{\mathsf{B}_i,\mathsf{I}}\text{ equals to negative constant} \right\}\,.
\end{empheq}
Together with eq.~\eqref{eq:H_nIF2}, this completes the generic recursive construction~\eqref{eq:H_ngeneric}. To show that the $\pmb{C}_3$ part indeed drops out when $X_{l_m,r_m}=0$, we first rewrite them into
%Similar to the previous cases,  we can understand $\pmb{C}_3$ by first rewriting them into
\begin{subequations}
\begin{align}\label{eq:C3a}
\pmb{C}_3^a: & & &X_{l_1,l_{\mathsf{I}}}+X_{l_m,r_m}=X_{l_\mathsf{I},r_m}+X_{l_1,l_m}+\sum_{l_m\leqslant\mathsf{J}<\mathsf{I}}\big(c_{\mathsf{J},r_m}+X_{l_\mathsf{J},l_{\mathsf{J}+1}}\big)\,,\\
\label{eq:lilI}
\pmb{C}_3^b: & & &X_{l_i,l_\mathsf{I}}+X_{l_m,r_m}=X_{l_m,l_\mathsf{I}}+X_{l_i,r_m}+\sum_{j=i}^{m-1}\sum_{\mathsf{I}\leqslant\mathsf{J}\leqslant\mathsf{B}_{m_s}}\left(c_{\mathsf{B}_j,\mathsf{J}}+X_{l_j,l_{j+1}}+X_{l_\mathsf{J},l_{\mathsf{J}+1}}\right),
\end{align}
\end{subequations}
where $\mathsf{J}{+}1$ denotes the sub-block coming right after $\mathsf{J}$ and $l_{\mathsf{J}+1}$ its first particle. If $\mathsf{J}=\mathsf{B}_{m_s}$ is the last sub-block, then $l_{\mathsf{J}+1}:=r_m$. In both equations, we have $\mathsf{I}\in\text{sub}[\mathsf{B}_m]$. The summation in eq.~\eqref{eq:C3a} is over all the sub-blocks before $\mathsf{I}$, including $l_m$. In eq.~\eqref{eq:lilI}, we have $2\leqslant i\leqslant m{-}1$, and the second summation is over all the sub-blocks between $\mathsf{I}$ and the last sub-block $\mathsf{B}_{m_s}$. At $X_{l_m,r_m}=L_m^2=0$, neither eq.~\eqref{eq:C3a} nor~\eqref{eq:lilI} impose any boundaries in the positive kinematic subspace since $X_{l_1,l_\mathsf{I}}$ and $X_{l_i,l_\mathsf{I}}$ cannot reach zero. We can thus drop the $\pmb{C}_3$ part and arrive at the desired factorization behavior
\begin{align}
\pmb{H}_n[\alpha]&\xrightarrow{X_{l_m,r_m}=0}\pmb{H}^{L}_{n-|\mathsf{B}_m|+2}[\beta,(L_m,r_m)]\cup\pmb{H}^{R}_{|\mathsf{B}_m|}[(-L_m,l_m),\mathsf{B}_{m_1},\ldots,\mathsf{B}_{m_s}]\,,
\end{align}
where $\pmb{H}^L$ and $\pmb{H}^R$ only depend on the left and right kinematic subspace respectively. This implies that the facet $X_{l_m,r_m}=0$ of the polytope $\mathcal{A}$ is a direct product of $\mathcal{A}^L$ and $\mathcal{A}^R$. The canonical form thus factorizes as $\Omega(\mathcal{A})\rightarrow\Omega(\mathcal{A}^L)\wedge\Omega(\mathcal{A}^R)$ and the amplitude as eq.~\eqref{eq:ALAR1}.

Similarly, to show that in eq.~\eqref{eq:H_ngenericalt} the constraints in $\pmb{C}_2$ and $\pmb{C}_3^b$ drop out automatically when $X_{l_1,l_m}=0$, we rewrite them as
%The constraints in $\pmb{C}_2$ and $\pmb{C}_3^b$ can be dropped when $X_{l_1,l_m}=0$, which can be shown by first rewriting them as 
\begin{subequations}
\begin{align}\label{eq:lirm}
&\pmb{C}_2:& X_{l_i,r_m}+X_{l_1,l_m}=&\;X_{l_1,l_i}+X_{l_m,r_m}+\sum_{j=i}^{m-1}\big(c_{\mathsf{B}_j,r_m}+X_{l_j,l_{j+1}}\big)\,,\\
\label{eq:lilI2}
&\pmb{C}_3^b:& X_{l_i,l_\mathsf{I}}+X_{l_1,l_m}=&\;X_{l_m,l_\mathsf{I}}+X_{l_1,l_i}+\sum_{j=1}^{m-1}\left(c_{\mathsf{B}_j,r_m}+X_{l_j,l_{j+1}}\right)\nonumber\\
& & &+\sum_{j=i}^{m-1}\sum_{\mathsf{I}\leqslant\mathsf{J}\leqslant\mathsf{B}_{m_s}}\left(c_{\mathsf{B}_j,\mathsf{J}}+X_{l_j,l_{j+1}}+X_{l_\mathsf{J},l_{\mathsf{J}+1}}\right),
\end{align}
\end{subequations}
where $2\leqslant i\leqslant m{-}1$ and $\mathsf{I}\in\text{sub}[\mathsf{B}_m]$. When $X_{l_1,l_m}=0$, clearly $X_{l_i,r_m}$ and $X_{l_i,l_\mathsf{I}}$ cannot reach zero such that they do not give rise to any constraints. These incompatible channels are also given in figure~\ref{fig:X1lm}. Therefore, we get
\begin{align}
\pmb{H}_n[\beta,\mathsf{B}_m]\xrightarrow{X_{l_1,l_m}=0}\pmb{H}^L_{n-|\mathsf{B}_m|+1}[\beta,P_m]\cup\pmb{H}^R_{|\mathsf{B}_m|+1}[-P_m,\mathsf{B}_m]\,,
\end{align}
which is the desired factorization behavior that leads to a direct product geometry $\mathcal{A}^L\times\mathcal{A}^R$ on the facet $X_{l_1,l_m}=0$ of $\mathcal{A}$. 

%These constraints are again automatically satisfied since when $X_{l_1,l_m}=0$ the incompatible channel $X_{l_i,l_\mathsf{I}}=0$ cannot be reached. This completes the proof of eq.~\eqref{eq:1lm} and the direct product geometry~\eqref{eq:1lmA}.
%\begin{align}\label{eq:lirm}
%X_{l_i,r_m}+X_{l_1,l_m}=X_{l_1,l_i}+X_{l_m,r_m}+\sum_{j=i}^{m-1}\big(c_{\mathsf{B}_j,r_m}+X_{l_j,l_{j+1}}\big)\,.
%\end{align}
%\begin{align}\label{eq:lilI2}
%X_{l_i,l_\mathsf{I}}+X_{l_1,l_m}=&\;X_{l_m,l_\mathsf{I}}+X_{l_1,l_i}+\sum_{j=1}^{m-1}\left(c_{\mathsf{B}_j,r_m}+X_{l_j,l_{j+1}}\right)\nonumber\\
%&+\sum_{j=i}^{m-1}\sum_{\mathsf{I}\leqslant\mathsf{J}\leqslant\mathsf{B}_{m_s}}\left(c_{\mathsf{B}_j,\mathsf{J}}+X_{l_j,l_{j+1}}+X_{l_\mathsf{J},l_{\mathsf{J}+1}}\right).
%\end{align}

We close this section by the explicit example of constructing the subspace constraints for $\alpha=[(1,2),(3,(4,5),6,7),(8,9,10,11)]$. Since the last block $(8,9,10,11)$ is an $\mathfrak{f}$-$\mathfrak{af}$ block with substructures, we use eq.~\eqref{eq:H_ngeneric} to write
%\begin{align}
%\pmb{H}_{11}[\beta,(8,9,10,11)]=&\;\pmb{H}_7[\beta]\cup\pmb{C}_1[\beta,(8,9,10,11)]\cup\pmb{C}_2[\beta,(8,9,10,11)]\nonumber\\
%&\cup\pmb{C}_3[(8,9,10,11)]\cup\pmb{H}_4[(11,8),9,10]\,,
%\end{align}
\begin{align}
\pmb{H}_{11}[\beta,(8,9,10,11)]=&\;\pmb{H}_9[\beta,(P_{8,9,10},11)]\cup\pmb{C}_3[\beta,(8,9,10,11)]\cup\pmb{H}_4[(-{P}_{8,9,10},8),9,10]\,,
\end{align}
where $\beta=[(1,2),(3,(4,5),6,7)]$. The $\pmb{C}_3$ part is given by eq.~\eqref{eq:C3_n}:
%The $\pmb{C}_1$, $\pmb{C}_2$ and $\pmb{C}_3$ part are given by eq.~\eqref{eq:C1C2_Bm} and~\eqref{eq:C3_n} respectively:
\begin{align}\label{eq:Cpart}
&\pmb{C}_3[\beta,(8,9,10,11)]=\left\{\begin{array}{c}
s_{8,11},s_{9,11},s_{3,4,5,6,7,9},s_{3,4,5,6,7,10} \\
\text{set to negative constants}
\end{array}\right\}\,,
\end{align}
where the constraints on $s_{8,11}$ and $s_{9,11}$ come from $\pmb{C}_3^a$ and the rest from $\pmb{C}_3^b$. Using eq.~\eqref{eq:H_nIF2}, we can write $\pmb{H}_9[\beta,(P_{8,9,10},11)]$ as
\begin{align}
\pmb{H}_9[\beta,(P_{8,9,10},11)]=\pmb{H}_7[\beta]\cup\pmb{C}_1[\beta,(P_{8,9,10},11)]\cup\pmb{C}_2[\beta,(P_{8,9,10},11)]\,,
\end{align}
where $\pmb{C}_1$ and $\pmb{C}_2$ are give by
\begin{align}\label{eq:Cpart2}
&\pmb{C}_1[\beta,(P_{8,9,10},11)]=\left\{\begin{array}{c}
s_{2,8,9,10,11},s_{3,8,9,10,11},s_{4,5,8,9,10,11},s_{6,8,9,10,11} \\
\text{set to negative constants}
\end{array}\right\},\nonumber\\
&\pmb{C}_2[\beta,(P_{8,9,10},11)]=\left\{s_{3,4,5,6,7,11}\text{ set to negative constant}\right\}.
\end{align}
Applying eq.~\eqref{eq:H_ngeneric} to $\pmb{H}_7[\beta]$ and $\pmb{H}_4[(-{P}_{8,9,10},8),9,10]$, we get
\begin{align}\label{eq:H7}
\pmb{H}_7[(1,2),(3,(4,5),6,7)]&=\left\{\begin{array}{c}
s_{3,6},s_{4,6},s_{3,7},s_{4,5,7} \\
\text{ set to negative constants}
\end{array}\right\}\,,\nonumber\\
\pmb{H}_4[(-{P}_{8,9,10},8),9,10]&=\left\{s_{8,10}\text{ set to negative constant}\right\}\,,
\end{align}
where we have used eq.~\eqref{eq:H_nISgeneric}, or for this case equivalently eq.~\eqref{eq:H_nIS}, to obtain $\pmb{H}_5$. Thus $\pmb{H}_{11}[(1,2),(3,(4,5),6,7),(8,9,10,11)]$ is given by the union of all the constraints in eq.~\eqref{eq:Cpart},~\eqref{eq:Cpart2} and~\eqref{eq:H7}. The result agrees with eq.~\eqref{eq:H_11}.

\subsection{Remarks}\label{sec:remarks}
%\textcolor{blue}{(not sure if needed)}
We have checked up to $n=20$ and a variety of block structures that if we pull-back the planar scattering form $\Omega_n[\alpha]$ to the polytope $\mathcal{A}_n[\alpha]$ given by the constraints $\pmb{\mathcal{A}}_n[\alpha]=\pmb{H}_n[\alpha]\cup\pmb{\Delta}_n[\alpha]$, where $\pmb{H}_n[\alpha]$ is constructed from eq.~\eqref{eq:H_ngeneric} and $\pmb{\Delta}_n[\alpha]$ is the union of $\pmb{P}_n[\alpha]$ and $\pmb{F}_n[\alpha]$ given in eq.~\eqref{eq:PnFn}, we get the correct amplitude,
\begin{align}
    \Omega_n[\alpha]\Big|_{\mathcal{A}_n[\alpha]}=A_n[\alpha]d^{n-3}X\,.
\end{align}
There are several interesting features in this open kinematic polytope that do not appear in the bi-adjoint case. Although the facet geometry of $X_{l_m,r_m}=0$ and $X_{l_1,l_m}=0$ is by construction a direct product of lower-dimensional polytopes carved out by the same process as eq.~\eqref{eq:H_ngeneric}, it is generally not true for other facets. When restricted to a generic facet $X_{i,j}=0$, we actually have
\begin{align}
    \mathcal{A}_n[\alpha]\Big|_{X_{i,j}=0}\cong \mathcal{A}^L[\alpha^L;a]\ltimes\mathcal{A}^R[\alpha^R]\,.
\end{align}
While $\mathcal{A}^R[\alpha^R]$ is given by eq.~\eqref{eq:H_ngeneric}, $\mathcal{A}^L[\alpha^L;a]$ is deformed by some linear combinations of the planar Mandelstam variables in the left kinematic subspace, where $a$ is a formal collection of the deformation parameters. They change the shape of the polytope but leave the expression of the canonical form unchanged after pull-back,
%In appendix~\ref{sec:defsoftlimit}, we will present a special class of such deformations. While the shape of the polytope is changed, the canonical form after pull-back is still has the same expression after the deformation,
\begin{align}
\Omega[\alpha^L]\Big|_{\mathcal{A}[\alpha^L;a]}=A^L[\alpha^L]d^{n-3}X\,.
\end{align}
In appendix~\ref{sec:defsoftlimit}, we will present a special class of such deformations on $\pmb{H}_n[\alpha]$.
Moreover, the facet geometry is a \emph{semi-direct product} ``$\ltimes$'' between $\mathcal{A}^L[\alpha^L;a]$ and $\mathcal{A}^R[\alpha^R]$: the $c$ constants in the constraints for $\mathcal{A}^L[\alpha^L;a]$ receive a linear shift by the planar variables in $\mathcal{A}^R$. The canonical form still factorizes nicely despite this shift, 
\begin{align}
\text{Res}_{X_{i,j}=0}\Omega_n[\alpha,\mathcal{A}_n]=\Omega_n[\alpha,\mathcal{A}_n|_{X_{i,j}=0}]=\Omega_n[\alpha,{\mathcal{A}}^L\ltimes\mathcal{A}^R]=\Omega[\alpha^L,{\mathcal{A}}^L]\wedge\Omega[\alpha^R,\mathcal{A}^R]\,,
\end{align}
since the top form $\Omega[\alpha^R,\mathcal{A}^R]$ removes all the linear shifts in the $c$ constants. We will give a few concrete factorization examples that manifest these features in appendix~\ref{sec:factorizationExamples}, and leave the detailed factorization analysis to a future work.

\section{``Color is Kinematics" for (Anti-)Fundamental States}\label{sec4}

We now turn to the positive geometry interpretation of full, color-dressed amplitudes with ($\mathfrak{a}$)$\mathfrak{f}$ states in more generic theories, not just partial amplitudes for scalar theories. We find a natural extension of the ``color is kinematics" philosophy of \cite{Arkani-Hamed:2017mur} to (anti-)fundamental amplitudes. In section \ref{sec:gensmallkinespace}, we first discuss a natural generalization of small kinematic space for (anti-)fundamental scattering amplitudes which is necessary for (anti-)fundamental color-kinematics duality. In section \ref{sec:dualdifforcolfac}, we show how the duality between differentials forms in kinematic space and color factors extends to (anti-)fundamental scattering forms. In section \ref{projscf}, we focus on connections between BCJ numerator relations and projectivity for (anti-)fundamental scattering forms. In section~\ref{sec:genDDMdec}, we show how Melia decomposition is dual to pulling back the scattering form to a specific sub-space, $H^{T}_{n}[\alpha]$. Interestingly, the $k>1$ planar scattering form does not need to be a top-form like in the bi-adjoint case. 

Since we will be dealing with ($\mathfrak{a}$)$\mathfrak{f}$ states without a definite ordering, we introduce a minor notation change from section \ref{posgeoinkinspace}. When referring to ($\mathfrak{af}$)$\mathfrak{f}$ states, we will use capital letters, where an underline (bar) indicates a $\mathfrak{f}$ ($\mathfrak{af}$) state.  

\subsection{(Anti-)Fundamental Small Kinematic Space}
\label{sec:gensmallkinespace}

We now define (anti-)fundamental kinematic space, $\mathcal{K}_{n}^{k}$. We start with big kinematic space, $\mathcal{K}_{n}^{\star}$, which is the same for (anti-)fundamental and adjoint amplitudes. Big kinematic space is defined as a vector space spanned by $S_{I}$ variables, which are indexed by subsets, $I\subset \{ 1,2,\ldots, n\}$, and obey 

\begin{itemize}
\item $S_{\bar{I}}=S_{I}$ where $\bar{I}$ is the complement of $I$, 
\item $S_{I}=0$ for $|I|=0,1,n-1,n$.
\end{itemize}

\noindent The dimension of big kinematic space is 

\begin{equation}
\dim(\mathcal{K}_{n}^{\star})=2^{n-1}-n-1 \ .
\end{equation}

\noindent As reviewed in section \ref{scatteringformsrevieew}, the reduction to small kinematic space for $k=0$ is done by imposing the seven-term identity on all four-point sub-graphs. 

\begin{figure}
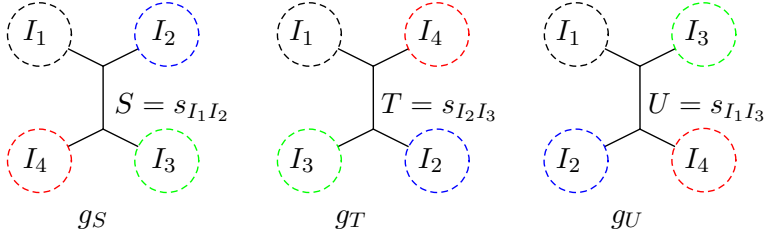

\begin{center}
\begin{overpic}[width=.8\linewidth]{ggst_gr7-eps-converted-to.pdf}
%\label{ggst7}
\put(12,26){$I_1$}\put(26,26){$I_2$}\put(26,12){$I_3$}\put(12,12){$I_4$}
\put(41,26){$I_1$}\put(55,26){$I_4$}\put(55,12){$I_2$}\put(41,12){$I_3$}
\put(70,26){$I_1$}\put(84,26){$I_3$}\put(84,12){$I_4$}\put(70,12){$I_2$}
%\put(43,26){$I_4$}\put(56,26){$I_1$}\put(56,12){$I_2$}
\put(22,18){$S=s_{I_1I_2}$}\put(51,18){$T=s_{I_2I_3}$}\put(80,18){$U=s_{I_1I_3}$}
\put(18,6){$g_S$}\put(46,6){$g_T$}\put(76,6){$g_U$}
\end{overpic}
\end{center}
\caption{A four-set partition $I_1\sqcup I_2\sqcup I_3\sqcup I_4$ of the external labels and the three corresponding channels. The three graphs  $g_S,g_T,g_U$  are identical except for a 4-point subgraph. This is the same as figure~\ref{figjacobi}, but reproduced here for the reader's convenience. }
\label{figjacobi2}
\end{figure}

For adjoint amplitudes, one can show that upon imposing the seven-term identity on all sub-graphs, any variable $s_{I}$ can be written as a sum of $s_{i,j}$, which can be identified as Mandelstam variables. $\mathcal{K}_{n}^{k=0}$ is therefore spanned by Mandelstam variables, $s_{i,j}$, implying that the seven-term identity is equivalent to imposing momentum conservation and that the dimension of the space is 

\begin{equation}
\dim(\mathcal{K}_{n}^{k=0})=\frac{n(n-3)}{2} \ .
\end{equation}

\noindent The process of reducing from $\mathcal{K}_{n}^{\star}$ to $\mathcal{K}^{k}_{n}$ when $k\neq 0,1$ is slightly modified. As in section \ref{posgeoinkinspace}, internal propagators which violate flavor conservation are truncated from the vector space, which leads to the \emph{(anti-)fundamental big kinematic space }$(\mathcal{K}_n^k)^{\star}$,
\begin{equation}
S_{I}=b_{I}, \quad \textrm{ if  }\vartheta_{I}=0 \ .  
\end{equation}
To further reach the \emph{(anti-)fundamental small kinematic space} $\mathcal{K}_n^k$, \emph{the seven-term identity is then imposed on all sub-graphs except those where all} $I_{i}$ \emph{correspond to ($\mathfrak{a}$)$\mathfrak{f}$ states,} 

\begin{equation}
S_{I_{1},I_{2}}+S_{I_{2},I_{3}}+S_{I_{1},I_{3}}=S_{I_{1}}+S_{I_{2}}+S_{I_{3}}+S_{I_{4}} \ .
\label{7termidentityrepro}
\end{equation}

\noindent This mirrors the color structure of tree graphs with (anti-)fundamental states, as visualized in figure~\ref{coloralgebraidentit}. We can interpret this generalization of kinematic space as a natural abstractification of momentum conservation where the seven-term identity is treated as fundamental. Due to the truncation procedure and fewer number of relations between remaining $S_{I}$, (anti-)fundamental small kinematic space has more non-trivial combinatorial structure than adjoint small kinematic space. For example, the dimension of (anti-)fundamental small kinematic space is 

\begin{equation}
\dim(\mathcal{K}_{n}^{k})=\frac{n(n-3)}{2}-\frac{k(k-1)}{2} \ .
\label{dimensmallkinespacealfla}
\end{equation}

\noindent Eq.~(\ref{dimensmallkinespacealfla}) is derived in appendix~\ref{sec:derivatofdim}. In the small kinematic space, we identify $S_I$ with the Mandelstam variable $s_I$ for each $I$. Furthermore, the two-particle Mandelstam variables $s_{i,j}$ do not always provide a complete basis for $\mathcal{K}_{n}^{k}$ after the truncation of those forbidden by the flavor structure. To see this, consider amplitudes with only (anti-)fundamental external states. There are only $k$ Mandelstam variables not forbidden by flavor conservation, which is smaller than the dimension of the space for $k>3$. Instead, a more natural basis is the planar variables of some ordering, $\alpha$, where all $\mathfrak{f}$-$\mathfrak{af}$ pairs are adjacent. The number of planar propagators for such orderings matches eq.~(\ref{dimensmallkinespacealfla}). \\
\indent As a simple example, consider the four-point amplitude ${\bf M}[\underline{A},\bar{A},\underline{B},\bar{B}]$. Big kinematic space is the same as the adjoint case:

\begin{equation}
\mathcal{K}_{n}^{\star}=\{ s_{\underline{A},\bar{A}},s_{\underline{A},\bar{B}},s_{\underline{A},\underline{B}} \}   \,. 
\end{equation}

\noindent To reduce to small kinematic space, the propagators forbidden by flavor conservation, $s_{\underline{A},\bar{B}}$ and $s_{\underline{A},\underline{B}}$, are truncated from the spectrum. Since the only possible four-point sub-graph corresponds to four external ($\mathfrak{a}$)$\mathfrak{f}$ states, there are no seven-term identities to impose. Therefore, (ant-)fundamental small kinematic space is 

\begin{equation}
\mathcal{K}_{n=4}^{k=2}=\{ s_{\underline{A},\bar{A}}\}=(\mathcal{K}_{n=4}^{k=2})^{\star} \ .
\end{equation}

\noindent Note that the dimension of $\mathcal{K}_{n=4}^{k=2}$ matches eq.~(\ref{dimensmallkinespacealfla}). \\
\indent Now consider the amplitude ${\bf M}[\underline{A},\bar{A},\underline{B},\bar{B},\underline{C},\bar{C}]$. To find small kinematic space, all forbidden propagators are first truncated from the spectrum, leaving:
\begin{align}\label{bigkinespacess}
(\mathcal{K}_{n=6}^{k=3})^{\star}=\Big\{S_{\underline{A}\bar{A}}\,,  S_{\underline{B}\bar{B}},  S_{\underline{C}\bar{C}}\,,S_{\underline{A}\underline{B}\bar{B}}\,,  S_{\underline{B}\underline{C}\bar{C}}\,,  S_{\underline{C}\underline{A}\bar{A}}\,,S_{\bar{A}\underline{B}\bar{B}}\,,  S_{\bar{B}\underline{C}\bar{C}}\,,  S_{\bar{C}\underline{A}\bar{A}}\Big\}
\end{align}
%\begin{equation}
%\begin{split}
%&S_{\underline{A}\bar{A}}, \quad S_{\underline{B}\bar{B}}, \quad S_{\underline{C}\bar{C}} \ ,\\
%&S_{\underline{A}\underline{B}\bar{B}}, \quad S_{\underline{B}\underline{C}\bar{C}}, \quad S_{\underline{C}\underline{A}\bar{A}} \ ,\\
%&S_{\bar{A}\underline{B}\bar{B}}, \quad S_{\bar{B}\underline{C}\bar{C}}, \quad S_{\bar{C}\underline{A}\bar{A}} \ .\\
%\end{split}
%\end{equation}
The seven-term identity is then imposed on all valid four-point sub-graphs, of which there are only three: 
\begin{equation}
\begin{split}
 &S_{I_{1},I_{2}}+S_{I_{2},I_{3}}+S_{I_{1},I_{3}}=S_{I_{1}}+S_{I_{2}}+S_{I_{3}}+S_{I_{4}}, \\
 &\indent\textrm{with } I_{1}=\{ \underline{A}\}, \ I_{2}=\{ \bar{A}\}, \ I_{3}=\{ \underline{B},\bar{B}\}, \ I_{4}=\{ \underline{C},\bar{C}\} \\
&(A \leftrightarrow B \leftrightarrow C) \ .
\end{split}
\label{lowerimdneioss}
\end{equation}
The (anti-)fundamental small kinematic space is therefore six dimensional, which again matches eq.~(\ref{dimensmallkinespacealfla}) for $k=3, n=6$. 

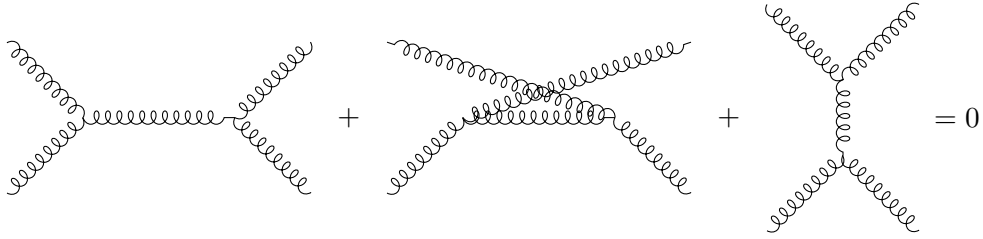
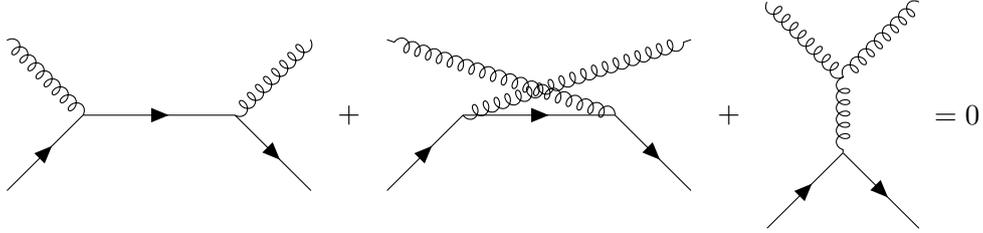
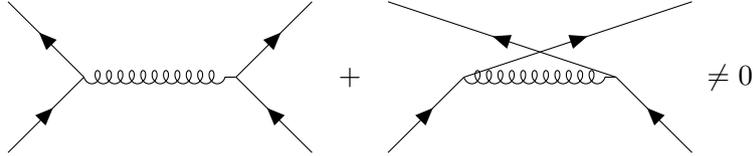
\begin{figure}
  \centering
  \begin{subfigure}[d]{1\textwidth}
  \centering
  \begin{tikzpicture}
    \begin{feynman}
      \vertex (aux1t) at (-6,0);
      \vertex (aux1b) at (-4,0);
      \vertex (exttop1a) at (-7,1);
      \vertex (exttop1b) at (-3,1);
      \vertex (extbot1a) at (-7,-1);
      \vertex (extbot1b) at (-3,-1);
      \vertex (aux2t) at (-1,0);
      \vertex (aux2b) at (1,0);
      \vertex (exttop2a) at (-2,1);
      \vertex (exttop2b) at (2,1);
      \vertex (extbot2a) at (-2,-1);
      \vertex (extbot2b) at (2,-1);
      \vertex (aux3t) at (4,0.5);
      \vertex (aux3b) at (4,-0.5);
      \vertex (exttop3a) at (5,1.5);
      \vertex (exttop3b) at (3,1.5);
      \vertex (extbot3a) at (5,-1.5);
      \vertex (extbot3b) at (3,-1.5);
      \draw (-2.5,0) node {$+$};
      \draw (2.5,0) node {$+$};
      \draw (5.5,0) node {$=0$};
      \diagram* {
        (extbot1a) -- [gluon](aux1t),
        (aux1t) -- [gluon](aux1b),
        (aux1b) -- [gluon](extbot1b),
        (aux1t) -- [gluon](exttop1a),
        (aux1b) -- [gluon](exttop1b),
        (extbot2a) -- [gluon](aux2t),
        (aux2t) -- [gluon](aux2b),
        (aux2b) -- [gluon](extbot2b),
        (aux2b) -- [gluon](exttop2a),
        (aux2t) -- [gluon](exttop2b),
        (exttop3a) -- [gluon](aux3t),
        (exttop3b) -- [gluon](aux3t),
        (aux3t) -- [gluon](aux3b),
        (extbot3b) -- [gluon](aux3b),
        (aux3b) -- [gluon](extbot3a)
      };
   \end{feynman}
  \end{tikzpicture}
  \caption{Color algebra in the adjoint representation.}
  \end{subfigure}
  \begin{subfigure}[d2]{1\textwidth}
  \centering
  \begin{tikzpicture}
    \begin{feynman}
      \vertex (aux1t) at (-6,0);
      \vertex (aux1b) at (-4,0);
      \vertex (exttop1a) at (-7,1);
      \vertex (exttop1b) at (-3,1);
      \vertex (extbot1a) at (-7,-1);
      \vertex (extbot1b) at (-3,-1);
      \vertex (aux2t) at (-1,0);
      \vertex (aux2b) at (1,0);
      \vertex (exttop2a) at (-2,1);
      \vertex (exttop2b) at (2,1);
      \vertex (extbot2a) at (-2,-1);
      \vertex (extbot2b) at (2,-1);
      \vertex (aux3t) at (4,0.5);
      \vertex (aux3b) at (4,-0.5);
      \vertex (exttop3a) at (5,1.5);
      \vertex (exttop3b) at (3,1.5);
      \vertex (extbot3a) at (5,-1.5);
      \vertex (extbot3b) at (3,-1.5);
      \draw (-2.5,0) node {$+$};
      \draw (2.5,0) node {$+$};
      \draw (5.5,0) node {$=0$};
      \diagram* {
        (extbot1a) -- [fermion](aux1t),
        (aux1t) -- [fermion](aux1b),
        (aux1b) -- [fermion](extbot1b),
        (aux1t) -- [gluon](exttop1a),
        (aux1b) -- [gluon](exttop1b),
        (extbot2a) -- [fermion](aux2t),
        (aux2t) -- [fermion](aux2b),
        (aux2b) -- [fermion](extbot2b),
        (aux2b) -- [gluon](exttop2a),
        (aux2t) -- [gluon](exttop2b),
        (exttop3a) -- [gluon](aux3t),
        (exttop3b) -- [gluon](aux3t),
        (aux3t) -- [gluon](aux3b),
        (extbot3b) -- [fermion](aux3b),
        (aux3b) -- [fermion](extbot3a)
      };
   \end{feynman}
  \end{tikzpicture}
  \caption{Color algebra in the fundamental representation that obeys a Jacobi-like identity.}
  \end{subfigure}
  \begin{subfigure}[d3]{1\textwidth}
  \centering
  \begin{tikzpicture}
    \begin{feynman}
      \vertex (aux1t) at (-6,0);
      \vertex (aux1b) at (-4,0);
      \vertex (exttop1a) at (-7,1);
      \vertex (exttop1b) at (-3,1);
      \vertex (extbot1a) at (-7,-1);
      \vertex (extbot1b) at (-3,-1);
      \vertex (aux2t) at (-1,0);
      \vertex (aux2b) at (1,0);
      \vertex (exttop2a) at (-2,1);
      \vertex (exttop2b) at (2,1);
      \vertex (extbot2a) at (-2,-1);
      \vertex (extbot2b) at (2,-1);
      \draw (-2.5,0) node {$+$};
      \draw (2.5,0) node {$\neq 0$};
      \diagram* {
        (extbot1a) -- [fermion](aux1t),
        (aux1t) -- [gluon](aux1b),
        (extbot1b) -- [fermion](aux1b),
        (aux1t) -- [fermion](exttop1a),
        (aux1b) -- [fermion]( exttop1b),
        (extbot2a) -- [fermion](aux2t),
        (aux2t) -- [gluon](aux2b),
        (extbot2b) -- [fermion](aux2b),
        (aux2b) -- [fermion](exttop2a),
        (aux2t) -- [fermion](exttop2b)
      };
   \end{feynman}
  \end{tikzpicture}
  \caption{Color algebra in the fundamental representation that does not obey a Jacobi-like identity.}
  \end{subfigure}
  \caption{Graphic representations of the color algebra for theories with (anti-)fundamental and adjoint states.}
  \label{coloralgebraidentit}
\end{figure}

\subsection{Duality between Differential Forms and Color Factors}
\label{sec:dualdifforcolfac}

The notion of a scattering form, which corresponds to the full, color-dressed amplitude, has a natural application to amplitudes with ($\mathfrak{a}$)$\mathfrak{f}$ states. Without losing any information, we can pullback the scattering form~\eqref{form} to the (anti-)fundamental big/small kinematic space, which leads to the \emph{(anti-)fundamental scattering form} 

\begin{equation}
\Omega_{n}^{k}[N]=
%\sum_{\textrm{cubic } g}
\sum_{g\in\Gamma_n}N(g|\alpha_{g})W(g|\alpha_{g}) \prod_{I\in g} \frac{\vartheta_{I}}{s_{I}}\ ,
\label{scatterinfoss}
\end{equation}

\noindent where the summation is effectively over all cubic graphs that respect flavor conservation since $\vartheta_{I}$ defined in eq.~\eqref{eq:theta} kills those graphs that violate it, and $W$ is a differential form 

\begin{equation}
W(g|\alpha_g)=\textrm{sign}(g|\alpha_{g})\bigwedge_{g\in I}ds_{I} \ . 
\label{dlogforgraph}
\end{equation}

\noindent Unlike the adjoint scattering forms, which live in $\mathcal{K}_{n}^{k=0}$, the (anti-)fundamental $d$log forms live in (anti-)fundamental small kinematic space, $\mathcal{K}_{n}^{k}$. \\
\indent The duality between differential forms and color factors generalizes to (anti-)fundamental scattering forms. In the (anti-)fundamental case, the color factor is a product of structure constants and fundamental representation matrices. Relations between color factors arise from both Jacobi and commutation identities 

\begin{equation}
\begin{split}
f^{abe}f^{cde}+f^{bce}f^{ade}+f^{cae}f^{bde}&=0 \,,\\
(T^{a}T^{b})^{i}_{j}-(T^{b}T^{a})^{i}_{j}-f^{abc}(T^{c})^{i}_{j}&=0\,.
\label{jacoblie}
\end{split}
\end{equation}

\noindent Unlike the adjoint case, the Jacobi-like identities for color factors in eq.~(\ref{jacoblie}) do not apply to all four-point sub-graphs. When all external states are ($\mathfrak{a}$)$\mathfrak{f}$ states, the color factors are 

\begin{equation}\label{eq:afcolor}
\begin{split}
C_{s}\propto (T^{a})^{i_{1}}_{i_{2}}(T^{a})^{i_{3}}_{i_{4}}, \quad &C_{t}\propto (T^{a})^{i_{1}}_{i_{4}}(T^{a})^{i_{3}}_{i_{2}}, \quad C_{u}=0 \ , \\
\end{split}
\end{equation}

\noindent which clearly do not obey a Jacobi-like relation. The relation 

\begin{equation}
C(g_S|I_1I_2I_3I_4)+C(g_T|I_1I_4I_2I_3)+C(g_U|I_1I_3I_4I_2)=0 \ ,
\label{anitfund}
\end{equation}

\noindent therefore applies to all four-point sub-graph except those where all $I_{i}$ correspond to ($\mathfrak{a}$)$\mathfrak{f}$ states. We now argue that the set of identities in eq.~(\ref{anitfund}) exactly mirror the seven-term identity. Taking the differential of eq.~(\ref{7termidentityrepro}) and repeating the argument in section \ref{scatteringformsrevieew}, we see that for any triplet of graphs corresponding to eq.~(\ref{anitfund}):

\begin{equation}
(ds_{I_{1},I_{2}}+ds_{I_{1},I_{3}}+ds_{I_{1},I_{4}})\wedge \bigwedge^{n-4}_{b=1}ds_{J_{b}}  \,, 
\end{equation}

\noindent where $s_{J_{b}}$ denotes the propagators shared by the triplet of graphs. It follows that,

\begin{equation}
W(g_S|I_1I_2I_3I_4)+W(g_T|I_1I_4I_2I_3)+W(g_U|I_1I_3I_4I_2)=0\,,
\label{anitfunddiff}
\end{equation}

\noindent only applies to desirable four-point sub-graphs and is therefore dual to eq.~(\ref{anitfund}). \\
\indent Consider the four-point amplitude ${\bf M}[\underline{A}^{\bar{i}}\bar{A}_{\underline{j}}\underline{B}^{\bar{m}}\bar{B}_{\underline{n}}]$ with two pairs of (anti-)fundamental particles. Here, we explicitly display the (anti-)fundamental representation indices of the particles.
%we have included additional indices that transform in the (anti-)fundamental representation. 
There is only a single Feynman diagram contribution, so there is only one color-factor dual to a one-form:

\begin{equation}
C_{s}=(T^{a})^{\bar{i}}_{\underline{j}}(T^{a})^{\bar{m}}_{\underline{n}} \leftrightarrow ds\,.
\end{equation}

\noindent There is no $U$-channel graph, because it is forbidden by flavor charge conservation. \\
\indent Now consider the four-point amplitude ${\bf M}[\underline{A}^{\bar{i}}\bar{A}_{\underline{j}}\phi^{a}\phi^{b}]$ in which $\phi^a$ and $\phi^b$ are two adjoint representation particles. There are three color factors dual to one forms:

\begin{equation}
\begin{split}
C_{s}=f^{abc}(T^{c})^{\bar{i}}_{\underline{j}} &\leftrightarrow ds \,,\\
C_{t}=(T^{b}T^{a})^{\bar{i}}_{\underline{j}} &\leftrightarrow dt \,, \\
C_{u}=-(T^{a}T^{b})^{\bar{i}}_{\underline{j}} &\leftrightarrow du\,. 
\end{split}
\end{equation}

\noindent The duality holds because there exists a valid seven-term identity that leads to the linear relation between the differentials,
%Since there is a valid seven-term identity, there is a linear relationship between the differentials

\begin{equation}
ds+dt+du=0    \,,
\end{equation}

\noindent which corresponds to the second line of eq.~(\ref{jacoblie}).  

\subsection{(Non-)Projectivity of the Flavored Scattering Form}
\label{projscf}

In this sub-section, we connect projectivity of the (anti-)fundamental scattering without flavor structure to generalized Jacobi identities for kinematic numerator factors. We review projectivity of the adjoint scattering form before examining two generalizations: scattering forms where all $\mathfrak{f}$-$\mathfrak{af}$ pairs have different flavor and scattering amplitudes where all $\mathfrak{f}$-$\mathfrak{af}$ pairs have the same flavor. \\
\indent We now review implications of projectivity for adjoint scattering forms. We consider the usual scattering form~\eqref{form} in big kinematic space $\mathcal{K}_{n}^{\star}$.\footnote{Note that the scattering form in  eq. ~\eqref{form} is defined in small kinematic space, $\mathcal{K}_{n}^{k}$, while we are now defining the same form in big kinematic space, $\mathcal{K}_{n}^{\star}$.}. Projectivity of a differential form is defined as invariance under a local GL($1$) transformation $S_{I}\rightarrow \Lambda(S)S_{I}$. Projectivity of the scattering form in $\mathcal{K}_{n}^{\star}$ could be used to derive Jacobi identities for kinematic numerator factors \cite{Arkani-Hamed:2017mur}. To see this, note that the $\Lambda$ dependence of the scattering form under a local GL($1$) transformation is 

\begin{equation}
\big[N(g_S|I_1I_2I_3I_4)+N(g_T|I_1I_4I_2I_3)+N(g_U|I_1I_3I_4I_2)\big] d\log(\Lambda)\wedge\left ( \bigwedge^{n-4}_{b=1}d\log(S_{J_{b}}) \right )+\ldots    \,,
\label{projectivitss}
\end{equation}

\noindent where $S_{J_{b}}$ denote the $(n{-}4)$ propagators shared by a triplet of graphs and the ``$\ldots$'' denotes the same expression for all other triplet. It is clear that the $\Lambda$-dependence only vanishes if the kinematic numerators obey Jacobi identities. It is key that we are only considering projectivity in big kinematic space, as the $d\log(s_{I})$ factors would be related by linear relations in small kinematic space. Such linear relations could lead to cancellations between triplets. \\
\indent In small kinematic space, projectivity of the adjoint scattering form implies BCJ relations between different partial amplitudes. Importantly, while the numerator Jacobi identity implies BCJ relations, BCJ relations do not imply the Jacobi numerator identity. In the language of positive geometry, while projectivity in big kinematic space implies projecitivty in small kinematic space, projectivity in small kinematic space does NOT imply projectivity in big kinematic space.  The corresponding generalizations of the Jacobi kinematic identity which obey BCJ relations were explored in \cite{BjerrumBohr:2010zs}. To derive BCJ relations from projectivity in $\mathcal{K}_{n}$, one takes a GL($1$) transformation of the scattering form in $\mathcal{K}_{n}$, finding an undesirable term proportional $d\log(\Lambda)$. One then pulls back this differential form back to various $(n-4)$-dimensional sub-spaces, finding what condition the vanishing of the pull-back imposes. For example, consider the $n=4$, $k=0$ scattering form:

\begin{equation}
\begin{split}
\Omega_{n=4}^{k=0}[N]&=N_s\frac{ds}{s}+N_t\frac{dt}{t}+N_u\frac{du}{u} \ , \\
&=-\left ( -\frac{N_t}{t}+\frac{N_s}{s} \right )dt+\left ( \frac{N_u}{u} -\frac{N_s}{s}\right )du \ , \\
&=-A_{4}[1,2,3,4]dt+A_{4}[1,2,4,3]du  \ . \\
\end{split}
\end{equation}

\noindent Under a $GL(1)$ transformation, $\Omega_{n=4}^{k=0}[N]$ transforms as 

\begin{equation}
\Omega_{n=4}^{k=0} \rightarrow  \Omega_{n=4}^{k=0}+ (-tA_{4}[1,2,3,4]+uA_{4}[1,2,4,3])d\log(\Lambda) \ .
\end{equation}

\noindent Since the term proportion to $d\log(\Lambda)$ is a scalar, there is no need to pull it back to any subspace. We simply find that projectivity implies the BCJ relation,

\begin{equation}
0=-tA_{4}[1,2,3,4]+uA_{4}[1,2,4,3] \ .   
\end{equation}

\noindent This argument generalizes to larger $n$. We now turn to generalizations of the above arguments for theories with flavor. \\
\indent We first discuss the scattering forms where all $\mathfrak{f}$-$\mathfrak{af}$ pairs have different flavor. Unlike the adjoint case, the (anti-)fundamental scattering form~\eqref{scatterinfoss} with $k>1$ distinct flavor $\mathfrak{f}$-$\mathfrak{af}$ pairs cannot be projective, in either $\mathcal{K}_{n}^{\star}$ or $\mathcal{K}_{n}^{k}$, even if the kinematic numerators obey color-kinematic duality. This is obvious already at four-points. A generic $k=2, \ n=4$ scattering form takes the form,

\begin{equation}
\Omega^{k=2}_{n=4}[N]=N(g_{S}|\underline{A}\bar{A}\underline{B}\bar{B})d\log(s) \ ,
\end{equation}

\noindent which clearly cannot be invariant under a local GL($1$) transform \footnote{Importantly, the lack of projectivity of the scattering form does not imply that the flavored scattering amplitudes do not obey color-kinematics duality, as seen in \cite{Brown:2018wss, Johansson:2015oia}. Given a flavored scattering amplitude with color-factors, one can still replace the color-factors with kinematic numerator factors to find the corresponding amplitude in some gravitational theory:

\begin{equation}
\mathcal{A}=\sum_{g\in \Gamma} \frac{C(g|\alpha_{g})N(g|\alpha_{g})}{\Pi_{I\in g}s_{I}} \rightarrow \mathcal{M}=\sum_{g\in \Gamma} \frac{\tilde{N}(g|\alpha_{g})N(g|\alpha_{g})}{\Pi_{I\in g}s_{I}} \ .
\end{equation}}. The flavored scattering form's lack of projectivity under GL($1$) transformation in $\mathcal{K}_{n}^{k}$ implies that the flavored scattering forms do not obey generic BCJ relations. Unlike the $k=0$ scattering forms, the term proportional to $d\log(\Lambda)$ does not vanish for generic GL($1$) transformation. However, although the flavored scattering forms do not obey all BCJ relations, they do obey some BCJ relations when $n>2k$:

\begin{equation}
\sum^{n-1}_{i=2}\Big(\sum^{i}_{j=2}s_{j,n}\Big)A[1,2,\ldots,i,n,i+1,\ldots,n-1]=0 \ . 
\label{equationsss}
\end{equation}

\noindent where $n$ is restricted to an $\mathfrak{adj}$ state. We suspect that flavored scattering forms should obey a weaker condition than projectivity which is equivalent to eq.~\eqref{equationsss}. \\
\indent We now turn to the case where $\mathfrak{f}$-$\mathfrak{af}$ pairs states transform in the same flavor representation. Although we have not studied the positive geometry of these scalar amplitudes, their scattering forms are interesting enough to bear mentioning. We will focus on projectivity of the scattering form in $\mathcal{K}_{n}^{\star}$. Interestingly, if all ($\mathfrak{a}$)$\mathfrak{f}$ states transform in the same flavor representation, the scattering form is projective. Consider the four-point example again, except now all states transform in the same flavor representation:

\begin{equation}
\Omega^{n_{f}=1,k=2}_{n=4}[N]=N(g_{S}|\underline{A}\bar{A}\underline{B}\bar{B})d\log(S_{1,2})+N(g_{T}|\underline{A}\underline{B}\bar{A}\bar{B})d\log(S_{2,3}) \ .
\end{equation}

\noindent The $\mathfrak{f}$-$\mathfrak{af}$ pairs $A$ and $B$ now have the same flavor so $s_{\underline{A}\bar{B}}$ is not forbidden by flavor conservation. Under a local GL($1$) transformation, the scattering form transforms as

\begin{equation}
\Omega^{n_{f}=1,k=2}_{n=4}[N] \rightarrow (N(g_{S}|\underline{A}\bar{A}\underline{B}\bar{B})+N(g_{T}|\underline{A}\underline{B}\bar{A}\bar{B}))d\log(\Lambda)+\Omega^{n_{f}=1,k=2}_{n=4}[N] \ .
\end{equation}

\noindent Therefore, the requirement that the scattering form is projective imposes that 
\begin{equation}\label{eq:2term}
N(g_{S}|\underline{A}\bar{A}\underline{B}\bar{B})+N(g_{T}|\underline{A}\underline{B}\bar{A}\bar{B})=0 \ .
\end{equation}
This relation generalizes to higher points, where projectivity of the $n_{f}=1$ scattering form in $\mathcal{K}_{n}^{\star}$ implies

\begin{equation}
N(g_S|I_1I_2I_3I_4)+N(g_T|I_1I_4I_2I_3)+N(g_U|I_1I_3I_4I_2)=0 \,,
\label{colorkinemdualitfull}
\end{equation}

\noindent for all sub-graphs which do not violate charge conservation. For sub-graphs corresponding to all $\mathfrak{f}$-$\mathfrak{af}$ external states, the kinematic factor associated with the propagator that violates charge conservation is simply zero and eq.~(\ref{colorkinemdualitfull}) reduces to a two-term identity.

%like eq.~\eqref{eq:2term}. \\
\indent Eq.~(\ref{colorkinemdualitfull}) is interesting for a number of reasons. For example, for sub-graphs corresponding to all $\mathfrak{f}$-$\mathfrak{af}$ external states, eq.~(\ref{colorkinemdualitfull}) does not correspond to any relationship that color factors obey as eq.~(\ref{anitfund}) does not apply to sub-graphs corresponding to all $\mathfrak{f}$-$\mathfrak{af}$ external states. Instead, we see that eq.~(\ref{colorkinemdualitfull}) can be considered a natural generalization of color-kinematics duality that emerges from requiring the scattering form to be projective. These two-term identities were noted in \cite{Johansson:2017bfl}, but not expanded on further as they were not necessary for their double-copy prescription. In addition, the applicablity of eq.~(\ref{colorkinemdualitfull}) to ALL sub-graphs implies that the original KLT relations can be applied to $n_{f}=1$ (anti-)fundamental amplitudes \cite{Bjerrum_Bohr_2011, BjerrumBohr:2010yc, Bjerrum_Bohr_2010}. The only difference between $n_{f}=1$ (anti-)fundamental amplitudes and adjoint amplitudes is that many of the kinematic numerator factors in the (anti-)fundamental amplitudes are trivially zero due to violating charge conservation. Finally, a natural direction for further research is studying what conditions the scattering form must obey to be projective in $\mathcal{K}_{n}^{k}$, not $\mathcal{K}_{n}^{\star}$, when all $\mathfrak{f}$-$\mathfrak{af}$ pairs have the same flavor. \\

\subsection{Melia Decomposition Dual to Pullback}
\label{sec:genDDMdec}

We now further explore the color-kinematics duality by examining how the Melia decomposition of the amplitude is dual to pulling back the scattering form to an appropriate subspace, $H^{T}_{n}[\alpha]$. Unlike the adjoint case, this sub-space generally has higher dimension than $(n-3)$, but the pulled-back scattering form only depends on the coordinates of the $(n-3)$-dimensional subspace, $H_{n}[\alpha]$. We will simply state the qualitative results here, leaving the technical details to Appendix \ref{explciitformofHa}.  \\
\indent In the case of (anti-)fundamental color-dressed amplitudes, the color-dressed amplituded can be decomposed into a sum of partial amplitudes using eq.~(\ref{jacoblie}) and requiring that the kinematic numerators, $N(g|\alpha)$, obey the same relations as their associated color factors \cite{Johansson:2015oia}. The Melia decomposition of the amplitude is 

\begin{equation}
{\bf M}_n[N]=\sum_{\sigma\in\textrm{Melia basis}}C'((1,2),\sigma)M[N;(1,2),\sigma] \ ,
\end{equation}

\noindent where the sum is over all valid Melia basis orderings. The color constants $C'((1,2),\sigma)$ are non-trivial color constants given explicitly in \cite{Johansson:2015oia} and $M_{n}[N;(1,2),\sigma]$ is the color-stripped partial amplitude:

\begin{equation}
M_{n}[N;\alpha]=\sum_{\alpha\textrm{-planar }g}N(g|\alpha_{g})\prod_{I\in g} \frac{\vartheta_{I}}{s_{I}}  \ . 
\label{partiamplitudess}
\end{equation}

\noindent For the dual scattering form, we claim that Melia decomposition of the partial amplitude is dual to pulling back the scattering form to a specific subspace. The partial amplitude, eq.~(\ref{partiamplitudess}), can be obtained by pulling back the scattering form, eq.~(\ref{scatterinfoss}), to a subspace $H_{n}^{T}[\alpha]$,\footnote{We focus on Melia decomposition, and not color-trace decomposition, because the duality between color factors and differentials is applicable to theories transforming in any gauge group, such as Sp($N$), and naive color trace decomposition is not. To see this, note that our derivation in the previous section only relied on the definition of the structure constants and commutation relations. We did not use any properties unique to SU($N$) or U($N$) groups.} where

\begin{equation}
W(g|\kappa)|_{H_{n}^{T}[\alpha]}=\left\{\begin{matrix}
(-1)^{\textrm{flip}(\kappa,\alpha)}dV[\alpha] &\hphantom{aa} & \textrm{if }g\textrm{ is compatible with $\alpha$} \\ 
0 &\quad & \textrm{otherwise}
\end{matrix}\right. \ .
\label{conjectdiffofr} 
\end{equation}

\noindent where $\text{flip}(\kappa,\alpha)$ is the number of vertex flips that relates $\kappa$ and $\alpha$. Moreover, here $\kappa$ can be any ordering of external states, unlike $\alpha$, for which the first two states must be a $\mathfrak{f}$-$\mathfrak{af}$ pair. Unlike the adjoint case, the (anti-)fundamental scattering form after the pullback is not a top-form of $H_{n}^{T}[\alpha]$ but only 
%has support on 
depends on the coordinates of the $(n{-}3)$-dimensional subspace $H_{n}[\alpha]\subset H_{n}^{T}[\alpha]$. The measure ``$dV[\alpha]$'' is a volume form of $H_{n}[\alpha]$, not $H_{n}^{T}[\alpha]$. This phenomena is a direct consequence of the fact that the planar scattering form is not a top-form of $\mathcal{K}_{n}^{k}$.\\

The pullback to $H_n^T[\alpha]$ can be understood as follows. We consider the planar scattering form, $\Omega[N;\alpha]$,

\be
\Omega[N;\alpha]:=\sum_{g\in \Gamma[\alpha]}~N(g|\alpha_{g})W(g|\alpha_{g})\prod_{I\in g}\frac{\vartheta_{I}}{s_{I}}\, ,
\ee

\noindent where $\Gamma[\alpha]\subset \Gamma_n$ is the set of all the graphs compatible with the planar ordering $\alpha$. According to eq.~(\ref{conjectdiffofr}), only the planar scattering form should survive upon pullback of the full scattering form, eq.~(\ref{eq:H_ngeneric}), to $H^{T}_{n}[\alpha]$. However, unless all $\mathfrak{f}$-$\mathfrak{af}$ pairs are adjacent in $\alpha$, the planar variables associated with $\Omega[N;\alpha]$ do not span $\mathcal{K}_{n}^{k}$. Therefore, since $\Omega[N;\alpha]$ only depends on the coordinates of the planar variables, we can say that $\Omega[N;\alpha]$ only has support on the affine subspace $\mathcal{K}_{n}^{k}[\alpha]$ in $\mathcal{K}_{n}^{k}$.\footnote{Our construction of $\mathcal{K}_{n}^{k}[\alpha]$ is very similiar to the construction of the affine subsapce $\mathcal{Y}[\mathcal{Z}]$ from general twistor space in section 9 of~\cite{Arkani_Hamed_2018}. The planar coordinates that form a complete basis for $\mathcal{K}_{n}^{k}[\alpha]$ are analogous to the $y^{i}_{\alpha}$ that span the affine subspace $\mathcal{Y}[\mathcal{Z}]$.} We denote the orthogonal complement to $\mathcal{K}_{n}^{k}[\alpha]$ as $\mathcal{D}_{n}^{k}[\alpha]$:

\begin{equation}
\mathcal{D}_{n}^{k}[\alpha]=(\mathcal{K}_{n}^{k}[\alpha])^{\perp} \ . 
\end{equation}

\noindent The full relationship between $\mathcal{D}_{n}^{k}[\alpha]$, $\mathcal{K}_{n}^{k}[\alpha]$, and $\mathcal{K}_{n}^{k}$ is 

\begin{equation}
\mathcal{K}_{n}^{k}=\mathcal{K}_{n}^{k}[\alpha] \otimes \mathcal{D}_{n}^{k}[\alpha] \ .
\label{fullrelationss}
\end{equation}

\noindent The planar scattering form only has support in $\mathcal{K}_{n}^{k}[\alpha]$, but the full scattering form $\Omega^{(n-3)}[N]$ has support in $\mathcal{D}_{n}^{k}[\alpha]$ as well. Based on eq. (\ref{fullrelationss}), we can decompose $H^{T}_{n}[\alpha]$ as

\begin{equation}
H_{n}^{T}[\alpha]=H_{n}[\alpha]\otimes H^{A}_{n}[\alpha], \quad H_{n}[\alpha]\subset \mathcal{K}_{n}^{k}[\alpha], \quad H^{A}_{n}[\alpha]\subset \mathcal{D}_{n}^{k}[\alpha] \ ,
\label{genersol}
\end{equation}

\noindent where $\dim (H_{n}[\alpha])=(n-3)$ and $\dim(H^{A}_{n}[\alpha])\geqslant 0$ for the auxiliary space $H_n^A[\alpha]$. The space $H_n[\alpha]$ is simply given by the restrictions $\pmb{H}_{n}[\alpha]$ in eq.~(\ref{eq:H_ngeneric}) from section~\ref{sec:reccontr}. This automatically validates the first line of eq.~(\ref{conjectdiffofr}) by construction. However, the restrictions from $\pmb{H}_{n}[\alpha]$ are not enough to get rid of all incompatible graphs for generic orderings with $k>3$. This is unsurprising as the incompatible graphs for generic $\alpha$ have support in $\mathcal{D}_{n}^{k}[\alpha]$ as well as $\mathcal{K}_{n}^{k}[\alpha]$, so additional restrictions from $\pmb{H}^{A}_{n}[\alpha]$ are necessary to remove these unwanted contributions. Due to the complexity of $\pmb{H}^{A}_{n}[\alpha]$, we leave computational results to appendix~\ref{explciitformofHa}, where a closed form for $\pmb{H}^{A}_{n}[\alpha]$ is provided in eq.~(\ref{setrestriequss}).

\section{Conclusion}

In this paper we initiate the study of positive geometry and scattering forms for amplitudes with matter particles, {\it i.e.} particles flavored in (anti-)fundamental representations. The original paper~\cite{Arkani-Hamed:2017mur}, which treats scattering amplitudes as differential forms in kinematic space, has focused on amplitudes with particles purely in adjoint representation; here we pinpoint the new ingredients to include matter particles in this geometric picture. As a toy model, we find that a class of the so-called open associahedra, {\it i.e.} associahedra with certain faces sent to infinity, underpin all tree-level amplitudes of the bi-color $\phi^3$ scalar theory, where the bi-adjoint scalars and bi-fundamental ones play the role of ``gluons" and ``quarks", respectively. For any flavor assignment and a given planar ordering, we obtain an open associahedron which is determined by a $(n{-}3)$-dim subspace in the kinematic space; the canonical form then gives the corresponding amplitudes, with forbidden poles sent to infinity. Moreover, we discuss ``color is kinematics", {\it i.e.} the duality between color factors and wedge-products for cubic diagrams now in the presence of matter particles, and the projectivity of the scattering forms when there is only a single flavor. 

There are many open questions suggested by our preliminary discussions. First, we would like to study further the construction of subspaces for bi-color amplitudes, {\it e.g.} how the inverse soft construction {\it etc.} can be generalized, and how to obtain other open polytopes which are relevant for scattering amptlidues such as the Cayley polytopes~\cite{Gao:2017dek,He:2018pue}. Moreover, it is straightforward to generalize our construction to off-diagonal bi-color amplitudes, $m[\alpha|\beta]$ for $\alpha\neq \beta$; the latter is given by the intersection of the diagonal cases with $\alpha$ and $\beta$ ordering ~\cite{Frost_2018,Mizera_2017}. We similarly conjecture that the bi-color amplitude can be obtained as the volume of intersection of the corresponding dual open associahedra. A related open question is how to obtain a inverse matrix which can be used as the KLT matrix for QCD amplitudes~\cite{Brown:2018wss,Johansson:2019dnu}. 

An alternative direction is considering different triangulations of the canonical forms of open associahedra, which would in turn yield recursion relations for bi-color theory. Due to facets at infinity, the recursions given in~\cite{Arkani-Hamed:2017mur, He:2018svj} initially seem somewhat impractical for efficient calculations. It would be interesting to see if the triangulation in~\cite{Salvatori:2019phs,Yang:2019esm} could be generalized to open associahedra, yielding a BCFW-like recursion for bi-color amplitudes. It is possible that the inverse soft construction of the amplitude would be intimately tied to any such recursion.  Another approach is considering triangulations of the dual polytope that are not trivially equivalent to the Feynman diagram expansion. It would also be interesting to study possible connections to CHY-like formulas for QCD amplitudes in four dimensions~\cite{He:2016dol}.

The construction we proposed reveal rich structures underlying such positive geometry which deserve further investigations by themselves. 
%In~\cite{song:2013ttaaa} 
In~\cite{Arkani-Hamed:2019vag} the ABHY associahedron is generalized to polytopes for other finite-type cluster algebra, where the classical cases correspond to bi-adjoint $\phi^3$ amplitudes through one loop. It would be interesting to extend that construction to open cases with facets at infinity. We note that the factorization channels used in our construction are similar to those for constructing mulit-quark color decomposition~\cite{Ochirov:2019mtf}. 

%Such a construction may be related to the procedure in \cite{Ochirov:2019mtf} for constructing multi-quark color decomposition at one-loop using analysis of factorization channels. The procedure for finding color decomposition's using analysis of factorization channels is remarkably similar to our inverse soft construction. 

Throughout the paper we have not discussed the worldsheet perspective (string theory and CHY) for bi-color amplitudes and open associahedra. It is not difficult to come up with CHY formulas for such amplitudes, and some of them coincide with CHY formulas for Cayley polytopes~\cite{Gao:2017dek}. However, for general case the one-to-one map from moduli space to kinematic space and pushforward for scattering forms are still missing. The proper framework to proceed is the stringy canonical forms of~\cite{Arkani-Hamed:2019mrd}, and it would be highly desirable to find such string-like integrals where the so-called Minkowski sum of Newton polytopes gives an open associahedron. We remark that this new picture leads to new, CHY-like formulas for bi-color amplitudes, and we leave it to future investigations.

%We remark that such this new picture of~\cite{Arkani-Hamed:2019vag} gives new CHY-like formulas for bi-color amplitudes (or even just for bi-adjoint amplitudes with $\alpha\neq \beta$). 

%Color is kinematics and scattering forms, pullback, projectivity; color-kinematics duality and double copy, relations to QCD {\it etc.}...

%\AH{Future work: further generalizations of inverse soft construction, open cayley polytopes, Stringy Canonical Form/generalized CHY, extended quiver constructions of open cluster polytopes }

\acknowledgments
We would like to thank Alfredo Guevara, Marios Hadjiantonis, Henrik Johansson, Callum R. T. Jones, Gregor K\"{a}lin, Alok Laddha, Stephen Naculich, Shruti Paranjape and Jaroslav Trnka  %\textcolor{blue}{(add other people)}
for inspiring discussions. AH would like to especially thank Henriette Elvang for instrumental support and discussion early in the project. 
SH's research is supported in part by the Thousand Young Talents program, the Key Research Program of Frontier Sciences of CAS under Grant No. QYZDBSSW-SYS014, Peng Huanwu center under Grant No. 11747601 and National Natural Science Foundation of China under Grant No. 11935013. FT is supported by the Knut and Alice Wallenberg Foundation under grant
KAW 2013.0235, and the Ragnar S\"{o}derberg Foundation (Swedish Foundations’ Starting
Grant). FT would also like to thank the hospitality of CAS key laboratory of theoretical physics and Leinweber Center for Theoretical Physics at the University of Michigan.

\appendix

\section{Possible Deformations on Constraints}
\label{sec:defsoftlimit}
The subspace constraints $\pmb{H}_n[\alpha]$ in eq.~\eqref{eq:H_ngeneric} are all of the form $-s_{\mathsf{A},\mathsf{C}}=c_{\mathsf{A},\mathsf{C}}>0$ with non-adjacent sets $\mathsf{A}$ and $\mathsf{C}$. However, we can introduce certain deformations to $\pmb{H}_n[\alpha]$, and thus the polytope, while keep the canonical form unchanged. In fact, certain facets of the kinematic polytope given by $\pmb{H}_n[\alpha]$ are characterized by deformed constraints. Thus deformations are essential to understand generic factorization behavior of the polytope. Here, we provide a special class of such deformations, and leave more generic discussions to a following work \cite{so2013ttaaa}.

%Deformations are relevant to our construction since the facets of the kinematic polytope given by eq.~\eqref{eq:H_ngeneric} 
%Viewing the constraints as a linear system of equations, we can act a general linear transformation on them without changing its solution. The linear rearrangement can be useful to understand the factorization behavior of the polytope.

For a given block $\mathsf{B}_i$, we rewrite the constraints on $s_{L_j,r_j,r_i}=s_{\mathsf{B}_j,r_i}$ with $j\geqslant i{+}2$ in favor of those on $s_{\mathsf{B}_j,\mathsf{B}_i}$. These constraints come from the $\pmb{C}_1$ part in the recursive process when the block $\mathsf{B}_j$ is added. We first use the seven-term identity~\eqref{eq:7term} to write
\begin{align}
-s_{\mathsf{B}_i,\mathsf{B}_j}=-s_{\mathsf{B}_j,r_i}+X_{l_i,l_j}-X_{r_i,l_k}+X_{l_{j+1},r_i}-X_{l_{j+1},l_i}-X_{l_i,l_{i+1}}\,.
\end{align}
With the help of the constraints on $\{s_{\mathsf{B}_j,l_i},s_{\mathsf{B}_j,\mathsf{I}}\,|\,\mathsf{I}\in\text{sub}[\mathsf{B}_i]\}$ also generated in $\pmb{C}_1$, the right hand becomes
\begin{align}\label{eq:C1alter}
-s_{\mathsf{B}_i,\mathsf{B}_j}&=c_{\mathsf{B}_j,r_i}+c_{\mathsf{B}_j,l_i}+X_{l_j,l_{j+1}}-X_{l_i,l_{i+1}}+\sum_{\mathsf{I}\in\text{sub}[\mathsf{B}_i]}\left(c_{\mathsf{B}_j,\mathsf{I}}+X_{l_{\mathsf{I}},l_{\mathsf{I}+1}}+X_{l_j,l_{j+1}}\right)\nonumber\\
&=c_{\mathsf{B}_j,r_i}+c_{\mathsf{B}_j,l_i}+(d_i+1)X_{l_j,l_{j+1}}-X_{l_i,l_{i+1}}+\sum_{\mathsf{I}\in\text{sub}[\mathsf{B}_i]}\left(c_{\mathsf{B}_j,\mathsf{I}}+X_{l_{\mathsf{I}},l_{\mathsf{I}+1}}\right),
\end{align}
where $d_i$ counts the number of sub-blocks in $\mathsf{B}_i$. If $\mathsf{I}$ is the last sub-block, then $l_{\mathsf{I}+1}=r_i$. The constraint~\eqref{eq:C1alter} is completely equivalent to $-s_{\mathsf{B}_j,r_i}=c_{\mathsf{B}_j,r_i}>0$. We can thus use the former instead of the latter in $\pmb{H}_n[\alpha]$ and the resultant polytope $\mathcal{A}_n[\alpha]$ is unchanged. Now suppose $\mathsf{B}_i$ actually contains $a$ additional sub-blocks but all of which are taken soft, one can check that we recover most of the constraints in $\pmb{H}_n[\alpha]$ while the only trace of these soft sub-blocks is in eq.~\eqref{eq:C1alter}: $d_i$ gets shifted by $a$,
\begin{align}
-s_{\mathsf{B}_i,\mathsf{B}_j}=c_{\mathsf{B}_j,r_i}+c_{\mathsf{B}_j,l_i}+(d_i+a+1)X_{l_j,l_{j+1}}-X_{l_i,l_{i+1}}+\sum_{\mathsf{I}\in\text{sub}[\mathsf{B}_i]}\left(c_{\mathsf{B}_j,\mathsf{I}}+X_{l_{\mathsf{I}},l_{\mathsf{I}+1}}\right).
\end{align}
If $\mathsf{B}_i=g_i$ is an $\mathfrak{adj}$ particle, then we can start with a generic block and take all the sub-blocks, including the pair $(l_i,r_i)$, to be soft, which leads to a similar formula:
\begin{align}\label{eq:CgiBj}
-s_{g_i,\mathsf{B}_j}=c_{g_i,\mathsf{B}_j}+aX_{l_j,l_{j+1}}\,.
\end{align}
Switching back to the constraint on $s_{\mathsf{B}_j,r_i}$ amounts to cast the shift $a X_{l_j,l_{j+1}}$ onto the constant $c_{\mathsf{B}_j,r_i}$ for $i+2\leqslant j\leqslant m$:
\begin{align}\label{eq:deformedH}
\pmb{H}_n[\alpha;a]&=\pmb{H}_n[\mathsf{B}_1,\ldots,\mathsf{B}_i,\ldots,\mathsf{B}_m;a]\nonumber\\
&=\pmb{H}_n[\alpha]\Big|_{\left\{-s_{\mathsf{B}_j,r_i}=c_{\mathsf{B}_j,r_i}\rightarrow -s_{\mathsf{B}_j,r_i}=c_{\mathsf{B}_j,r_i}+aX_{l_j,l_{j+1}}\text{ for }i+2\leqslant j\leqslant m\right\}}\,.
\end{align}
We note that $\mathsf{B}_i=g_i$ being an $\mathfrak{adj}$ particle is also covered by the above replacement because we identify $l_i=r_i=g_i$. If all the blocks are $\mathfrak{adj}$ particles, these shifts vanish identically since $X_{l_j,l_{j+1}}=0$ for all cases. For example, the constraints
\begin{align}
\pmb{H}_7[(1,2),q,(5,6),(7,8)]=\left\{\begin{array}{c}
s_{2,5,6},s_{5,6,8},s_{q,6},s_{q,8},s_{2,7,8},s_{5,7,8},s_{q,7,8} \\
\text{set to negative constants}
\end{array}\right\}
\end{align}
allow the following deformation on $s_{q,7,8}$ according to eq.~\eqref{eq:deformedH},
\begin{align}\label{eq:H_7def}
\pmb{H}_7[(1,2),q,(5,6),(7,8);a]=&\left\{\begin{array}{c}
s_{2,5,6},s_{5,6,8},s_{q,6},s_{q,8},s_{2,7,8},s_{5,7,8} \\
\text{set to negative constants}
\end{array}\right\}\nonumber\\
&\cup\{-s_{q,7,8}=c_{q,7,8}+aX_{1,7}\}\,.
\end{align}
This deformation is relevant to the $X_{3,5}=q^2=0$ facet of $\mathcal{A}_8[(1,2),(3,4),(5,6),(7,8)]$, as we will see in the next appendix.

We denote the polytope obtained from $\pmb{H}_n[\alpha;a]$ as $\mathcal{A}_n[\alpha;a]$. We have checked a variety of examples to high enough multiplicity that the canonical form after the pullback has the same expression,
\begin{align}
\Omega_n[\alpha]\Big|_{\mathcal{A}_n[\alpha]}=A_n[\alpha]d^{n-3}X\,,& &\Omega_n[\alpha]\Big|_{\mathcal{A}_n[\alpha,a]}=A_n[\alpha]d^{n-3}X\,,\qquad (a\geqslant 0)\,.
\end{align}
Therefore, we have a continuous equivalent class of subspaces that yields the same canonical form. The soft limit argument is a physical way to understand why this equivalence class exists. \\
\indent We can also perform an additional check that the canonical form remains unchanged using the dual polytope picture. Instead of directly calculating the canonical form, we calculate the canonical rational function,

\begin{equation}
A_n=\underline{\Omega}[\mathcal A]=\text{vol}[\mathcal A^{\star}] \ ,
\end{equation}

\noindent where $A^{\star}$ is the dual polytope. The canonical rational function associated with given facet vectors can be calculated using a vertex expansion of the rational function:

\begin{equation}
\underline{\Omega}=\sum_{v\in \textrm{vertices}}\frac{\langle \Pi_{I\in v}W_{I}\rangle}{(Y\cdot W^{\star})\Pi_{I\in v} \langle (Y\cdot W_{I})\rangle } \ ,
\label{fess}    
\end{equation}

\noindent where $\langle \ldots \rangle$ denotes a determinant. To derive eq.~(\ref{fess}), first remember that the canonical rational function is equivalent to the volume of the dual polytope and each vertex in the original polytope is mapped to a facet in the dual polytope. Each term in eq.~(\ref{fess}) corresponds to the volume of a simplex in the dual polytope bounded by a facet and reference vector, $W^{\star}$. If $W^{\star}=(1,0,\ldots,0)$, one finds $\langle \Pi_{I\in v}W_{I}\rangle=\pm 1$ and that this expansion is equivalent to the Feynman diagram expansion of the canonical rational function. To show that the Feynman diagram is still equivalent after some deformation, one must show that the numerator, $\langle \Pi_{I\in v}W_{I}\rangle$, is unchanged. We checked a number of examples and found this property held under the deformation in eq.~(\ref{eq:CgiBj}).

\section{Explicit Factorization Examples}
\label{sec:factorizationExamples}
As we have briefly discussed in section~\ref{sec:remarks}, the facets of the kinematic polytope $\mathcal{A}_n[\alpha]$ given by the constraints $\pmb{H}_n[\alpha]$ following the procedure~\eqref{eq:H_ngeneric} are in general a semi-direct product of lower dimensional polytopes bounded by a deformed version of constraints. 

%In this appendix, we work out some concrete examples. 

In this appendix, we study the polytope $\mathcal{A}_8[(1,2),(3,4),(5,6),(7,8)]$ as a concrete example. It is carved out by the constraints~\eqref{eq:H8b}, which can be written in terms of the planar Mandelstam variables as
\begingroup
\allowdisplaybreaks
\begin{align}\label{eq:constraints2}
& c_{2,5,6}=X_{3,7}+X_{2,5}-X_{5,7}-X_{3,5}-X_{2,7}\,,& & c_{3,5,6}=X_{4,7}+X_{3,5}-X_{5,7}-X_{3,7}\,,\nonumber\\
& c_{3,4,6}=X_{3,6}+X_{5,7}-X_{3,5}-X_{3,7}\,,& & c_{2,7,8}=X_{2,7}+X_{1,3}-X_{1,7}-X_{3,7}\,,\nonumber\\
& c_{3,7,8}=X_{3,7}+X_{1,4}-X_{1,7}-X_{4,7}-X_{1,3}\,,& & c_{4,7,8}=X_{4,7}+X_{1,5}-X_{1,7}-X_{5,7}-X_{1,4}\,,\nonumber\\
& c_{5,7,8}=X_{1,6}+X_{5,7}-X_{1,7}-X_{1,5}\,,& & c_{3,4,8}=X_{3,8}+X_{1,5}-X_{3,5}-X_{5,8}-X_{1,3}\,,\nonumber\\
& c_{5,6,8}=X_{1,7}+X_{5,8}-X_{5,7}-X_{1,5}\,.
\end{align}
\endgroup
We first show that the facet $X_{3,5}=p_{34}^2=0$ is given by the deformed constraints~\eqref{eq:H_7def},
 \begin{align}
\mathcal{A}_8[(1,2),(3,4),(5,6),(7,8)]\Big|_{X_{3,5}=0}=\mathcal{A}_7[(1,2),p_{34},(5,6),(7,8);a=1]\,,
\end{align}
Following eq.~\eqref{eq:C1alter}, we can replace the constraint $c_{3,4,8}$ by
\begin{align}
-s_{3,4,7,8}=c_{3,7,8}+c_{4,7,8}+X_{1,7}-X_{3,5}\,.
\end{align}
When $X_{3,5}=p_{34}^2=0$, it becomes 
\begin{align}
-s_{p_{34},7,8}=c_{3,7,8}+c_{4,7,8}+X_{1,7}=c_{p_{34},7,8}+X_{1,7}\,,
\end{align}
which reproduces the second line of eq.~\eqref{eq:H_7def} with $a=1$. For the rest of eq.~\eqref{eq:constraints2}, the constraint $c_{3,5,6}$ and $c_{3,7,8}$ are automatically satisfied when $X_{3,5}=0$ and thus drop out, while all the others are trivially inherited by eq.~\eqref{eq:H_7def}.

We next show that the facet $X_{3,7}=p_{3456}^2=0$ is a semi-direct product,
\begin{align}
\mathcal{A}_8[(1,2),(3,4),(5,6),(7,8)]\Big|_{X_{3,7}=0}=\mathcal{A}^L[(1,2),p_{3456},(7,8)]\ltimes\mathcal{A}^R[-p_{3456},(3,4),(5,6)]\,.
\end{align}
Among the original constraints shown in eq.~\eqref{eq:constraints2}, $c_{3,5,6}$ and $c_{3,4,6}$ together give $\mathcal{A}^R$, which is the same as obtained from eq.~\eqref{eq:H_ngeneric}. We can rewrite the constraint $c_{3,4,8}$ into
\begin{align}
-s_{3,4,5,6,8}=c_{3,4,8}+c_{5,6,8}+X_{3,5}+X_{5,7}-X_{3,7}
\end{align}
such that on the facet $X_{3,7}=p_{3456}^2=0$ it reduces to
\begin{align}\label{eq:sp8}
-s_{p_{3456},8}=c_{3,4,8}+c_{5,6,8}+X_{3,5}+X_{5,7}=c_{p_{3456},8}+X_{3,5}+X_{5,7}\,.
\end{align}
Together with the constraint $c_{2,7,8}$, it carves out $\mathcal{A}^L$. We may view eq.~\eqref{eq:sp8} as $$-s_{p_{3456},8}=\tilde{c}_{p_{3456},8}$$ but the constant $\tilde{c}$ is linearly shifted by variables in $\mathcal{A}^R$, and hence the semi-direct product. The linear shift does not affect the factorization of the canonical form since $\Omega(\mathcal{A}^R)$ is always a top form.

\section{Derivation of \texorpdfstring{$\dim(\mathcal{K}_{n}^{k})$}{dimKnk}}
\label{sec:derivatofdim}

In this section, we show that the dimension of $\mathcal{K}_{n}^{k}$ is 

\begin{equation}
\dim(\mathcal{K}_{n}^{k})=\frac{n(n-3)}{2}-\frac{k(k-1)}{2}\ . 
\label{dimessnbigkineamticsp}
\end{equation}

\noindent We argue that $\mathcal{K}_{n}^{k}$ is spanned by the planar variables of some ordering $\alpha$ where all $\mathfrak{f}$-$\mathfrak{af}$ pairs are adjacent. We will assume without proof that the planar variables are themselves orthogonal like in the $k=0$ case. By orthogonal, we mean that no planar variable of a given ordering, $\alpha$, can be written as a linear combination of the other planar variables of the same ordering. Therefore, since the number of planar variables for such an ordering is eq.~(\ref{dimessnbigkineamticsp}), this implies the dimension of the space is eq.~(\ref{dimessnbigkineamticsp}). \\
\indent Using the 7-term identity, one can directly prove that any $s_{I}$ with $p\leqslant 4$ can either be written as a sum of $X_{i,j}$ variables or $s_{I'}$ variables with $|I'|<|I|$. Such a direct proof is tedious, but straightforward, so we will not reproduce it here. We now prove that any $s_{I}$ with $p=|I|>4$ can be written as a sum of planar variables and $s_{I'}$, with $p'=|I'|<p$. Given any $s_{I}$, we isolate two elements $I_{1}$ and $I_{2}$, where $I_{i}$ is either a single $\mathfrak{adj}$ state or a $\mathfrak{f}$-$\mathfrak{af}$ pair. We then define $K:=I \setminus \{ I_{1},I_{2} \}$ and use the 7-term identity to write 

\begin{equation}
S_{I}=S_{I_{1},I_{2}}+S_{I_{1},K}+S_{I_{2},K}-S_{K}-S_{I_{1}}-S_{I_{2}}    \,,
\end{equation}

\noindent where every term on the right hand side takes the form of an $S_{I'}$ with $p'=|I'|<p$. It is always possible to do this if $p>4$, which is why direct proofs for $p\leqslant 4$ are necessary. Therefore, by induction, we can write any $S_{I}$ as $S_{I}=\sum X$ where the summation is over the planar variables of an ordering where all $\mathfrak{f}$-$\mathfrak{af}$ pairs are adjacent.

\section{Explicit Form of \texorpdfstring{$H_{n}^{A}[\alpha]$}{HnA}}
\label{explciitformofHa}

\indent In this Appendix, we will first provide a number of examples before giving the explicit form of $H_{n}^{A}[\alpha]$. For convenience, general $\mathfrak{f}$-$\mathfrak{af}$ blocks will now be denoted using the positions of the $l_{i}$ and $r_{i}$ states: $B_{l_{i},r_{i}}$. \\
\indent We first examine two examples where $\pmb{H}_{n}^{A}[\alpha]=\emptyset$. Consider a 6-point partial amplitude with the ordering $\alpha=[\underline{A},\bar{A},\underline{B},\bar{B},\underline{C},\bar{C}]$. Since all $\mathfrak{f}$-$\mathfrak{af}$ pairs are adjacent in $\alpha$, we find that $\mathcal{K}_{n}^{k}[\alpha]=\mathcal{K}_{n}^{k}$. Therefore, $\mathcal{D}_{n}^{k}[\alpha]$ is the null set and only restriction equations from $\pmb{H}_{n}[\alpha]$ are necessary to remove all incompatible diagrams:

\begin{equation}
\pmb{H}[\alpha]=\{ -s_{\underline{B},\underline{C},\bar{C}}=c_{\underline{B},\underline{C},\bar{C}}, \ -s_{\bar{A},\underline{C},\bar{C}}=c_{\bar{A},\underline{C},\bar{C}}, \ -s_{\underline{B},\bar{B},\bar{C}}=c_{\underline{B},\bar{B},\bar{C}} \} \ .
\label{exampleonedecom}
\end{equation}

\noindent For example, consider the diagram associated with the d$\log$ form:

\begin{equation}
W(g|\kappa)=ds_{\underline{A},\bar{A}}\wedge ds_{\underline{A},\bar{A}, \bar{B}} \wedge d s_{\underline{A},\bar{A}, \underline{B}, \bar{B}} \ .
\label{exampledifferent}
\end{equation}

\noindent The graph associated with eq.~(\ref{exampledifferent}) is clearly inconsistent with $\kappa=[\underline{A},\bar{A},\underline{B},\bar{B},C,\bar{C}]$, so the associated differential form must go to 0 on the pullback. On the pull-back defined by eq.~(\ref{exampleonedecom}), we see that $ds_{\underline{A},\bar{A},\bar{B}}=d(c_{\underline{C},\bar{C},\underline{B}})=0$, so eq.~(\ref{exampledifferent}) goes to 0 as expected. \\
\indent As a more non-trivial example, now consider a 6-point partial amplitude with ordering $\alpha=[\underline{A},\bar{A},\underline{B},\underline{C},\bar{C},\bar{B}]$. The planar scattering form is no longer top dimensional so $\mathcal{D}_{n}^{k}\neq \emptyset$. Since $\dim(K_{n}^{k}[\alpha])=4$, $H_{n}$ only contributes one restriction equation,

\begin{equation}
\pmb{H}_{n}[\alpha]=\{ -s_{\underline{B},\bar{B}}=c_{\underline{B},\bar{B}} \}  \ .
\label{singelrestreuq}
\end{equation}

\noindent Although $\mathcal{D}_{n}^{k}[\alpha]\neq\emptyset$, $H^{A}_{n}[\alpha]$ does not need to contribute any constraints for eq.~(\ref{conjectdiffofr}) to hold. For example, the differential corresponding to one incompatible diagram is 

\begin{equation}
dW=ds_{\underline{A},\bar{A}}\wedge ds_{\underline{B},\bar{B}} \wedge ds_{\underline{C},\bar{C}}
\end{equation}

\noindent which is 0 as $ds_{\underline{B},\bar{B}}=d(c_{\underline{B},\bar{B}})=0$. In fact, the corresponding $W(g|\alpha)$ factor of every incompatible diagram includes a factor of $ds_{\underline{B},\bar{B}}$, which goes to 0 on the support of eq.~(\ref{singelrestreuq}). \\
\indent For $k>3,n>6$, we need additional restrictions from $\pmb{H}^{A}_{n}[\alpha]$. Consider a partial amplitude with ordering  $\alpha=[\underline{A}\bar{A}\underline{B}\underline{C}\underline{D}\bar{D}\bar{C}\bar{B}]$. Since not all $\mathfrak{f}$-$\mathfrak{af}$ pairs are adjacent, $\mathcal{K}_{n}^{k}[\alpha]$ is not top dimensional. The restriction equations of $H_{n}[\alpha]$ are 

\begin{equation}
\pmb{H}_{n}[\alpha]=\{ -s_{\underline{B},\bar{B}}=c_{\underline{B},\bar{B}}, \  -s_{\underline{C},\bar{C}}=c_{\underline{C},\bar{C}} \} \ .
\label{firstround}
\end{equation}

\noindent Some incompatible diagrams survive the pullback to $H_{n}[\alpha]$, showing that these restriction equations are not enough to remove all incompatible Feynman diagrams. The differentials corresponding to these surviving incompatible diagrams are 

\begin{equation}
\begin{split}
dW_{\underline{A}}&=ds_{\underline{A},\bar{A}}\wedge ds_{\underline{A},\bar{A},\underline{C}}\wedge ds_{\underline{A},\bar{A},\underline{C},\bar{C}}\wedge ds_{\underline{A},\bar{A},\underline{B},\underline{C},\bar{C}}\wedge ds_{\underline{A},\bar{A},\underline{B},\underline{C},\bar{C},\bar{B}}\ ,\\
dW_{\bar{A}}&=ds_{\underline{A},\bar{A}}\wedge ds_{\underline{A},\bar{A},\underline{C}}\wedge ds_{\underline{A},\bar{A},\underline{C},\bar{C}}\wedge ds_{\underline{A},\bar{A},\underline{C},\bar{C},\bar{B}}\wedge ds_{\underline{A},\bar{A},\underline{B},\underline{C},\bar{C},\bar{B}} \ , \\
dW_{\underline{B}}&=ds_{\underline{A},\bar{A}}\wedge ds_{\underline{A},\bar{A},\bar{C}}\wedge ds_{\underline{A},\bar{A},\underline{C},\bar{C}}\wedge ds_{\underline{A},\bar{A},\underline{B},\underline{C},\bar{C}}\wedge ds_{\underline{A},\bar{A},\underline{B},\underline{C},\bar{C},\bar{B}}\ , \\
dW_{\underline{C}}&=ds_{\underline{A},\bar{A}}\wedge ds_{\underline{A},\bar{A},\bar{C}}\wedge ds_{\underline{A},\bar{A},\underline{C},\bar{C}}\wedge ds_{\underline{A},\bar{A},\underline{C},\bar{C},\bar{B}}\wedge ds_{\underline{A},\bar{A},\underline{B},\underline{C},\bar{C},\bar{B}} \ .
\end{split}
\end{equation}

\noindent To get rid of these incompatible diagrams, one additional restriction from $\pmb{H}^{A}_{n}[\alpha]$ is necessary:

\begin{equation}
\pmb{H}^{A}_{n}[\alpha]=\{ -s_{\underline{D},\bar{D},\bar{B}}=c_{\underline{D},\bar{D},\bar{B}}\}   \ . 
\label{restric}
\end{equation}

\noindent The first and third differential vanish as 

\begin{equation}
ds_{\underline{A},\bar{A},\underline{B},\underline{C},\bar{C}}=-d(c_{\underline{D},\bar{D},\bar{B}})=0 \ .
\end{equation}

\noindent It is less obvious that the second and fourth differentials vanish under the support of eq.~(\ref{restric}), but they do nonetheless.\footnote{To check this, first write out the $ds_{I}$ variables using a complete basis of $H_{n}^{T}[\alpha]$. For example, planar variables of an ordering where all $\mathfrak{f}$-$\mathfrak{af}$ pairs are adjacent form a complete basis. We then write each $ds_{I}$ variable as a vector in this basis. Checking that the differential vanishes amounts to showing that these vectors are not linearly independent.} \\
\indent We now consider the general form of $\pmb{H}^{A}_{n}[\alpha]$. The first type of restriction takes the form

\begin{equation}
-s_{\mathsf{B},\mathsf{B}'}=c_{\mathsf{B},\mathsf{B}'} \ ,
\label{firsttypeofrestri}
\end{equation}

\noindent where $\mathsf{B}$ and $\mathsf{B}'$ are $\mathfrak{f}$-$\mathfrak{af}$ blocks or $\mathfrak{adj}$ states.\footnote{By $\mathfrak{f}$-$\mathfrak{af}$ blocks, we are also including blocks at all levels in $\alpha$.} We impose the additional restrictions that $\mathsf{B}\cap \mathsf{B}'=\phi$ and that $\mathsf{B}$ and $\mathsf{B}'$ are separated by at least two flavor lines. The number of flavor lines separating $\mathsf{B}$ and $\mathsf{B}'$ is the number of flavor lines that cross a line connecting $l_{i} \in \mathsf{B}_{l_i,r_i}$ to $l_{j}'\in \mathsf{B}_{l_j, r_j}'$.\footnote{In the case that $\mathsf{B}$ and/or $\mathsf{B}'$ is an adjoint state, $l$ is the adjoint state.} The second class of restrictions take the form 

\begin{equation}
-s_{\mathsf{B},r}=c_{\mathsf{B},r} \ ,
\label{sectypeofrestri}
\end{equation}

\noindent where $r$ is any $\mathfrak{af}$ state except for $r=2$. We again impose the restriction that $\mathsf{B}\cap r=\phi$. We impose the restriction that the line connecting the vertex associated with the $r$ state and the vertex associated with $l_{i} \in \mathsf{B}_{l_{i},r_{i}} $ must cross at least one flavor line other than the flavor line associated with the $r$ state. Imposing that $\mathsf{B}$ and $r$ ($\mathsf{B}'$) are separated by at least 2 (1) flavor lines ensures that these restrictions are orthogonal to planar variables. These restraints can be summarized as:

\begin{equation}
\begin{split}
\pmb{H}^{A}_{n}[\alpha]:=&\{ s_{\mathsf{B},Y}=-c_{\mathsf{B},Y}, \quad Y=\mathsf{B}' \textrm{ or }r \ | \  \textrm{when }\emptyset=\mathsf{B} \cap Y,  \\
&\indent \textrm{If }Y=\mathsf{B}', \ \mathsf{B} \textrm{ and } \mathsf{B}' \textrm{ are separated by at least two flavor lines} \ ,\\
&\indent \textrm{If }Y=r, \ \mathsf{B} \textrm{ and } r \textrm{ are separated by at least one flavor line} \\
&\indent \indent \textrm{other than the flavor line associated with $r$} \ , \\
&\indent \textrm{and } \mathsf{B}, \mathsf{B}' \neq B_{1,2}, \  r \neq 2 \}  \,.
\end{split}
\label{setrestriequss}
\end{equation}

\noindent Eq. (\ref{setrestriequss}) was numerically checked for all possible orderings up to $n=10$. After providing some examples for $\pmb{H}^{A}_{n}[\alpha]$ below, we sketch a proof for eq.~(\ref{setrestriequss}). A more rigorous proof, which requires a more systemic analysis of factorization channels, will be presented in \cite{so2013ttaaa}.  \\

\begin{figure}[t]
\centering
\begin{tikzpicture}[every node/.style={font=\footnotesize,},dir/.style={decoration={markings, mark=at position \halfway with {\arrow{Latex}}},postaction={decorate}}]
\coordinate (p1) at (0:2);
\coordinate (p2) at (45:2);
\coordinate (p3) at (90:2);
\coordinate (p4) at (135:2);
\coordinate (p5) at (180:2);
\coordinate (p6) at (225:2);
\coordinate (p7) at (270:2);
\coordinate (p8) at (315:2);
\node at (p1)  [vertex,label={[label distance=-2pt]0:{$1$}}] {};
\node at (p2)  [vertex,label={[label distance=-2pt]45:{$8$}}] {};
\node at (p3) [vertex,label={[label distance=-1pt]90:{$7$}}] {};
\node at (p4) [vertex,label={[label distance=-2pt]135:{$6$}}] {};
\node at (p5) [vertex,label={[label distance=-2pt]180:{$5$}}] {};
\node at (p6) [vertex,label={[label distance=-1pt]225:{$4$}}] {};
\node at (p7) [vertex,label={[label distance=-1pt]270:{$3$}}] {};
\node at (p8) [vertex,label={[label distance=-1pt]315:{$2$}}] {};
\path (p1.center) -- (p7.center) node (d1) [pos=0.5] {};
\path (p2.center) -- (p6.center) node (d2) [pos=0.5] {};
\path (p3.center) -- (p5.center) node (d3) [pos=0.5] {};
\path (p1.center) -- (p2.center) node (f1) [pos=0.5] {};
\path (p2.center) -- (p3.center) node (f2) [pos=0.5] {};
\path (p3.center) -- (p4.center) node (f3) [pos=0.5] {};
\path (p4.center) -- (p5.center) node (f4) [pos=0.5] {};
\path (p5.center) -- (p6.center) node (f5) [pos=0.5] {};
\path (p6.center) -- (p7.center) node (f6) [pos=0.5] {};
\path (p7.center) -- (p8.center) node (f7) [pos=0.5] {};
\path (p8.center) -- (p1.center) node (f8) [pos=0.5] {};
\draw [thick,gray] (f3.center) to [bend left=20] (f4.center) ;
\draw [thick,gray] (f5.center) to [bend right=5] (f2.center) ;
\draw [thick,gray] (f6.center) to [bend left=5] (f1.center) ;
\draw [thick,gray] (f8.center) to [bend right=20] (f7.center) ;
\draw [thick,purple] (p5.center) to (p2.center) ;
\draw [thick] (p1.center) -- (p2.center) -- (p3.center) -- (p4.center) -- (p5.center) -- (p6.center) -- (p7.center) -- (p8.center) -- cycle;
%\draw [thick] (p1.center) to [bend left=30] (p3.center) (p1.center) to [bend right=30] (p3.center);
\node at (0,-2.7) [] {$\alpha=[(1,2),(3,(4,(5,6),7),8)]$};
\begin{scope}[xshift=6cm]
\coordinate (p1) at (0:2);
\coordinate (p2) at (36:2);
\coordinate (p3) at (72:2);
\coordinate (p4) at (108:2);
\coordinate (p5) at (144:2);
\coordinate (p6) at (180:2);
\coordinate (p7) at (216:2);
\coordinate (p8) at (252:2);
\coordinate (p9) at (288:2);
\coordinate (p10) at (324:2);
\node at (p1)  [vertex,label={[label distance=-2pt]0:{$1$}}] {};
\node at (p2)  [vertex,label={[label distance=-2pt]36:{$10$}}] {};
\node at (p3) [vertex,label={[label distance=-1pt]72:{$9$}}] {};
\node at (p4) [vertex,label={[label distance=-2pt]108:{$8$}}] {};
\node at (p5) [vertex,label={[label distance=-2pt]144:{$7$}}] {};
\node at (p6) [vertex,label={[label distance=-1pt]180:{$6$}}] {};
\node at (p7) [vertex,label={[label distance=-1pt]216:{$5$}}] {};
\node at (p8) [vertex,label={[label distance=-1pt]252:{$4$}}] {};
\node at (p9) [vertex,label={[label distance=-1pt]288:{$3$}}] {};
\node at (p10) [vertex,label={[label distance=-1pt]324:{$2$}}] {};
\path (p2.center) -- (p8.center) node (d1) [pos=0.5] {};
\path (p3.center) -- (p7.center) node (d2) [pos=0.5] {};
\path (p4.center) -- (p6.center) node (d3) [pos=0.6] {};
\path (p1.center) -- (p2.center) node (f1) [pos=0.5] {};
\path (p2.center) -- (p3.center) node (f2) [pos=0.5] {};
\path (p3.center) -- (p4.center) node (f3) [pos=0.5] {};
\path (p4.center) -- (p5.center) node (f4) [pos=0.5] {};
\path (p5.center) -- (p6.center) node (f5) [pos=0.5] {};
\path (p6.center) -- (p7.center) node (f6) [pos=0.5] {};
\path (p7.center) -- (p8.center) node (f7) [pos=0.5] {};
\path (p8.center) -- (p9.center) node (f8) [pos=0.5] {};
\path (p9.center) -- (p10.center) node (f9) [pos=0.5] {};
\path (p10.center) -- (p1.center) node (f10) [pos=0.5] {};
\draw [thick,gray] (f9.center) to [bend left=20] (f10.center) ;
\draw [thick,gray] (f1.center) to [bend left=20] (f2.center) ;
\draw [thick,gray] (f8.center) to (f3.center) ;
\draw [thick,gray] (f7.center) to (f4.center) ;
%\draw [thick,gray] (f5.center) to [bend right=20] (f3.center);
\draw [thick,gray] (f6.center) to [bend right=20] (f5.center) ;
\draw [thick,red] (p7.center) to (p3.center) ;
\draw [thick,purple] (p6.center) to (p3.center) ;
\draw [thick,purple] (p5.center) to (p3.center) ;
\draw [thick,purple] (p7.center) to (p4.center) ;
\draw [thick,purple] (p7.center) to (p2.center) ;
\draw [thick,purple] (p8.center) to (p2.center) ;
\draw [thick] (p1.center) -- (p2.center) -- (p3.center) -- (p4.center) -- (p5.center) -- (p6.center) -- (p7.center) -- (p8.center) -- (p9.center) -- (p10.center) -- cycle;
\node at (0,-2.7) [] {$\alpha=[(1,2)(3,(4,(5,6),7),8)(9,10)]$};
\end{scope}
\end{tikzpicture}
\caption{The flavor lines given by the $\mathfrak{f}$-$\mathfrak{af}$ pairs are shown in gray. The red and purple lines are associated with constraints of the form $s_{\mathsf{B},\mathsf{B}'}$ and $s_{\mathsf{B},r}$ respectively. Note that each purple line crosses at least one flavor line and each red line crosses at least two flavor lines.}
%Note that in the left panel, the $\mathfrak{f}$-$\mathfrak{af}$ pair $(1,2)$ and $(3,5)$ correspond to the same diagonal but counted as two different flavor lines.
\label{fig:flavorseparations}
\end{figure}
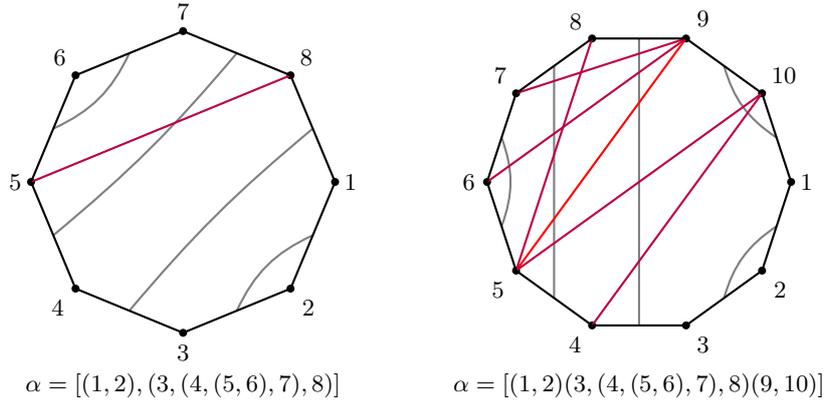

\indent Due to the complexity of eq.~(\ref{setrestriequss}), we consider some examples of how to calculate $\pmb{H}^{A}_{n}[\alpha]$. First, consider the ordering $$\pmb{H}^{A}_{n}[(1,2),(3,(4,(5,6),7),8)]\ .$$ The set of all relevant blocks is 

\begin{equation}
\{ \mathsf{B}_{3,8}, \ \mathsf{B}_{4,7}, \ \mathsf{B}_{5,6} \}   
\label{setsubblockss}
\end{equation}

\noindent and the set of relevant $\mathfrak{af}$ states is

\begin{equation}
\{ 6, \ 7, \ 8 \} \ . 
\end{equation}

\noindent Note that we have not included the $\mathsf{B}_{1,2}$ block or $r=2$. Furthermore, we have included all sub-blocks in eq.~(\ref{setsubblockss}). The set of all $s_{\mathsf{B},Y}$ with $\emptyset=\mathsf{B} \cap A$ is  

\begin{equation}
\{ s_{\mathsf{B}_{5,6},7},\ s_{\mathsf{B}_{5,6},8}, \ s_{\mathsf{B}_{4,7},8} \} \ .
\end{equation}

\noindent We now subtract all $s_{\mathsf{B},Y}$ which are not separated by enough flavor lines. A visualization of the surviving constraints is provided in figure \ref{fig:flavorseparations}.

\begin{equation}
\begin{split}
\pmb{H}^{A}_{n}[(1,2),(3,(4,(5,6),7),8)]=\{-s_{\mathsf{B}_{5,6},8}=c_{\mathsf{B}_{5,6},8} \} \ .
\end{split}
\end{equation}

\noindent As a slightly more non-trivial example, consider the ordering  $$\pmb{H}^{A}_{n}[(1,2)(3,(4,(5,6),7),8),(9,10)]\ . $$ The set of all blocks relevant for eq.~(\ref{setrestriequss}) is now

\begin{equation}
\{ \mathsf{B}_{3,8}, \ \mathsf{B}_{4,7}, \ \mathsf{B}_{5,6}, \ \mathsf{B}_{9,10}\}    
\end{equation}

\noindent and the set of relevant $\mathfrak{af}$ states is

\begin{equation}
\{ 6, \ 7, \ 8, \ 10\} \ . 
\end{equation}

\noindent The set of all $s_{\mathsf{B},Y}$ with $\emptyset=\mathsf{B} \cap Y$ is 
\begin{equation}
\begin{split}
\{ &s_{\mathsf{B}_{3,8}, \mathsf{B}_{9,10}}, \ s_{\mathsf{B}_{4,7},\mathsf{B}_{9,10}}, \ s_{\mathsf{B}_{5,6},\mathsf{B}_{9,10}},\ s_{\mathsf{B}_{5,6},7}, \ s_{\mathsf{B}_{5,6},8}, \\
& s_{\mathsf{B}_{5,6},10}, \ s_{\mathsf{B}_{4,7},8}, \ s_{\mathsf{B}_{4,7},10}, \ s_{\mathsf{B}_{3,8},10}, \ s_{\mathsf{B}_{9,10},6}, \ s_{\mathsf{B}_{9,10},7}, \ s_{\mathsf{B}_{9,10},8} \} \ .
\end{split}
\end{equation}

\noindent Removing the $s_{\mathsf{B},Y}$ which are not separated by enough flavor lines leaves

\begin{equation}
\pmb{H}^{A}_{n}[(1,2)(3,(4,(5,6),7),8)(9,10)]=\left\{\begin{array}{c}
-s_{\mathsf{B}_{5,6},\mathsf{B}_{9,10}}=c_{\mathsf{B}_{5,6},\mathsf{B}_{9,10}}, \ -s_{\mathsf{B}_{5,6},10}=c_{\mathsf{B}_{5,6},10}, \\
-s_{\mathsf{B}_{9,10},6}=c_{\mathsf{B}_{9,10},6}, \ -s_{\mathsf{B}_{9,10},7}=c_{\mathsf{B}_{9,10},7},  \\
-s_{\mathsf{B}_{4,7},10}=c_{\mathsf{B}_{4,7},10},-s_{\mathsf{B}_{5,6},8}=c_{\mathsf{B}_{5,6},8}
\end{array}\right\}\, .
\label{exampleconstrainttwo}
\end{equation}

\noindent A visualization of the surviving constraints in eq.~(\ref{exampleconstrainttwo}) is provided in Fig. \ref{fig:flavorseparations}. Note that $-s_{B_{3,8},10}=c_{B_{3,8},10}$ is NOT a constraint although the associated line intersects a flavor line, because the intersecting flavor line is associated with the $r=10$ state. \\
\indent We now sketch a general argument of eq.~(\ref{conjectdiffofr}). The goal of the argument is to show by induction that 

\begin{equation}
W(g|\kappa)|_{H^{T}_{n}[\alpha]}=0
\label{goalequ}
\end{equation}

\noindent if $W(g|\kappa)$ is inconsistent with $\alpha$. The strategy will be to manipulate $W(g|\kappa)$ of an incompatible graph and ordering into the form 

\begin{equation}
W(g|\kappa)=ds_{I}\wedge W(g'|\kappa')
\label{refereq}
\end{equation}

\noindent where $s_{I}$ is a constant if $I$ is not compatible with $\alpha$. Therefore, for $W$ to not immediately vanish, $I$ must be compatible with $\alpha$. We can then apply inductive arguments to $W'$, which corresponds to the differential of a reduced graph where states in $I$ are combined into a single state. The relevant ordering and restriction equations are now $\kappa'$ and $H[\alpha']$, which correspond to the factorization channel $s_{I}\rightarrow 0$. \\
\indent Given an arbitrary diagram, there must be at least one propagator of the form 

\begin{equation}
s_{l_{j}r_{j}}, \quad s_{i,k}, \quad s_{i,r_{j}}, \quad  s_{i_{1},i_{2}\ldots i_{m},l_{j}}
\label{allcasess}
\end{equation}

\noindent where $l_{j},r_{j}\neq 1,2$ in $\alpha$. For the first 3 propagators in eq.~(\ref{allcasess}), $s_{I}=\textrm{constant}$ unless $I$ is consistent with $\alpha$. Therefore, $W(g|\kappa)$ can be written as 

\begin{equation}
W(g|\kappa)=ds_{I}\wedge W(g'|\kappa')
\label{facotirchannels}
\end{equation}

\noindent where $W'$ corresponds to the reduced graph where states in $I$ are combined into a single external state. The case where $I=\{ i_{1},\ldots i_{m},l_{k} \}$ requires additional analysis as it is not immediatly obvious that $s_{I}$ is a constant if $I$ is inconsistent with $\alpha$. In the original graph, we assume that the $m$ $\mathfrak{adj}$ states directly coupled to the $l_{j}$ flavor line. If any of the $\mathfrak{adj}$ states have self-couplings, we can simply apply the inductive argument for $s_{i,j}$. We then apply the seven-term identity to the 4-point sub-graph with internal propagator $s_{ i_{1}\ldots i_{m}l_{k} }$. This leads to two differentials corresponding to two graphs: one with a propagator of the form $s_{i_{m}r_{j}}$ and the other with a propagator $s_{i_{1}i_{2}\ldots i_{m-1}l_{j}}$. Either both of the new diagrams are consistent with $\alpha$ or neither of them are. If both diagrams are consistent with $\alpha$, then $ds_{i_{m},r_{j}}|_{H_{n}[\alpha]}=-ds_{i_{1}i_{2} \ldots i_{m-1}l_{j}}|_{H_{n}[\alpha]}$ and the two terms cancel. Otherwise, we apply the previous argument to the diagram with the propagator $s_{i_{m}r_{j}}$, so it vanishes. Therefore, we are left with the differential associated to a diagram containing the propagator $s_{i_{1},i_{2}\ldots i_{m-1}l_{j}}$, which is again inconsistent with $\alpha$. We apply this procedure repeatedly, until the differential vanishes or we eventually have a differential of the form 
    
$$W'=ds_{l_{j}r_{j}}\wedge W''$$
    
\noindent where $W'$ is the differential of a diagram inconsistent with $\alpha$ ordering. We then apply the inductive assumption for $W''$, concluding the argument. \\
\indent There are a number of holes in the above argument. For example, we have not proven that restrictions on the factorization channel, $H_{n}^{T}[\alpha']$, have the same properties as $H^{T}_{n}[\alpha]$. This point was trivial in the bi-adjoint case, as the factorization channels were simply a direct product of closed associahedra. However, as discussed in section \ref{sec:remarks}, the geometry associated with factorization channels is more complicated for open associahedra. A more rigorous proof, with the required analysis of factorization channels, will be presented in~\cite{so2013ttaaa}.

\bibliographystyle{JHEP}
\bibliography{reference}

\end{document}